\documentclass[numsec,namedate,webpdf,traditional,medium]{supplementary}  %simplified .cls for supplementary

\onecolumn % for one column layouts

\graphicspath{{}}

\usepackage{graphicx}%
\usepackage{multirow}%
\usepackage{amsmath,amssymb,amsfonts}%
\usepackage{amsthm}%
\usepackage{mathrsfs}%
\usepackage{xcolor}%
\usepackage{textcomp}%
\usepackage{manyfoot}%
\usepackage{booktabs}%
\usepackage{algorithm}%
\usepackage{algorithmicx}%
\usepackage{algpseudocode}%
\usepackage{listings}%
\theoremstyle{thmstyleone}%
\newtheorem{proposition}{Proposition}[section]% to get separate numbers for theorem and proposition etc.
\newtheorem{remark}{Remark}[section]%
\usepackage{bm}
\usepackage{bbm}
\usepackage{dsfont}

\usepackage{adjustbox}
\usepackage{multirow}
\usepackage{multicol}
\usepackage{makecell}

\usepackage{mwe}
\newlength{\tempheight}
\newlength{\tempwidth}
\newcommand{\rowname}[1]% #1 = text
{\rotatebox{90}{\makebox[\tempheight][c]{{#1}}}}
\newcommand{\columnname}[1]% #1 = text
{\makebox[\tempwidth][c]{{#1}}}

\usepackage[figurename=Figure]{caption}
\usepackage{subcaption}

\usepackage{hyperref}
\hypersetup{
    colorlinks=true,
    linkcolor=blue,    
    urlcolor=darkgray,
    citecolor=blue
    }

\usepackage{setspace}


\begin{document}

\journaltitle{Biostatistics}
\DOI{}
\copyrightyear{}
\pubyear{}
\access{}
\appnotes{Article}

\firstpage{1}

\title[]{Regularized $k$-POD: Sparse $k$-means clustering for high-dimensional missing data}

\author[1,2,$\ast$]{Xin Guan}
\author[1,3]{Yoshikazu Terada}

%\authormark{Author Name et al.}

\address[1]{\doublespacing \orgdiv{Graduate School of Engineering Science}, \orgname{Osaka University}, \orgaddress{\street{1-3, Machikaneyamacho, Toyonaka}, \postcode{560-0043}, \state{Osaka}, \country{Japan}}}

\address[2]{\orgdiv{Graduate School of Information Sciences}, \orgname{Tohoku University}, \orgaddress{\street{6-3-09, Aramaki-Aza-Aoba-ku, Sendai}, \postcode{980-8579}, \state{Miyagi}, \country{Japan}}}

\address[3]{\orgdiv{AIP}, \orgname{RIKEN}, \orgaddress{\street{1-4-1, Nihonbashi, Chuo-ku}, \postcode{103-0027}, \state{Tokyo}, \country{Japan}}}

\corresp[$\ast$]{Corresponding author: \href{email:email-id.com}{Email: guan.xin.c5@tohoku.ac.jp}}

\abstract{
\begin{center}
    \bf ABSTRACT
\end{center}
\doublespacing 
The classical $k$-means clustering, based on distances computed from all data features, cannot be directly applied to incomplete data with missing values. A natural extension of $k$-means to missing data, namely $k$-POD, uses only the observed entries for clustering and is both computationally efficient and flexible. 
However, for high-dimensional missing data including features irrelevant to the underlying cluster structure, the presence of such irrelevant features leads to the bias of $k$-POD in estimating cluster centers, thereby damaging its clustering effect. 
Nevertheless, the existing $k$-POD method performs well in low-dimensional cases, highlighting the importance of addressing the bias issue. 
To this end, in this paper, we propose a regularized $k$-POD clustering method that applies feature-wise regularization on cluster centers into the existing $k$-POD clustering. Such a penalty on cluster centers enables us to effectively reduce the bias of $k$-POD for high-dimensional missing data. 
To the best of our knowledge, our method is the first to mitigate bias in $k$-means-type clustering for high-dimensional missing data, while retaining the computational efficiency and flexibility. 
Simulation results verify that the proposed method effectively reduces bias and improves clustering performance. 
Applications to real-world single-cell RNA sequencing data further show the utility of the proposed method. 
}
\keywords{Clustering; high-dimensional data; $k$-means; missing data}

\maketitle
\doublespacing

\section{Introduction}

The $k$-means clustering is one of the most widely-used clustering methods, which gives a partition based on the nearest cluster centers of each data point. However, the requirement for a full observed dataset of $k$-means limits its capacity for missing data. 
For example, in single-cell RNA sequence (scRNA-seq) data applications, cells need to be clustered to identify distinct cell subtypes or different stages of cell differentiation \citep{andrews2021tutorial}, while these data often contain missing values due to technical and sampling issues, making directly applying $k$-means clustering infeasible. 
A usual strategy is to apply common missing data handling techniques (such as complete-case analysis or multiple imputation), followed by performing the $k$-means clustering \citep{hathaway2001fuzzy}, whereas, such a two-stage strategy is not always effective, particularly when the number of complete data points is too small or the hidden probabilistic model about missingness is complicated and unknown \citep{morvan2021sa,niang2023clustering}. Moreover, multiple imputation such as \textit{mice} \citep{buuren2011mice} and \textit{Amelia} \citep{honaker2011amelia} can be computationally inefficient. 
In addition, primitively developed methods for $k$-means on missing data by \cite{wagstaff2004clustering} and \cite{datta2018clustering} use the partial distances that involve only the observed dimensions to modify the Euclidean distance used in classical $k$-means clustering, whereas the modified measurements for distance may not reflect the true structure based on all dimensions and may not even be a distance measure. 

Another intuitive strategy is to minimize the $k$-means loss over observed entries only, known as $k$-~POD (i.e., $k$-means for partial observed data) proposed by \cite{chi2016k}. 
In specific, consider a data matrix $\bm{X}=(x_{ij})_{n\times p}$ containing $n$ data points $\bm{x}_i$ in $\mathbb{R}^p$, and a set of indexes $\Omega\subset \{1,\dots,n\}\times \{1,\dots,p\}$ indicating all observed entries. A projection $\mathcal{P}$ onto an index set $\Omega$ is introduced to replace the missing entries with zero, that is, $[\mathcal{P}_{\Omega}(\bm{X})]_{ij}=x_{ij}$ if $(i,j)\in\Omega$, 0 otherwise. Then, the $k$-POD clustering is given by 
\begin{align}
\label{eq_rkpod_kpodloss}
    \min_{\bm{U},\bm{M}} \|\mathcal{P}_{\Omega}(\bm{X}-\bm{UM})\|_F^2
    \quad \text{such that}\quad \bm{U}\in\{0,1\}^{n\times k}\quad \text{and} \quad \sum_{l=1}^{k}u_{il}=1 \quad (i=1,\dots,n), 
\end{align}
where $\bm{U}=(u_{il})_{n\times k}$ is for cluster membership, where $u_{il}=1$ if $\bm{x}_i$ belongs to the $l$-th cluster, and $\bm{M}=(\mu_{lj})_{k\times p}$ takes the $l$-th cluster center as its $l$-th row, and $\|\bm{A}\|_F=\left(\sum_{i=1}^{n}\sum_{j=1}^{p}a_{ij}^2\right)^{1/2}$ is the Frobenius norm of a matrix $\bm{A}=(a_{ij})_{n\times p}$. 
When $\Omega=\{1,\dots,n\}\times \{1,\dots,p\}$, then Eq.~(\ref{eq_rkpod_kpodloss}) is equivalent to classical $k$-means clustering. 
Moreover, the minimization problem Eq.~(\ref{eq_rkpod_kpodloss}) can be solved by a simple and fast majorization-minimization algorithm \citep{hunter2004tutorial}, each iteration of which consists of an imputation step (imputing missing entries by the center of the cluster to which the corresponding data point belongs) and a clustering step (performing $k$-means on imputed data).
Therefore, the $k$-POD clustering is regarded as a natural and efficient extension of $k$-means to missing data, and has received much attention. 
For example, \cite{lithio2018efficient} proposed a variant of $k$-POD by using the Hartigan-Wong algorithm \citep{hartigan1979} to speed up the original Lloyd's algorithm \citep{Lloyd1982}, and \cite{aschenbruck2023imputation} extended the application of $k$-POD to mixed type data with missingness, and \cite{agliz2025joint} adopted the idea of $k$-POD for dimensional reduction. It should be mentioned that an equivalent expression of Eq.~(\ref{eq_rkpod_kpodloss}) was ever independently proposed by \cite{wang2019k}.  

However, \cite{terada2025a} theoretically show the inconsistency of $k$-POD even under the missing completely at random (MCAR) assumption. 
For a limited sample, this fundamental limitation implies a biased estimator of cluster centers by $k$-POD, making its clustering result unreliable. 
The essential reason for this bias lies in the fact that the loss of $k$-POD is equivalent to a weighted sum of the losses of $k$-means on all possible feature subsets (See Eq.~(\ref{eq_rkpod_kpodkm_weightedsum})). Some component losses have biased minimizers due to the existence of features irrelevant to the true cluster structure (noise features), resulting in the minimizer of $k$-POD loss no longer coinciding with the true centers. 
Particularly in high-dimensional cases, the ubiquitous presence of numerous noise features \citep{li2023sparse,zong2025model} usually causes a more serious bias in $k$-POD, thereby heavily damaging the performance of clustering. 

\begin{figure}[!h]
    \centering
    \begin{subfigure}{0.2352\textwidth}
        \includegraphics[width=\textwidth]{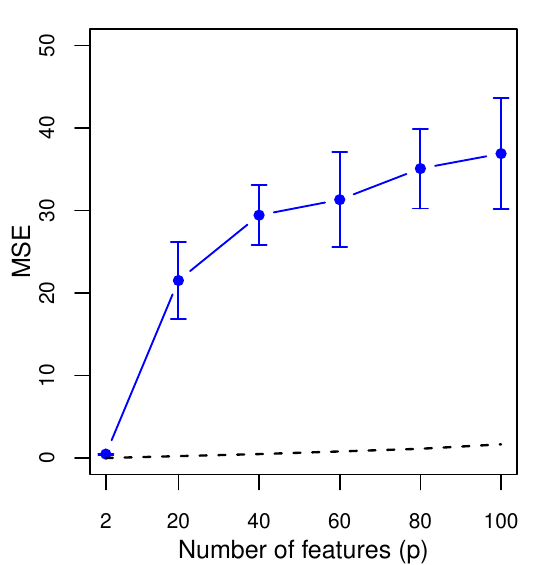}
        \caption*{(a) MSE of $k$-POD}
    \end{subfigure}
    \begin{subfigure}{0.2352\textwidth}
        \includegraphics[width=\textwidth]{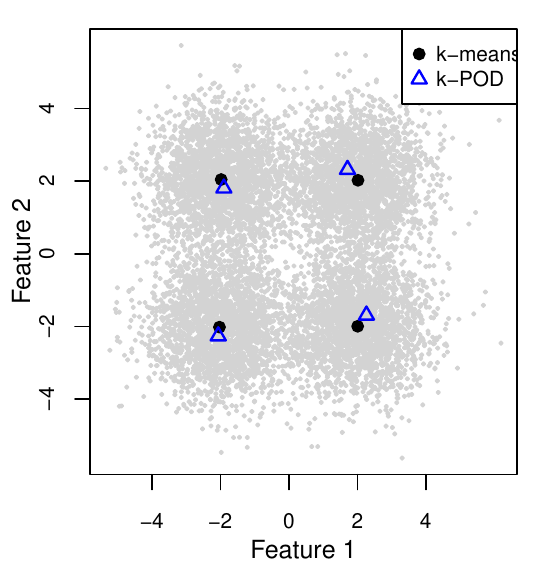}
        \caption*{(b) Estimated centers ($p=2$)}
    \end{subfigure}
    \begin{subfigure}{0.48\textwidth}
        \includegraphics[width=0.49\textwidth]{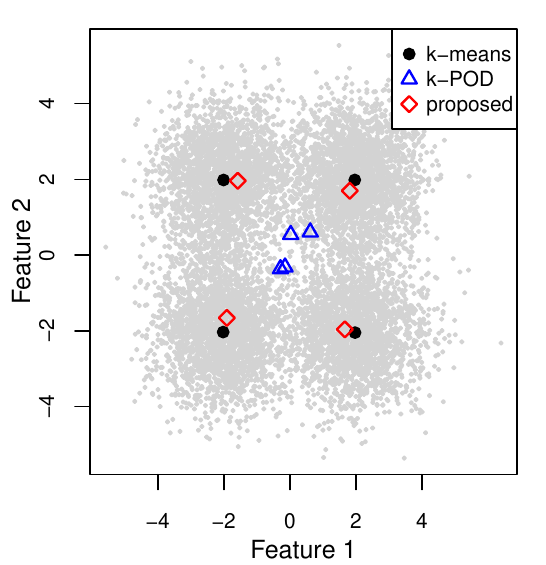}
        \includegraphics[width=0.49\textwidth]{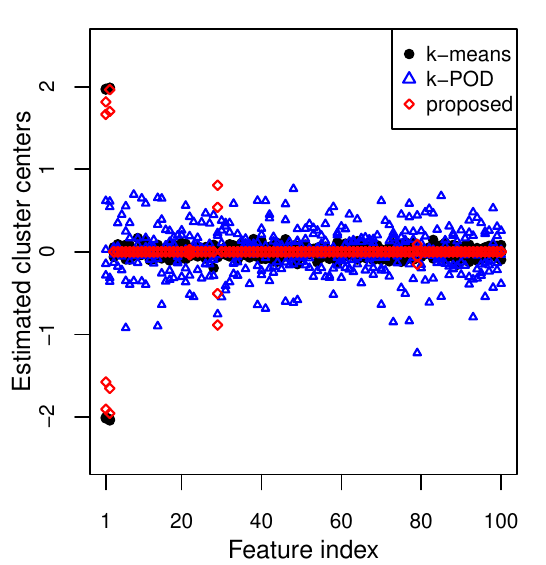}
        \caption*{(c) Estimated centers ($p=100$)}
    \end{subfigure}
    %\begin{subfigure}{0.24\textwidth}
    %    \includegraphics[width=\textwidth]{fig_p100k4_IntroEg_0715_kpod+rkpodg_estimator_in_R2.pdf}
    %    \caption*{(c) Estimator of $p=100$ \\ (first 2 features)}
    %\end{subfigure}
    %\begin{subfigure}{0.24\textwidth}
    %    \includegraphics[width=\textwidth]{fig_p100k4_IntroEg_0715_kpod+rkpodg_mu_lj.pdf}
    %    \caption*{(d) Estimator of $p=100$}
    %\end{subfigure}
    \caption{\doublespacing\normalfont (a) The MSEs of estimated centers of $k$-POD (blue line) for data with different numbers of features. The black dashed line is that of performing $k$-means on original full observed dataset. 
    (b) Estimated centers in the case of $p=2$, where the two axes are the first two features, and gray dots are original full observed data. 
    (c) Estimated centers in the case of $p=100$. Left: Two axes are the first two features. Right: The horizontal axis is the feature index and the vertical axis is the estimated value of centers. 
    }
    \label{rkpod_fig_introEG_p100_onlykpod}
\end{figure}

Figure~\ref{rkpod_fig_introEG_p100_onlykpod} illustrates the bias issue of $k$-POD in high-dimensional missing data, where the dataset contains $n=10000$ data points in $\mathbb{R}^{p}$ and consists of 4 clusters. Data points in each cluster are drawn independently from a Gaussian distribution. The 4 true cluster centers are given by $(\pm 2,\pm 2,0,\dots,0)$, implying only the first two features are relevant features while the rest $p-2$ are noise features. Whether each entry $x_{ij}$ is missing is completely at random and with the same missing probability of 30\%. 
The bias is measured by Mean-Squared Error (MSE) of estimated centers to the true centers.  
Figure~\ref{rkpod_fig_introEG_p100_onlykpod}(a) demonstrates the bias of $k$-POD on incomplete data with different numbers of features ($p$), where the average and standard deviation of 30 repetitions are reported. The results of estimated centers of $p=2$ and $p=100$ are illustrated in Figure~\ref{rkpod_fig_introEG_p100_onlykpod}(b) and (c), respectively. 
It can be seen that when $p=2$ (no noise features), the bias of $k$-POD is actually nearly zero, and the estimation result is very close to that of $k$-means on original full observed data. However, as the number of noise features increases, the bias of $k$-POD becomes larger. When $p=100$ (98 noise features), there exists a dramatic bias in almost every feature, implying that one can hardly obtain a reliable clustering result based on such biased estimated centers. 

Despite such bias in high-dimensional cases, the $k$-POD method performs well in the absence of noise features, and its computational efficiency and flexibility are attractive in real-world applications. 
Therefore, it is necessary to address the bias issue and improve the performance of $k$-POD. 
To this end, as suggested by the results of Figure~\ref{rkpod_fig_introEG_p100_onlykpod}, the key point is to eliminate those noise features. 
Fortunately, for classical $k$-means clustering, many effective methods have already been proposed to deal with noise features. 
For example, the framework of sparse $k$-means \citep{witten2010framework,chang2018sparse} uses the sparse estimation of feature weights to exclude noise features from clustering process, whereas it does not estimate cluster centers explicitly.  
Moreover, regularization techniques have also been applied to $k$-means \citep{sun2012regularized, raymaekers2022regularized}, which penalizes cluster centers feature by feature when performing $k$-means and explicitly yields a feature-sparse estimator of centers, thereby ensuring a smaller or even zero bias in noise features. 
Since such a penalty is not related to missingness, it shows us the possibility of bias mitigation for $k$-POD clustering through adopting similar regularization techniques. 

In this paper, we propose a novel clustering method by applying regularization to penalize cluster centers feature by feature while performing the $k$-POD clustering, named regularized $k$-POD clustering. 
Specifically, we add a penalty on cluster centers to the $k$-POD loss, consisting of the sum of $p$ regularization terms, where the $j$-th regularization term is the $l_0$ or $l_2$ norm of the $k$ cluster centers in the $j$-th feature. 
By selecting an appropriate regularization parameter, we obtain a feature-sparse estimator for cluster centers, and the optimization can be efficiently solved by the majorization-minimization algorithm. 
Consequently, as illustrated in Figure~\ref{rkpod_fig_introEG_p100_onlykpod}(c), in the case of high-dimensional missing data, our method (red diamonds) has less bias, and thus effectively improves the performance of existing $k$-POD clustering and meanwhile retains its computational efficiency and flexibility. 
Furthermore, to the best of our knowledge, our method is the first to mitigate bias in $k$-means-type clustering for high-dimensional missing data. 

The rest of this paper is organized as follows. In Section~\ref{rkpod_sec_methodology}, we propose the regularized $k$-POD clustering in Section~\ref{rkpod_sec_proposedmethod}, where two types of regularization are considered. The optimization of the proposed method is given in Section~\ref{rkpod_sec_optimization}, including the iterative algorithm, initialization strategy, and selection of tuning parameter. We also provide theoretical properties about the regularized $k$-POD clustering solution in Section~\ref{rkpod_sec_property}. 
In Section~\ref{rkpod_sec_experiments}, we verify the improved performance of the proposed method by comparing with existing methods via simulations. In Section~\ref{rkpod_sec_application}, we demonstrate the utility of the proposed method to real-world datasets, including two scRNA-seq datasets from neuronal cells subtypes and differentiation studies. 
In Section~\ref{rkpod_sec_discussion}, we provide the conclusion and discussions.

\clearpage
\section{Methods}
\label{rkpod_sec_methodology}

Let $\bm{X}=(x_{ij})_{n\times p}\in \mathbb{R}^{n\times p}$ be a matrix containing $n$ data points of dimension $p$. Write $\bm{x}_i=(x_{i1},\dots,x_{ip})\in \mathbb{R}^{1\times p}$ for the $i$-th row of $\bm{X}$ (i.e., the $i$-th data point), and write $\bm{x}_{(j)}=(x_{1j},\dots,x_{nj})^T\in \mathbb{R}^{n\times 1}$ for the $j$-th column of $\bm{X}$. 
Let $\bm{M}=(\mu_{lj})_{k\times p} \in \mathbb{R}^{k\times p}$ be a matrix containing $k$ cluster centers of dimension $p$. Write $\bm{\mu}_l=(\mu_{l1},\dots,\mu_{lp})\in \mathbb{R}^{1\times p}$ for the $l$-th row of $\bm{M}$ (i.e., the $l$-th cluster center), and write $\bm{\mu}_{(j)}=(\mu_{1j},\dots,\mu_{kj})^T\in \mathbb{R}^{k\times 1}$ for the $j$-th column of $\bm{M}$.
Let $\bm{U}=(u_{il})_{n\times k}\in \{0,1\}^{n\times k}$ be a matrix indicating the assignment relationship between data points and clusters, where $u_{il}=1$ if and only if the $i$-th data point $\bm{x}_i$ is assigned to the $l$-th cluster. Since one data point is assigned to a unique cluster, through this paper, we suppose $\bm{U}\bm{1}_k=\bm{1}_n$, where $\bm{1}_k=(1,\dots,1)^T \in \mathbb{R}^{k\times 1}$ and $\bm{1}_n=(1,\dots,1)^T \in \mathbb{R}^{n\times 1}$. 
Let $\bm{R}=(r_{ij})_{n\times p}\in \{0,1\}^{n\times p}$ be a matrix indicating the missingness of each entry, where $r_{ij}=1$ if $x_{ij}$ is observed, 0 otherwise. The $i$-th row and the $j$-th column of $\bm{R}$ are denoted by $\bm{r}_{i}$ and $\bm{r}_{(j)}$, respectively. 
Denote by $\|\bm{a}\|$ the $l_2$ norm of a vector $\bm{a}=(a_1,\dots,a_p)$, that is, $\|\bm{a}\|=\left(\sum_{j=1}^{p} a_j^2\right)^{1/2}$. 
Denote by $\|\bm{A}\|_F$ the Frobenius norm of a matrix $\bm{A}=(a_{ij})_{n\times p}$, that is, $\|\bm{A}\|_F=\left(\sum_{i=1}^{n}\sum_{j=1}^{p}a_{ij}^2\right)^{1/2}$. 
The $\mathds{1}(\cdot)$ is the indicator function, and $\bm{I}_k$ is the identity matrix with size $k\times k$. 
Through this paper, we suppose that the number of clusters $k\geq 2$ is fixed and known, and the $k$ true cluster centers in $\mathbb{R}^p$ denoted by $\bm{\mu}_1^{\ast},\dots,\bm{\mu}_k^{\ast}$ are determined by the minimizer of expected loss function of $k$-means, and thus only rely on the true distribution of data, but rather unrelated to the sampled dataset. 
We call a feature $j$ ($j=1,\dots,p$) noise feature, if $\mu_{1j}^{\ast}=\dots=\mu_{kj}^{\ast}=0$.

\subsection{The proposed method}
\label{rkpod_sec_proposedmethod}

We define the loss function of regularized $k$-POD clustering with respect to membership $\bm{U}\in \{0,1\}^{n\times k}$, $\bm{U}\bm{1}_k=\bm{1}_n$, and cluster centers $\bm{M}\in \mathbb{R}^{k\times p}$ to be
\begin{align}
    \widehat{L}_n(\bm{U},\bm{M})=\|\mathcal{P}_{\Omega}(\bm{X}-\bm{UM})\|_F^2 + \lambda\cdot J(\bm{M}).
    \label{eq_rkpod_loss}
\end{align}
The first term is the loss of the $k$-POD clustering, and $J(\bm{M})$ is a regularization function with respect to $\bm{M}$. To shrink the estimated cluster centers feature-wisely, we consider two types of $J(\bm{M})$: 
\begin{align*}
    J_0(\bm{M})=\sum_{j=1}^{p}\mathds{1}(\|\bm{\mu}_{(j)}\|>0) \quad \textnormal{ and }\quad
   J_1(\bm{M})=\sum_{j=1}^p w_j \|\bm{\mu}_{(j)}\|. 
\end{align*}
Both types of $J(\cdot)$ are column-wised, which means that all elements of $\bm{\mu}_{(j)}$, that is $\{\mu_{1j},\dots,\mu_{kj}\}$ would be shrunk together. The $l_0$ type $J_0(\cdot)$ constrains the number of non-zero columns of $\bm{M}$, and the group lasso type $J_1(\cdot)$ constrains the weighted sum of $l_2$ norm of $\bm{M}$ in each feature. 

Since the regularization parameter $\lambda$ and the weights $w_j$ used in $J_1(\bm{M})$ play a crucial role in recognizing noise features, we provide two criteria for choosing $\lambda$ in Section~\ref{rkpod_sec_method_tuning} and introduce the construction of $w_j$ here. 
We note that in the framework of group lasso regression, a common choice for $w_j$ is based on the square root of the size of the $j$-th group \citep{yuan2006model, yang2015fast}, which means a uniform weight $w_j=\sqrt{k}$ in our case. 
However, as shown in Figure~\ref{rkpod_fig_introEG_p100_onlykpod}, the bias of the $k$-POD estimator in each feature is different, which implies that the adaptive weights are more reasonable. 
Specifically, we here consider the weights based on the $k$-POD estimator $\widetilde{\bm{M}}$, that is, $w_j=\|\widetilde{\bm{\mu}}_{(j)}\|^{-1}$. 
If the estimated cluster centers of the $k$-POD clustering in a feature are relatively concentrated, the corresponding weight would be relatively large, which makes the group lasso estimator in the corresponding feature more likely to be zero. 

In addition, when the data matrix $\bm{X}$ is fully observed, then $\Omega=\{1,\dots,n\}\times\{1,\dots,p\}$ and the loss of the proposed method is equivalent to that of the regularized $k$-means clustering \citep{sun2012regularized,raymaekers2022regularized}. Therefore, the proposed method can also be viewed as an extension of the regularized $k$-means clustering to missing data. 
Moreover, the formulation of Eq.~(\ref{eq_rkpod_loss}) is similar to the loss function of low-rank matrix completion \citep{jain2013low}, which aims to recover the partially observed matrix, whereas, we emphasize that our primary concern is clustering and estimating cluster centers, instead of imputing, and we also need the factor matrices to have special structures.

\begin{remark}
    The capability of the proposed method to cope with bias caused by noise features can be interpreted as follows.  
    In fact, the loss of $k$-POD can be expanded as
\begin{align}
    \widehat{L}_n^{\textnormal{(KPOD)}}(\bm{M})=\frac{1}{n}\|\mathcal{P}_{\Omega}(\bm{X}-\bm{UM})\|_F^2
    %&=\frac{1}{n}  \sum_{i=1}^{n}\left\{ \sum_{l=1}^{k} u_{il} \sum_{j:r_{ij}=1}  (x_{ij} - \mu_{lj})^2 \right\} \notag\\
    &=\sum_{\bm{\xi}\in \{0,1\}^p } \frac{n_{\bm{\xi}}}{n} \underbrace{  \left[  \frac{1}{n_{\bm{\xi}}} \sum_{i:\bm{r}_i=\bm{\xi}}  \left\{  \min_{l=1,\dots,k} \sum_{j:\xi_j=1}  (x_{ij} - \mu_{lj})^2 \right\} \right] }_{\widehat{L}_n^{\text{(KM)}}(\bm{M}\;|\;\bm{\xi}) }  ,
    \label{eq_rkpod_kpodkm_weightedsum}
\end{align}
where $\bm{\xi}=(\xi_1,\dots,\xi_p)\in \{0,1\}^p$ is a vector corresponding to a missing pattern, indicating a subset of observed features by its non-zero elements. 
For a fixed $\bm{\xi}$, the $n_{\bm{\xi}}/n=n^{-1}\sum_{i=1}^{n} \mathds{1}(\bm{r}_i=\bm{\xi})$ is the ratio of data points satisfying the missing pattern given by $\bm{\xi}$, where $\bm{r}_i=(r_{i1},\dots,r_{ip})\in \{0,1\}^p$ satisfies $r_{ij}=1$ if $x_{ij}$ is observed.  
The $\widehat{L}_n^{\text{(KM)}}(\bm{M}\;|\;\bm{\xi})$ is the loss function of $k$-means on subset of features defined by non-zero components of $\bm{\xi}$. 
For a $p$-dimensional data, there exist $2^p$ different missing patterns. 
As such, Eq.~(\ref{eq_rkpod_kpodkm_weightedsum}) means that the $k$-POD loss on a $p$-dimensional data is equivalent to a weighted sum of $2^p$ $k$-means losses on different subsets of features defined by different missing patterns. 
In the presence of noise features, some component losses have biased minimizers, such as losses on feature subsets that contain only noise features, which leads to the biased minimizer of $k$-POD. 
However, our method essentially considers a constraint for Eq.~(\ref{eq_rkpod_kpodkm_weightedsum}), as the regularization penalty term in Eq.~(\ref{eq_rkpod_loss}) is equivalent to a constraint on cluster centers. 
Such constraint allows us to correct those component losses that have biased minimizers, that is, de-bias $k$-means on those feature subsets that contain noise features, which has theoretical guarantees. Consequently, we are able to mitigate the bias and improve the effect of the existing $k$-POD method. 
\end{remark}

\subsection{Optimization}
\label{rkpod_sec_optimization}

\subsubsection{Algorithms}

We apply the majorization-minimization algorithm (MM algorithm) \citep{hunter2004tutorial} to minimize the proposed loss function Eq.~ (\ref{eq_rkpod_loss}). 
The MM algorithm constructs a majorization function $g(\theta \mid \theta^{(t)})$ for the original objective function $L(\theta)$ at the current value $\theta^{(t)}$, $t\in \mathbb{N}$. The majorization means that the domination condition $g(\theta \mid \theta^{(t)})\geq L(\theta)$ and the tangency condition $g(\theta^{(t)}\mid \theta^{(t)})=L(\theta^{(t)})$ are satisfied. Then update $\theta^{(t+1)}$ by minimizing $g(\theta \mid \theta^{(t)})$ instead of $L(\theta)$, which also guarantees $L(\theta^{(t+1)})\leq L(\theta^{(t)})$. 

Our goal is to minimize $\widehat{L}_n(\bm{U},\bm{M})$ of Eq.~(\ref{eq_rkpod_loss}) with respect to $(\bm{U},\bm{M})$. We define the following function at current value $(\bm{U}^{(t)},\bm{M}^{(t)})$, $t\in \mathbb{N}$: 
\begin{align*}
        g( \bm{U},\bm{M} \mid \bm{U}^{(t)},\bm{M}^{(t)} ) = \|\mathcal{P}_{\Omega}(\bm{X}-\bm{U}\bm{M} )\|_F^2 + \lambda\cdot J(\bm{M}) + \|\mathcal{P}_{\Omega^c}( \bm{U}\bm{M} - \bm{U}^{(t)}\bm{M}^{(t)}  ) \|_F^2, 
\end{align*}
where $\Omega^c$ is the complement set of $\Omega$. 
Because of the non-negativity of $\|\cdot\|_F^2$, the function $g( \bm{U},\bm{M} \mid \bm{U}^{(t)},\bm{M}^{(t)} )$ is a majorization function of $\widehat{L}_n(\bm{U},\bm{M} )$ in the sense that
\begin{align*}
    g( \bm{U},\bm{M} \mid \bm{U}^{(t)},\bm{M}^{(t)} )&\geq \widehat{L}_n(\bm{U},\bm{M}) &\text{(domination condition)}\\
    g( \bm{U}^{(t)},\bm{M}^{(t)} \mid \bm{U}^{(t)},\bm{M}^{(t)} )&=\widehat{L}_n(\bm{U}^{(t)},\bm{M}^{(t)}) & \text{(tangency condition)}
\end{align*}
are both satisfied. 
If we use the notation $\widehat{\bm{X}}= \mathcal{P}_{\Omega}(\bm{X} ) + \mathcal{P}_{\Omega^c}(\bm{U}^{(t)}\bm{M}^{(t)})$, then we have $g( \bm{U},\bm{M} \mid \bm{U}^{(t)},\bm{M}^{(t)} ) = \| \widehat{\bm{X}} - \bm{U}\bm{M} \|_F^2 + \lambda\cdot J(\bm{M})$. Notice that the matrix $\widehat{\bm{X}}$ is complete, then $g( \bm{U},\bm{M} \mid \bm{U}^{(t)},\bm{M}^{(t)} )$ is actually the loss function of regularized $k$-means clustering on the data matrix $\widehat{\bm{X}}$. We then minimize the majorization function $g( \bm{U},\bm{M} \mid \bm{U}^{(t)},\bm{M}^{(t)} )$ to update $(\bm{U}^{(t+1)},\bm{M}^{(t+1)})$. 

Therefore, we propose Algorithm \ref{alg_rkpod} for regularized $k$-POD clustering. 
Specifically, given current $\bm{U}^{(t)}$ and $\bm{M}^{(t)}$, $t\in \mathbb{N}$, the $(t+1)$-th iteration consists of two steps. Step 1 imputes missing entries of $\bm{X}$ by the corresponding entries of multiplication matrix of current $\bm{U}^{(t)}$ and $\bm{M}^{(t)}$, so that we can get a new complete data matrix $\widehat{\bm{X}}^{(t+1)}$. Step 2 updates $\bm{U}^{(t+1)}$ and $\bm{M}^{(t+1)}$ by applying regularized $k$-means clustering on the imputed data matrix $\widehat{\bm{X}}^{(t+1)}$, the details of which is discussed later. Repeat the iteration until the loss (Eq.~ (\ref{eq_rkpod_loss})) converges. 
Note that Algorithm \ref{alg_rkpod} is a general framework for any type of $J(\cdot)$, and the difference in results comes from Step 2.

\begin{algorithm}
\doublespacing
\caption{\; Regularized $k$-POD clustering}
\textbf{Input}: incomplete data matrix $\bm{X}$, set of observed positions $\Omega$, number of clusters $k$. \\
\textbf{Parameters}: regularization parameter $\lambda$, weights $\{w_j\}$ 
\begin{algorithmic}
 \State Initialize $\bm{U}^{(0)}$ and $\bm{M}^{(0)}$ 
 \While{Loss function Eq.~(\ref{eq_rkpod_loss}) does not converge}
  \State 1: Impute $\widehat{\bm{X}}^{(t+1)}=\mathcal{P}_{\Omega}(\bm{X})+\mathcal{P}_{\Omega^c}(\bm{U}^{(t)}\bm{M}^{(t)})$
  \State 2: Update $\bm{U}^{(t+1)}$ and $\bm{M}^{(t+1)}$ by applying Algorithm~\ref{alg_rkm} on $\widehat{\bm{X}}^{(t+1)}$ 
 \EndWhile
\end{algorithmic}
\textbf{Output}: $\bm{U}^{(t+1)}$ and $\bm{M}^{(t+1)}$
\label{alg_rkpod}
\end{algorithm}

The convergence of Algorithm \ref{alg_rkpod} to a local minima is guaranteed by the downhill trend 
\begin{align*}
        \widehat{L}_n(\bm{U}^{(t+1)},\bm{M}^{(t+1)})\leq \widehat{L}_n(\bm{U}^{(t)},\bm{M}^{(t)})
    \end{align*}
for any $t\in \mathbb{N}$. 
This is the immediate consequence of the domination condition, tangency condition, and the definition of $(\bm{U}^{(t+1)},\bm{M}^{(t+1)})$, which implies that
\begin{align*}
    g(\bm{U}^{(t+1)},\bm{M}^{(t+1)} \mid \bm{U}^{(t)},\bm{M}^{(t)})  \leq g(\bm{U}^{(t)},\bm{M}^{(t)} \mid \bm{U}^{(t)},\bm{M}^{(t)}).     
\end{align*}
According to our numerical experiments, the necessary number of iterations to convergence of the proposed method is generally comparable with that of the $k$-POD clustering.

Next, we introduce more details of Step 2 of Algorithm~\ref{alg_rkpod}, where we apply regularized $k$-means clustering on imputed data matrix $\widehat{\bm{X}}^{(t+1)}$. For the simplification of notations, we here omit the superscript $(t+1)$ and focus on the general imputed complete data matrix $\widehat{\bm{X}}$. The goal of Step 2 of Algorithm~\ref{alg_rkpod} is to solve 
\begin{align}
    \min_{\bm{U},\bm{M}} \|\widehat{\bm{X}} - \bm{U}\bm{M}\|_F^2 + \lambda\cdot J(\bm{M}),
    \label{eq_rkm_loss}
\end{align}
with respect to $\bm{U}\in \{0,1\}^{n\times k}$, $\bm{U}\bm{1}_k=\bm{1}_n$ and $\bm{M}\in \mathbb{R}^{k\times p}$. 
Since it is not necessarily convex, an alternatively iterative procedure similar to Lloyd's algorithm \citep{Lloyd1982} for classical $k$-means clustering can be used. 
Therefore, we propose Algorithm~\ref{alg_rkm} for this problem, which updates $\bm{U}$ and $\bm{M}$ separately. 
Specifically, given current $\bm{M}^{(r)}$, $r\in \mathbb{N}$, the membership $\bm{U}^{(r+1)}$ is determined by the distance between data points $\hat{\bm{x}}_i$ and cluster centers $\bm{\mu}_l^{(r)}$, that is, $u_{il^{\ast}}^{(r+1)}=1$ if $l^{\ast}=\arg\min_{1\leq l \leq k} \| \hat{\bm{x}}_{i} - \bm{\mu}_l^{(r)} \|^2 $, 0 otherwise. Then, given $\bm{U}^{(r+1)}$, updating $\bm{M}^{(r+1)}$ depends on the types of $J(\cdot)$.  

For $J=J_0$, the $l_0$ type, applying the KKT condition immediately leads to an explicit solution given by Eq.~(\ref{rkpod_eq_alg2_0}), where $\bm{0}_k$ is the all-zero vector in $\mathbb{R}^k$. This is a truncated version of the cluster means associated with current membership $\bm{U}^{(r+1)}$. 
For $J=J_1$, the group lasso type, since it is hard to derive an explicit expression, we apply the MM algorithm again to get $\bm{M}^{(r+1)}$. Denote by $f(\bm{M})$ the objective function in Eq.~(\ref{eq_rkm_loss}) with $\bm{U}=\bm{U}^{(r+1)}$ fixed and $J=J_1$, that is, 
\begin{align*}
    f(\bm{M})=\|\widehat{\bm{X}} - \bm{U}^{(r+1)}\bm{M}\|_F^2 + \lambda \sum_{j=1}^{p}w_j \|\bm{\mu}_{(j)}\|. 
\end{align*}
At current $\bm{M}^{(r)}$, we define the following function:  
\begin{align*}
        h(\bm{M} \mid  \bm{M}^{(r)})=\|\widehat{\bm{X}} - \bm{U}^{(r+1)} \bm{M}\|_F^2 + \lambda \sum_{j=1}^{p} w_j \left( \frac{\|\bm{\mu}_{(j)}\|^2}{ 2\|\bm{\mu}_{(j)}^{(r)}\|  }  +  \frac{1}{2}\|\bm{\mu}_{(j)}^{(r)}\| \right).
\end{align*}
It can be proved that $h(\bm{M} \mid  \bm{M}^{(r)})$ is a majorization of $f(\bm{M})$ at $\bm{M}^{(r)}$. Moreover, the solution of minimizing $h(\bm{M} \mid  \bm{M}^{(r)})$ is explicit and given by Eq.~(\ref{rkpod_eq_alg2_g}), where 
$\bm{I}_k$ is the identical matrix with the size of $k\times k$. This can be viewed as a ridge version of the cluster means associated with the given membership $\bm{U}^{(r+1)}$, and we use this solution as the update $\bm{M}^{(r+1)}$. 

\begin{algorithm}
\doublespacing
\caption{\ Regularized $k$-means clustering}
\textbf{Input}: complete data matrix $\widehat{\bm{X}}$, number of clusters $k$. \\
\textbf{Parameters}: regularization parameter $\lambda$, weights $\{w_j\}$
\begin{algorithmic}
 \State Initialize $\bm{M}^{(0)}$ 
 \While{Loss function Eq.~(\ref{eq_rkm_loss}) does not converge}
  \State a: Given $\bm{M}^{(r)}$, update $\bm{U}^{(r+1)}$ by: for any $i=1,\dots,n$
  \begin{align*}
      u_{il^{\ast}}^{(r+1)}=\left\{
      \begin{array}{ll}
          1 & \text{  if  }  l^{\ast}=\arg\min_{1\leq l \leq k} \| \hat{\bm{x}}_{i} - \bm{\mu}_l^{(r)} \|^2 \\
          0 & \text{  else}
      \end{array}
      \right. 
  \end{align*}
  \State b: Given $\bm{U}^{(r+1)}$, update $\bm{M}^{(r+1)}$ by: for any $j=1,\dots,p$ 
  \begin{align}
        (J=J_0)\quad & \bm{\mu}_{(j)}^{(r+1)}=\left\{ 
      \begin{array}{ll}
           \bm{v}_{(j)} & \text{  if  } \|\hat{\bm{x}}_{(j)}\|^2 > \| \hat{\bm{x}}_{(j)} - \bm{U}^{(r+1)}\bm{v}_{(j)} \|^2 + \lambda  \\
           \bm{0}_k & \text{  else  }
      \end{array}
      \right.   \label{rkpod_eq_alg2_0}\\ 
      &\text{where  }\bm{v}_{(j)}=\left( \bm{U}^{(r+1),T}\bm{U}^{(r+1)} \right)^{-1}  \bm{U}^{(r+1),T}  \hat{\bm{x}}_{(j)} \notag\\
     (J=J_1)\quad & \bm{\mu}_{(j)}^{(r+1)}=\left( \bm{U}^{(r+1),T}\bm{U}^{(r+1)} + \frac{\lambda w_j}{2 \|\bm{\mu}_{(j)}^{(r)} \|}\cdot \bm{I}_k \right)^{-1}  \bm{U}^{(r+1),T}  \hat{\bm{x}}_{(j)} 
     \label{rkpod_eq_alg2_g}
  \end{align}
 \EndWhile
\end{algorithmic}
\textbf{Output}: $\bm{U}^{(r+1)}$ and $\bm{M}^{(r+1)}$
\label{alg_rkm}
\end{algorithm}

We give the following remarks for the update of $\bm{M}^{(r+1)}$ when $J=J_1$ and leave the technical details of Algorithm~\ref{alg_rkm} in Section~A of Supplementary materials. 
\begin{remark}
    The standard way to get $\bm{M}^{(r+1)}$ by MM algorithm is to do another iteration, that is, minimize $h(\bm{M} \mid  \bm{M}^{(r_s)})$ on a sequence $\{\bm{M}^{(r_0)},\bm{M}^{(r_1)},\dots,\bm{M}^{(r_s)}\}$ about $s\in\mathbb{N}$ until convergence, which largely increases the computational cost. However, the multiple iteration for $s$ is not necessary, since an update $\bm{M}^{(r+1)}$ that reduces $f(\bm{M})$ is enough. Therefore, we can directly define $h(\bm{M} \mid  \bm{M}^{(r)})$ based on current $\bm{M}^{(r)}$, and take the solution of minimizing $h(\bm{M} \mid  \bm{M}^{(r)})$ to be the update $\bm{M}^{(r+1)}$. The optimality as well as majorization immediately implies $f(\bm{M}^{(r+1)})\leq f(\bm{M}^{(r)})$. In this way, we can decrease the number of embedded loops and speed up the whole algorithm. 
\end{remark}
\begin{remark}
    The minimization problem for $f(\bm{M})$ can be viewed as a group lasso regression of $\widehat{\bm{X}}$ on $\bm{U}^{(r+1)}$. Some existing literature that also considers MM algorithm uses the majorization based on a quadratic upper bound of $\|\widehat{\bm{X}} - \bm{U}^{(r+1)} \bm{M}\|_F^2$ (e.g.: \cite{yang2015fast}). Instead, we here use the upper bound of the penalty term $\lambda \sum_{j=1}^{p} w_j \|\bm{\mu}_{(j)}\|$ based on the basic inequality. 
    According to comparisons provided in Supplementary materials, the performance of these two methods is quite similar. Refer to Section~A.3 of Supplementary materials for more details. 
\end{remark}

Finally, we analyze the computation complexity of the proposed algorithm. In Step~1 of Algorithm~\ref{alg_rkpod}, imputing missing entries requires a complexity of $O(nkp+np(1-q))$, where $q$ is the proportion of observed entries. In Step 2, updating $\bm{U}$ and $\bm{M}$ has the same complexity as the classical $k$-means clustering, i.e., $O(nkp\tau)$, where $\tau$ is the total number of iterations in Algorithm~\ref{alg_rkm}. Therefore, the asymptotic complexity of each iteration of the proposed algorithm is nearly $O(nkp\tau)$.

\subsubsection{Initialization}

Although the proposed algorithm has the guarantee to converge to some local minima, the multiple initialization should be considered, since the loss function of the proposed method is not necessarily convex with respect to $\bm{U}$ and $\bm{M}$. 
In this paper, we consider two strategies to generate random initialization of $(\bm{U}^{(0)},\bm{M}^{(0)})$. 

The first strategy is based on the complete cases, which is referred to as {\fontfamily{qcr}\selectfont comp}. Specifically, we apply the $k$-means++ clustering \citep{arthur2007k} on the submatrix of $\bm{X}$ that only includes complete rows to obtain initial cluster centers $\bm{M}^{(0)}$. Then, the initial membership $\bm{U}^{(0)}$ is based on the Euclidean distances between data points and initial cluster centers. It should be noted that only the observed features are used to calculate the distance. 

The second strategy is based on imputation, which is referred to as {\fontfamily{qcr}\selectfont impt}. Specifically, we pre-impute the incomplete data matrix $\bm{X}$ by column-wised sample means without considering missing entries. Then, we randomly sample $k$ rows from the pre-imputed data matrix as the initial cluster centers $\bm{M}^{(0)}$. The initial membership $\bm{U}^{(0)}$ is based on the Euclidean distances between data points and initial cluster centers. It should be noted that if there are duplicated rows in $\bm{M}^{(0)}$, some small noise is added to it to ensure $k$ unique cluster centers. 

\begin{remark}
    The two strategies use unique $k$ random points to be initial $k$ cluster centers and initialize membership based on them. According to our experiments, the empirical choice for the number of initialization is at least 100 to get more stable results. In the case of high-dimension or a large proportion of missingness, to reduce the computation cost, the sparse initialization \citep{raymaekers2022regularized} can be used as an alternative. For example, based on the estimator by $k$-POD clustering, we can get several sparse submatrices of it by remaining columns with leading $l_1$ norms and letting others be zero, and then use these sparse submatrices to be initial cluster centers. Refer to Section~D.2 of Supplementary materials for more details. 
\end{remark}

\subsubsection{Selection of tuning parameters}
\label{rkpod_sec_method_tuning}

To select the tuning parameter, that is, the regularization parameter $\lambda$, we consider two kinds of criteria. 

The first criterion is the instability of clustering \citep{wang2010consistent}.  
The main idea is that a good value for the tuning parameter should yield a stable clustering in response to minor disruption to the sample. 
The instability of a clustering algorithm $\bm{\psi}$ with tuning parameter $\lambda$ is defined as 
\begin{align*}
    s(\bm{\psi};\lambda)=\mathbb{E}_{\bm{\mathrm{X}},\bm{\mathrm{X}}'}\left[ D\left( \bm{\psi}(\bm{\mathrm{X}};\lambda), \bm{\psi}(\bm{\mathrm{X}}';\lambda)
  \right) \right],
\end{align*}
where $\bm{\mathrm{X}}$ and $\bm{\mathrm{X}}'$ are two independent samples from the same distribution, and $\bm{\psi}(\bm{\mathrm{X}};\lambda)$ and $\bm{\psi}(\bm{\mathrm{X}}';\lambda)$ are two clustering trained on $\bm{\mathrm{X}}$ and $\bm{\mathrm{X}}'$, respectively. 
The notation $D(\cdot,\cdot)$ is the distance between two clusterings, which is given by the probability of the disagreement between them, that is, 
\begin{align*}
    D(\psi_1,\psi_2)=\text{Pr}\left[\mathds{1}(\psi_1(\bm{\mathrm{x}})=\psi_1(\bm{\mathrm{x}}'))+\mathds{1}(\psi_2(\bm{\mathrm{x}})=\psi_2(\bm{\mathrm{x}}'))=1\right],
\end{align*}
where $\bm{\mathrm{x}}$ and $\bm{\mathrm{x}}'$ are two random variables independently sampled from the same distribution, $\psi_1$ and $\psi_2$ are two clusterings, and $\psi(\bm{x})$ indicates the cluster that the data point $\bm{x}$ is assigned to.  
The instability index is calculated as follows. First, the dataset with sample size $n$ is randomly divided into three subsets, two of which consist of $m$ data points as training sets and the third one as validation set. Second, we conduct the proposed clustering method with some $\lambda$ on both training sets separately and obtain two estimators of cluster centers. Third, based on the two estimators, we predict corresponding labels for the validation set. Finally, we calculate the disagreement between two prediction results. Repeat the procedure several times, the instability index for the corresponding value of $\lambda$ is given by the averaged disagreement. 
In addition, when the sample size $n$ is small, the random division would make training sets too small. The bootstrap sampling can be an alternative to generate training and validation sets \citep{fang2012selection}. 

The second criterion is the BIC index. Inspired by \cite{raymaekers2022regularized, hofmeyr2020degrees}, we use the following formulation:  
\begin{align}
    \text{BIC}(\lambda)=\| \mathcal{P}_{\Omega}(\bm{X}- \widehat{\bm{U}}\widehat{\bm{M}})  \|_F^2 + \log(n)\cdot k\cdot d,
\end{align}
where $\widehat{\bm{U}}$ and $\widehat{\bm{M}}$ are estimators based on $\lambda$ and $d=\sum_{j=1}^{p}\mathds{1}(\|\hat{\bm{\mu}}_{(j)}\|>0)$ is the number of non-zero columns of $\widehat{\bm{M}}$. The first term corresponds to the log-likelihood according to \cite{fraley2002model}, while the second term is the degree of freedom, for which we use the number of independent parameters $kd$ since the membership can be determined by cluster centers. More details are provided in Section~B of Supplementary materials. 

For a set of values for $\lambda$, we select the best one with the smallest instability or BIC.

\subsection{Theoretical properties}
\label{rkpod_sec_property}
In this section, we further analyze some properties of the proposed method. For simplification, we assume that whether each entry is missing is completely at random. Then, by using $\bm{R}=(r_{ij})_{n\times p}\in \{0,1\}^{n\times p}$ to indicate whether the $(i,j)$-th entry is observed, the incomplete data matrix can be expressed by $\bm{X}\circ \bm{R}$, where $\circ$ is the entry-wised multiplication. 
The loss function of regularized $k$-POD clustering can be rewritten as 
\begin{align}
\label{eq_rkpod_loss2}
    \widehat{L}_n(\bm{M})=\sum_{i=1}^{n} \min_{l=1,\dots,k} \|\bm{x}_i\circ \bm{r}_i - \bm{\mu}_l \circ \bm{r}_i  \|^2 + \lambda\cdot J(\bm{M}). 
\end{align}
We note that this expression regards the loss as a function only with respect to $\bm{M}$. 

Write $\widehat{\bm{M}}$ for the minimizer of Eq.~(\ref{eq_rkpod_loss2}). Then, we can define the corresponding partition of the sample $\{\bm{x}_1,\dots,\bm{x}_n\}$ in the following way. We first define a subset of $\{\bm{x}_i\}_{i=1}^{n}$ by 
\begin{align*}
    W_{l}=\{\bm{x}_i \mid \|\bm{x}_i\circ \bm{r}_i - \hat{\bm{\mu}}_l \circ \bm{r}_i \| \leq \|\bm{x}_i\circ \bm{r}_i - \hat{\bm{\mu}}_{l'} \circ \bm{r}_i \|,\; \forall l'\neq l \}. 
\end{align*}
Since it is possible that $W_l \cap W_{l'}\neq \emptyset$ for some $l$, $l'$ $\in \{1,\dots,k\}$, then $\{W_1,\dots,W_k\}$ is not a partition of $\{\bm{x}_i\}_{i=1}^{n}$. We instead define a sequence of subsets $\widehat{C}_l = W_l \setminus \left(\bigcup_{l'<l} W_{l'}  \right)$. 
Then $\widehat{\mathcal{C}}=\{\widehat{C}_1,\dots,\widehat{C}_k\}$ forms a partition of $\{\bm{x}_i\}_{i=1}^{n}$. 
Associated with $\widehat{\mathcal{C}}$, we define the membership matrix $\widehat{\bm{U}}$ by $\hat{u}_{il}=\mathds{1}(\bm{x}_i \in \widehat{C}_l)$ for any $i=1,\dots,n$ and $l=1,\dots,k$. 
Furthermore, write $\hat{q}_j$ for the proportion of observed entries in the $j$-th feature, and write $\bar{\bm{\mu}}_{(j)}=(\bar{\mu}_{1j},\dots,\bar{\mu}_{kj})^T$ and $\bar{\sigma}_j^2$ for the sample mean and variance in the $j$-th feature ignoring missing entries, that is, 
    \begin{align*}
        \hat{q}_j=\frac{1 }{n}\sum_{i=1}^{n} r_{ij} \; ,\quad \bar{\mu}_{lj}=\frac{1}{\sum_{i=1}^{n}\hat{u}_{il}r_{ij}} \sum_{i=1}^{n}\hat{u}_{il}r_{ij} x_{ij} \;, \quad \bar{\sigma}_j^2=\frac{1}{\sum_{i=1}^{n} r_{ij}} \sum_{i=1}^{n} r_{ij}x_{ij}^2 .
    \end{align*}
Moreover, define the \textit{within-cluster sum-of-square} associated with $\widehat{\mathcal{C}}$ in the $j$-th feature to be 
\begin{align*}
    \text{WCSS}_j(\widehat{\mathcal{C}})=\frac{1}{n} \sum_{i=1}^{n} \sum_{l=1}^{k} \mathds{1}(\bm{x}_i \in \widehat{C}_l) r_{ij}(x_{ij}-\bar{\mu}_{lj})^2. 
\end{align*}
Let $\widehat{Q}_j$ be the minima of the function $Q_j$ with respect to $\bm{\mu}_{(j)}$, which is given by 
    \begin{align*}
        Q_j(\bm{\mu}_{(j)})=\frac{1}{n}\sum_{i=1}^{n} \min_{l=1,\dots,k} r_{ij}(x_{ij}-\mu_{lj})^2. 
    \end{align*}
The following proposition shows the sparsity of the estimated cluster centers $\widehat{\bm{M}}$ with different types of $J(\cdot)$, the proof of which is provided in Section~C of Supplementary materials. 
\begin{proposition}
\label{rkpod_proposition}
    (a) For $J(\cdot)=J_0(\cdot)$, if $\hat{q}_j \bar{\sigma}_j^2 - \text{WCSS}_j(\widehat{\mathcal{C}}) \leq \lambda/n$, 
    then $\hat{\bm{\mu}}_{(j)}=(0,0,\dots,0)^T$. Otherwise, $\hat{\bm{\mu}}_{(j)}\neq (0,0,\dots,0)^T$ and has the $l$-th component $\hat{\mu}_{lj}$ ($l=1,\dots,k$) satisfying $\hat{\mu}_{lj}=\bar{\mu}_{lj}$. 
    
    (b) For $J(\cdot)=J_1(\cdot)$ with weights $\{w_j\}_{j=1}^{p}$, if $\sqrt{\hat{q}_j \bar{\sigma}_j^2 - \widehat{Q}_j } < (\lambda w_j)/(2n)$, 
    then $\hat{\bm{\mu}}_{(j)}=(0,0,\dots,0)^T$. 
    Otherwise, $\hat{\bm{\mu}}_{(j)}\neq (0,0,\dots,0)^T$ and has the $l$-th component $\hat{\mu}_{lj}$ ($l=1,\dots,k$) satisfying 
    \begin{align*}
        \hat{\mu}_{lj}=\left( 1 + \frac{\lambda w_j}{2\cdot \|\hat{\bm{\mu}}_{(j)}\| \cdot \sum_{i=1}^{n}\hat{u}_{il} r_{ij} } \right)^{-1}\cdot \bar{\mu}_{lj}.
    \end{align*}
\end{proposition}
\begin{remark}
    For $J=J_0$, those features in which the gap between total variance and WCSS is larger than a uniform threshold would be selected, and cluster centers in selected features are equal to the sample means. 
    For $J=J_1$, the sparsity of cluster centers is determined by the weights, and cluster centers in selected features are a shrunk version of the sample means. 
    Moreover, if there is no missing, this result coincides with that of regularized $k$-means clustering derived by \cite{raymaekers2022regularized} and \cite{Levrard2018}. 
\end{remark}

\section{Simulations}
\label{rkpod_sec_experiments}
In this section, we empirically evaluate the performance of the proposed method. 
The incomplete datasets used in this section are constructed by artificially setting missing on original complete datasets. 
The structure of this section is as follows: 
(a) We describe the experimental setup in Section~\ref{rkpod_sec_setup}, including the generation of original complete data and the missingness mechanisms. 
(b) Focusing on the proposed method, we compare the effects of different strategies of initialization in Section~\ref{rkpod_sec_initialization}. 
(c) We compare the effects of different criteria on the tuning parameter in Section~\ref{rkpod_sec_tuning}. 
(d) The comparisons with other methods are summarized in Section~\ref{rkpod_sec_syn_exp}. 
(e) We further evaluate the effect of reducing bias of the proposed method via simulations on a complete real-world high-dimensional dataset with artificial missingness in Section~\ref{rkpod_sec_realdata_complete}. 

\subsection{Experimental setup}
\label{rkpod_sec_setup}
\subsubsection{Complete data}
For the original complete datasets, we consider synthetic datasets on which the $k$-means clustering performs well in the absence of missing data. The Gaussian mixture model of $k$ components with equal weight $1/k$ and the same diagonal covariance matrix $\bm{\Sigma}$ is used, where the mean vector of the $l$-th component is denoted by $\bm{\mu}_l^{\ast}$, $l=1,\dots,k$.  
Specifically, the synthetic complete data points $\bm{x}_i\in \mathbb{R}^p$, $i=1,\dots,n$, are generated as follows. 
For each $i$, we first uniformly sample $z_i$ from $\{1,\dots,k\}$ as the true cluster label. Then $\bm{x}_i$ is generated from a Gaussian distribution $\mathcal{N}(\bm{\mu}_l^{\ast},\bm{\Sigma})$ if $z_i=l$. 

Through this section, we fix the sample size $n=3000$ and the number of clusters $k=4$, and the following $\bm{\mu}_l^{\ast}$'s are used: 
\begin{align*}
    \begin{pmatrix}
        \bm{\mu}_1^{\ast}\\
        \bm{\mu}_2^{\ast}\\
        \bm{\mu}_3^{\ast}\\
        \bm{\mu}_4^{\ast}
    \end{pmatrix}=
    \begin{pmatrix}
        a\bm{1}_{d/2}^T & a\bm{1}_{d/2}^T & \bm{0}_{p-d}^T\\
        a\bm{1}_{d/2}^T & -a\bm{1}_{d/2}^T & \bm{0}_{p-d}^T\\
        -a\bm{1}_{d/2}^T & a\bm{1}_{d/2}^T & \bm{0}_{p-d}^T\\
        -a\bm{1}_{d/2}^T & -a\bm{1}_{d/2}^T & \bm{0}_{p-d}^T
    \end{pmatrix}.
\end{align*}
Since each $\bm{\mu}_l^{\ast}$ consists of $d$ informative values and $p-d$ zeros and the covariance matrix is diagonal, for complete data matrix $\bm{X}$, the first $d$ features are relevant to cluster structure, while the other $p-d$ features are noise features.  
To make most peer methods applicable for comparison, through this section, we consider two cases of features:
\begin{itemize}
    \item $p=10$ and $d=2$, where $a=2$ and $\bm{\Sigma}=\text{diag}(1,1,4,\dots,4)$
    \item $p=100$ and $d=10$, where $a=1$ or $a=0.8$ and $\bm{\Sigma}=\text{diag}(1,\dots,1,2,\dots,2)$.
\end{itemize}

\subsubsection{Missingness mechanism}
The mechanism of missingness is the cause of the missing values. There are three main types: missing completely at random (MCAR), missing at random (MAR), and missing not at random (MNAR) \citep{little2019statistical}. The MCAR requires that the missingness of $\bm{X}$ is independent with $\bm{X}$ itself, and the MAR requires that the missingness is only dependent on the observed part of $\bm{X}$. Otherwise, it is called MNAR. 
To match different missingness mechanisms, through this section, we consider four types of procedures for generating missingness: 
\begin{itemize}
    \item MCAR: The missing probability is set to be a constant. 
    For any $i=1,\dots,n$ and $j=1,\dots,p$, 
    \begin{align*}
        \text{Pr}(x_{ij}\text{ is missing})=\tau. 
    \end{align*}
    Different $\tau$ is to meet the total proportion of missingness from 10\% to 50\%. 
    \item MAR: We fix the first column of $\bm{X}$ to be observed and the missingness of the other $p-1$ columns is dependent on the first column. For any $i=1,\dots,n$ and $j=2,\dots,p$, 
    \begin{align*}
        \text{Pr}(x_{ij}\text{ is missing})=\frac{1}{1+\exp(-\psi_1(x_{i1}-\psi_2))}.
    \end{align*}
    Different $(\psi_1,\psi_2)$ are selected to meet the total proportion of missingness from 10\% to 30\%, which is provided in Section~D.1 of Supplementary materials. 
    \item MNAR1 (Self-masked \citep{josse2020imputation}): The missing probability is determined by the value of the data itself. For any $i=1,\dots,n$ and $j=1,\dots,p$,
    \begin{align*}
        \text{Pr}(x_{ij} \text{ is missing})=\frac{1}{1+\exp(-\phi_1 (x_{ij}-\phi_2))}.
    \end{align*}
    Different $(\phi_1,\phi_2)$ are selected to meet the total proportion of missingness from 10\% to 30\%, which is provided in Section~D.1 of Supplementary materials. 
    \item MNAR2 \citep{chi2016k}: In each column of $\bm{X}$, entries in the bottom 10\%, 20\% and 30\% quantiles are set to be missing. 
\end{itemize}

\subsubsection{Evaluation indexes}

Since we focus on the estimation of cluster centers, we use the mean-squared error (MSE) of the estimated cluster centers as the main index for evaluation. Specifically, denote $\widehat{\bm{M}}$ to be the estimated cluster centers, and $\bm{M}^{\ast}$ to be the underlying true cluster centers. The MSE is defined as 
\begin{align*}
    \text{MSE}(\widehat{\bm{M}},\bm{M}^{\ast})=\sum_{l=1}^{k}\min_{l'=1,\dots,k} \|\hat{\bm{\mu}}_l - \bm{\mu}^{\ast}_{l'}\|^2.
\end{align*}
Since for the $k$-means clustering, $\bm{M}^{\ast}$ is defined by the minimizer of the loss function in the population level, it is often unknown. However, based on the consistency of the $k$-means clustering, we can substitute it with the estimator under a sufficiently large sample size. That is, we generate a complete dataset with sample size $N=10^5$ following the same distribution as the original complete data, and apply the $k$-means clustering on it. The output cluster centers would be used as the substitute of $\bm{M}^{\ast}$. 

Moreover, to compare the performance of clustering, we use the classification error rate (CER) as the index for evaluation. Specifically, denote $\widehat{\bm{U}}$ to be the estimated membership matrix, of which the associated partition of data points is denoted by $\widehat{\mathcal{C}}$. Denote $\mathcal{C}^{\ast}$ to be the true partition of data points. The CER is defined as 
\begin{align*}
    \text{CER}(\widehat{\mathcal{C}},\mathcal{C}^{\ast})=\frac{1}{\tbinom{n}{2}}\sum_{i>i'} \big| \mathds{1}_{\widehat{\mathcal{C}}(i,i')} - \mathds{1}_{\mathcal{C}^{\ast}(i,i')} \big|,
\end{align*}
where $\mathds{1}_{\mathcal{C}(i,i')}=1$ if the $i$-th and $i'$-th data points are assigned to the same cluster according to the partition $\mathcal{C}$, 0 otherwise. 

In addition, we further compare the influence of the estimated cluster centers on predicting the partition of a validation dataset. Specifically, we generate a validation dataset that is complete with sample size $n_0=400$ and follows the sample distribution as the original complete data, and calculate the partition of it based on the estimated cluster centers. We use the classification error rate of the predictive partition to the true partition as the index for evaluation, and we call it \textit{predictive CER} for short.

\subsection{Effects of different initialization strategies}
\label{rkpod_sec_initialization}
For both the $k$-POD clustering and the proposed method, we consider two strategies for random initialization. One is based on complete data points ({\fontfamily{qcr}\selectfont comp} for short), while another is based on imputation ({\fontfamily{qcr}\selectfont impt} for short). 
Table~\ref{table_initialization} illustrates the averaged values of MSE (with standard deviation in bracket) of different methods using different initialization strategies. Here we only use the case of $p=10$, since for $p=100$, there is no complete data point left. 
It can be seen that the {\fontfamily{qcr}\selectfont impt} strategy generally performs better than the {\fontfamily{qcr}\selectfont comp} strategy for both $k$-POD clustering and the proposed method. 
Moreover, although the {\fontfamily{qcr}\selectfont comp} strategy can give smaller MSE for the proposed method when there is 10\% missing, it becomes less effective when the missing proportion gets large because there are fewer available complete data points for initialization. 

\begin{table*}[htbp]
\centering
\caption{\normalfont MSE (standard deviation in brackets) using different strategies for random initialization}
\label{table_initialization}
\begin{adjustbox}{center}
\resizebox{\columnwidth}{!}{
\begin{tabular}{@{}cccccccc@{}}
%\begin{tabular*}{\textwidth}{@{\extracolsep\fill}cccccccc}
\toprule
 \multirow{2}{*}{\makecell{Missing\\mechanism}} &  \multirow{2}{*}{\makecell{Missing\\proportion}} &  \multicolumn{2}{c}{$k$-POD}  &  \multicolumn{2}{c}{\makecell{Reg. $k$-POD (group lasso)}} & \multicolumn{2}{c}{\makecell{Reg. $k$-POD ($l_0$)}} \\
\cmidrule(r){3-4}\cmidrule(r){5-6}\cmidrule(r){7-8}
& & {\fontfamily{qcr}\selectfont impt} & {\fontfamily{qcr}\selectfont comp} & {\fontfamily{qcr}\selectfont impt} & {\fontfamily{qcr}\selectfont comp} & {\fontfamily{qcr}\selectfont impt} & {\fontfamily{qcr}\selectfont comp} \\
\midrule
 MCAR & 10\% & \textbf{1.994 (0.90)} & 2.454 (0.87)  & 0.118 (0.03) & \textbf{0.094 (0.03)}  & 0.038 (0.01)  & \textbf{0.025 (0.01)}\\
  & 20\% & \textbf{6.419 (2.11)} & 10.598 (4.17)  & \textbf{0.872 (0.57)} & 3.401 (3.73)  & \textbf{0.079 (0.03)} & 0.460 (1.25) \\
  & 30\% & \textbf{16.665 (4.74)} & 21.647 (4.42)  & \textbf{1.853 (0.71)} & 8.943 (6.01)  & \textbf{0.097 (0.03)} & 5.830 (7.05) \\
  & 40\% & \textbf{26.030 (4.24)} & 30.941 (5.81)  & \textbf{3.160 (0.88)} & 12.480 (8.39)  & \textbf{1.139 (2.46)} & 18.454 (11.26) \\[8pt]
 MAR & 10\% & \textbf{2.631 (1.05)} & 11.928 (4.66) & \textbf{0.364 (0.24)} & 2.715 (5.81) & \textbf{0.203 (0.05)} & 13.310 (3.79) \\
  & 20\% & \textbf{5.887 (1.83)} & 27.540 (4.80) & \textbf{0.298 (0.07)} & 21.233 (8.44) & \textbf{0.117 (0.04)} & 28.059 (4.35) \\
  & 30\% & \textbf{6.835 (1.90)} & 28.343 (5.49) & \textbf{0.484 (0.31)} & 13.322 (11.76) & 
  \textbf{0.115 (0.03)} & 28.778 (5.42) \\[8pt]
 MNAR1 & 10\% & \textbf{5.959 (0.65)} & 6.260 (0.75)  & 1.151 (0.10) & \textbf{1.083 (0.10)}  & \textbf{0.462 (0.05)}  & 0.637 (0.75)\\
  & 20\% & \textbf{15.740 (4.12)} & 17.191 (2.67)  & 3.932 (0.33) & \textbf{3.706 (0.33)}  & \textbf{0.283 (0.05)} & 8.979 (5.00) \\
  & 30\% & \textbf{21.314 (3.29)} & 24.917 (4.58)  & \textbf{2.301 (0.35)} & 4.797 (5.94)  & \textbf{0.210 (0.07)} & 9.252 (7.50) \\[8pt]
 MNAR2 & 10\% & \textbf{6.481 (0.39)} & 6.696 (0.46)  & 2.006 (0.12) & \textbf{1.942 (0.11)}  & 0.691 (0.07)  & \textbf{0.676 (0.07)}\\
  & 20\% & \textbf{21.531 (1.02)} & 23.848 (2.03)  & \textbf{4.901 (0.24)} & 5.458 (1.31)  & \textbf{2.346 (0.15)} & 7.491 (4.78) \\
  & 30\% & \textbf{47.923 (3.07)} & 52.439 (5.00)  & \textbf{24.829 (0.44)} & 24.975 (0.72)  & \textbf{9.733 (4.99)} & 21.930 (8.02) \\
\bottomrule
\end{tabular}}
\end{adjustbox}
\end{table*}

In addition, we found that the $l_0$ type of the proposed method is more sensitive to the initialization than the group lasso type. We thus need more random initialization points, which however increases computation cost. An alternative for random initialization is the sparse initialization, which has comparable performance and needs fewer initialization points. We provide more details in Section~D.2 of Supplementary materials.

\subsection{Selection of regularization parameter}
\label{rkpod_sec_tuning}
In this section, we compare the instability and BIC criteria for selecting the regularization parameter. 
We take the case of $p=100$, $d=10$, and $a=1$ as an example. We let the regularization parameter $\lambda$ vary in a grid of 20 candidate values given by $10^{-3+(4s/19)}$ for $s=0,1,\dots,19$, and calculate the corresponding values of instability and BIC criteria. For the instability criterion, we use 30 repetitions of random division. Note that only the {\fontfamily{qcr}\selectfont impt} strategy of initialization is used here.  

Table~\ref{table_tuning_p100} illustrates the averaged values of MSE (with the averaged number of active features in brackets) based on the $\lambda$ selected by BIC and instability. 
It can be seen that for both types of the proposed method, under MCAR and MAR mechanisms, the $\lambda$ selected by instability gives smaller MSE but larger/comparable number of active features than that selected by BIC. Under MNAR mechanisms, the instability criterion performs much better than the BIC criterion, especially for the $l_0$ type of proposed method. 
The main reason is that deriving the expression of BIC is based on the assumption that missingness is independent to the complete data. However, the instability follows the spirit of cross-validation and is defined by the clustering alignment. 
We provide more details of comparison in the case of $p=10$ and how the regularization parameter influences the performance of the proposed method in Section~D.3 of Supplementary materials.

\begin{table*}[htpb]
\centering
\caption{\normalfont MSE (number of active features in brackets) of proposed method using different criteria for selecting~$\lambda$}
\label{table_tuning_p100}
\begin{adjustbox}{center}
\resizebox{0.75\columnwidth}{!}{
\begin{tabular}{@{}cccccc@{}}
\toprule
 \multirow{2}{*}{\makecell{Missing\\mechanism}} &  \multirow{2}{*}{\makecell{Missing\\proportion}}  &  \multicolumn{2}{c}{\makecell{Reg. $k$-POD (group lasso)}} & \multicolumn{2}{c}{\makecell{Reg. $k$-POD ($l_0$)}} \\
\cmidrule(r){3-4}\cmidrule(r){5-6} 
 & &  Instability & BIC & Instability & BIC  \\
\midrule
 MCAR & 10\%  & \textbf{0.126 (47)} & 0.187 (14) & \textbf{0.109 (10)}  & 0.114 (10) \\
& 20\% & \textbf{0.206 (29)} & 0.458 (11) & \textbf{0.156 (10)} & 0.161 (10) \\
& 30\% & \textbf{0.407 (29)} & 0.743 (12) & 0.305 (10) & \textbf{0.280 (10)} \\
& 40\% & 1.934 (15) & \textbf{1.918 (16)} & \textbf{2.675 (13)} & 10.412 (12)\\
& 50\% & \textbf{5.546 (20)} & 9.018 (13) & 25.895 (22) & \textbf{25.073 (23)} \\[8pt]
MAR & 10\% & \textbf{0.150 (19)} & 0.175 (10) & \textbf{0.131 (10)} & 0.152 (10) \\
& 20\% & \textbf{0.140 (18)} & 0.182 (10) & \textbf{0.126 (10)} & 0.434 (16) \\
& 30\% & \textbf{0.204 (12)} & 0.228 (10) & 0.166 (10) & \textbf{0.164 (10)} \\[8pt]
MNAR1 & 10\% & \textbf{3.073 (98)} & 25.418 (100) & \textbf{1.873 (10)} & 26.062 (100) \\
& 20\% & \textbf{3.109 (77)} & 33.044 (100) & \textbf{1.738 (10)} & 33.559 (100) \\
& 30\% & \textbf{2.139 (85)} & 20.032 (100) & \textbf{1.324 (10)} & 30.417 (100) \\[8pt]
MNAR2 & 10\% & \textbf{4.696 (78)} & 29.490 (100) & \textbf{2.693 (10)} & 31.177 (100) \\
& 20\% & \textbf{40.286 (100)} & 96.354 (100) & \textbf{99.507 (100)} & 99.540 (100) \\
\bottomrule
\end{tabular}}
\end{adjustbox}
\end{table*}

\subsection{Comparison with other methods}
\label{rkpod_sec_syn_exp}
In this section, we compare the proposed method with other methods on synthetic incomplete datasets. We consider the following peer methods:  
\begin{itemize}
    \item Complete-case analysis. We delete all rows that includes missingness and then apply the classical $k$-means clustering to estimate the cluster centers. 
    It should be noted that we only report the result of this method for the case of $p=10$ since there are almost no complete data points left in the case of $p=100$. 
    \item Multiple imputation. We impute the missing entries via the popular mice model \citep{buuren2011mice}. The R package {\fontfamily{qcr}\selectfont mice} is used to get several complete data matrices after imputation. Then we pool the imputed data using element-wise mean to combine the multiple imputations into a single dataset, on which the classical $k$-means clustering is used to estimate the cluster centers. 
    \item The $k$-POD clustering. To compare the effects of different initialization strategies, we use a modified version of the original R package {\fontfamily{qcr}\selectfont kpodclustr} \citep{chi2016k}, and report the better result. 
\end{itemize}
For both group lasso and $l_0$ types of the proposed method, we consider two strategies of random initialization ({\fontfamily{qcr}\selectfont impt} and {\fontfamily{qcr}\selectfont comp}) and two criteria for selecting $\lambda$ (instability and BIC), and then report the best result. 

We apply these methods on all synthetic incomplete datasets to estimate cluster centers $\bm{M}$ and membership matrix $\bm{U}$, and then calculate the corresponding MSE, CER and predictive CER.  
Table~\ref{rkpod_table_syn_exp_mse}, Table~\ref{rkpod_table_syn_exp_cer} and Table~\ref{rkpod_table_syn_exp_predictcer} are results of different methods on different synthetic incomplete datasets, respectively. We report the results of $a=0.8$ for $p=100$ here and leave that of $a=1$ in Section~D.4 of Supplementary materials for the sake of space. The reported values are averaged indexes of 30 repetitions with standard deviations in the brackets. The bold font indicates the best results. 

It can be seen that the proposed method outperforms other methods for estimating cluster centers and clustering. 
Specifically, the $l_0$ type of proposed method performs better when $p$ is small, the missingness proportion is small and the mechanism is simple. 
The group lasso type of proposed method is stable against large $p$, large missingness proportion and complicated mechanisms. 
The main reason is that the solution of the $l_0$ type is based on a truncated expression, while the solution of the group lasso type would adjust the selected features as well, which improves the performance even though the $k$-POD clustering performs poorly in some complicated cases. 

It should be noted that in the case of $p=100$ with MCAR mechanism and missingness proportion larger than 40\%, the proposed method is less effective than the multiple imputation method Mice. It is because in this case, the MAR assumption of Mice is satisfied, and moreover, the relevant features are highly related, which makes the imputation of missing entries by Mice more accurate. 
Moreover, the MNAR2 mechanism is hard for all methods, which is because the missingness of each entry does not follow a probabilistic model and the reasonable imputation is more challenging.

\begin{figure}[!h]
\captionsetup[subfigure]{justification=centering}
    \centering
    \begin{subfigure}{0.24\textwidth}
        \includegraphics[width=\textwidth]{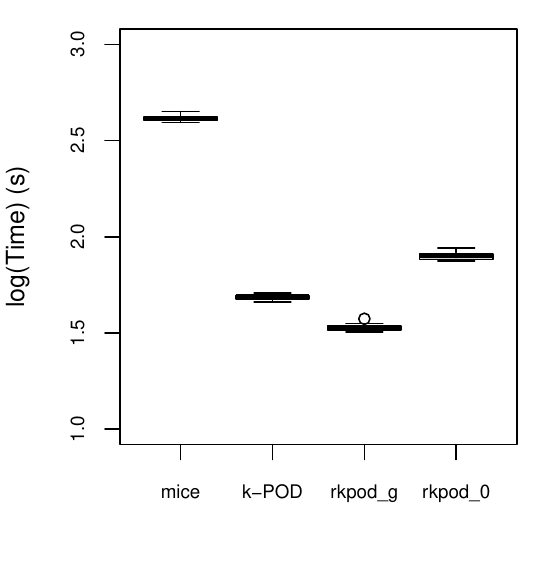}
        \vspace{-10pt}
        \caption*{MCAR (30\%)}
    \end{subfigure}
    \begin{subfigure}{0.24\textwidth}
        \includegraphics[width=\textwidth]{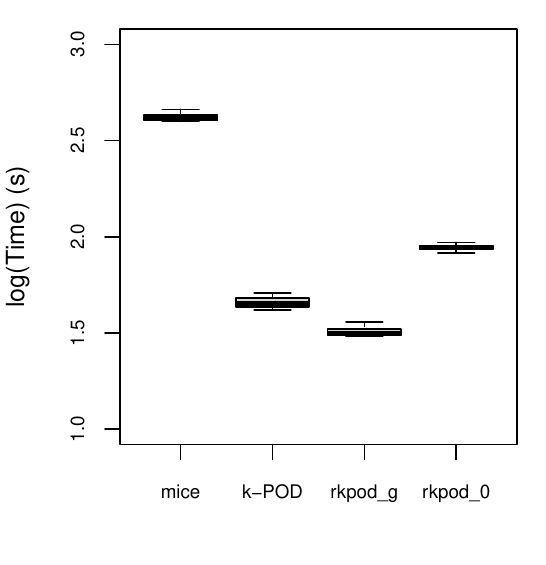}
        \caption*{MAR (20\%)}
    \end{subfigure}
    \begin{subfigure}{0.24\textwidth}
        \includegraphics[width=\textwidth]{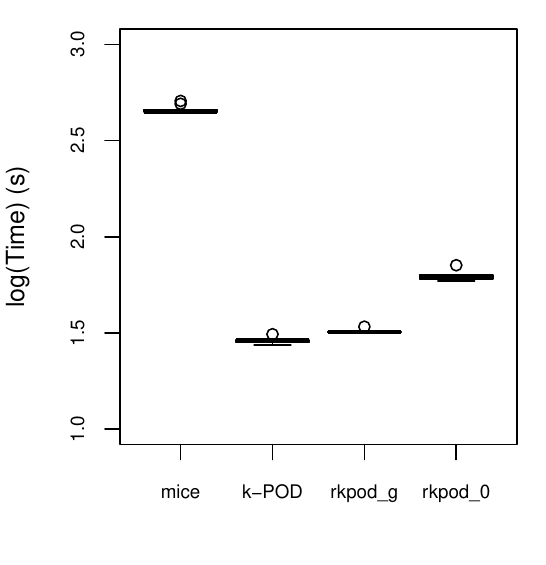}
        \caption*{MNAR1 (10\%)}
    \end{subfigure}
    \begin{subfigure}{0.24\textwidth}
        \includegraphics[width=\textwidth]{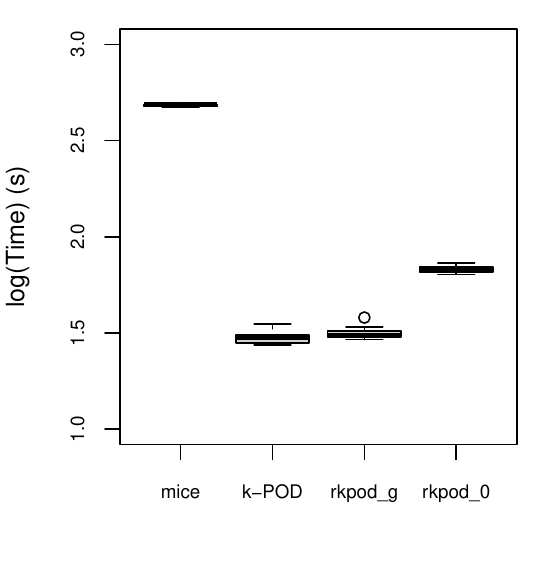}
        \caption*{MNAR2 (10\%)}
    \end{subfigure}
    \caption{\normalfont The box plot of computation time of different methods in the case of $p=100$. The group lasso and $l_0$ types of proposed method are denoted by \textit{rkpod\underline{ }g} and \textit{rkpod\underline{ }0} for short, respectively. The four panels from left to right are MCAR with 30\% missingness, MAR with 20\% missingness, MNAR1 and MNAR2 mechanisms with 10\% missingness, respectively. }
    \label{rkpod_fig_timecompare}
\end{figure}

Furthermore, we compare the computation time of different methods. Figure~\ref{rkpod_fig_timecompare} illustrates the results in the case of $p=100$ under MCAR mechanism with 30\% missingness, MAR mechanism with 20\% missingness, MNAR1 and MNAR2 mechanisms with 10\% missingness. We can see that the computation time of the proposed method is comparable to that of the $k$-POD clustering. However, the multiple imputation method Mice costs significantly more time, which coincides with the results of \cite{chi2016k}. 
In addition, the $l_0$ type of proposed method is more time-consuming than the group lasso type. It is because in Step~b of Algorithm~\ref{alg_rkm} with $l_0$ penalty, comparing the variance and the within-cluster sum-of-squares is needed, which costs more time.

\begin{table*}[htbp]
\centering
\caption{\normalfont MSE (standard deviation in brackets) of different methods}
\label{rkpod_table_syn_exp_mse}
\begin{adjustbox}{center}
\resizebox{\columnwidth}{!}{
\begin{tabular}{@{}lccccccccc@{}}
\toprule
 &\makecell{Missing\\mechanism}  &  \makecell{Missing\\proportion}  &  \makecell{Complete-case\\analysis} &  Mice & $k$-POD & \makecell{Reg. $k$-POD\\(group lasso)} & \makecell{Reg. $k$-POD \\($l_0$)} \\
\midrule
$p=10$ & MCAR & 10\% & 1.733 (1.15) & 1.129 (0.75)  & 1.994 (0.90) & 0.094 (0.03)  & \bf{0.025 (0.01)} \\
 &  & 20\% & 14.970 (5.08) & 4.954 (2.24)  & 6.419 (2.11) & 0.872 (0.57)  & \bf{0.079 (0.03)} \\
 &   & 30\% & 30.986 (5.36) & 9.447 (2.30)  & 16.665 (4.74) & 1.853 (0.71)  & \bf{0.097 (0.03)} \\
 &  & 40\% & 58.352 (12.80) & 12.612 (2.23)  & 26.030 (4.24) & 3.160 (0.88)  & \bf{1.139 (2.46)} \\
 &  & 50\% & - & 16.466 (2.20)  & 31.939 (5.47) & \bf{4.732 (0.77)}  & 22.601 (6.93) \\[8pt]
 & MAR & 10\% & 33.430 (1.13) & 0.767 (0.23) & 2.631 (1.05) & 0.364 (0.24) & \bf{0.203 (0.05)} \\
 &  & 20\% & 46.392 (1.60) & 2.221 (1.53) & 5.887 (1.83) & 0.298 (0.07) & \bf{0.117 (0.04)} \\
 &  & 30\% & 52.864 (5.71) & 3.138 (1.98) & 6.835 (1.90) & 0.484 (0.31) & \bf{0.115 (0.03)} \\[8pt]
& MNAR1 & 10\% & 5.032 (0.76) & 5.454 (0.85)  & 5.959 (0.65) & 1.083 (0.10)  & \bf{0.462 (0.05)} \\
 &  & 20\% & 19.881 (3.59) & 17.046 (1.39)  & 15.740 (4.12) & 3.706 (0.33)  & \bf{0.283 (0.05)} \\
 &   & 30\% & 33.241 (6.39) & 17.385 (1.50)  & 21.314 (3.29) & 2.301 (0.35)  & \bf{0.210 (0.07)} \\[8pt]
 & MNAR2 & 10\% & 6.329 (0.67) & 6.276 (0.33)  & 6.481 (0.39) & 1.942 (0.11)  & \bf{0.676 (0.07)} \\
 &  & 20\% & 24.454 (2.49) & 23.048 (2.41)  & 21.531 (1.02) & 4.901 (0.24)  & \bf{2.356 (0.15)} \\
 &   & 30\% & 55.481 (7.27) & 45.937 (1.78)  & 47.923 (3.07) & 24.829 (0.44)  & \bf{9.733 (4.99)} \\
\\
$p=100$ & MCAR & 10\% & - & 1.916 (0.20) & 2.558 (0.28) & 0.153 (0.02) & \bf{0.134 (0.02)}\\
 &  & 20\% & - & 2.239 (0.16) & 4.612 (0.64) & 0.162 (0.02) & \bf{0.153 (0.03)}\\
 &  & 30\% & - & 2.768 (0.26) & 15.475 (2.25) & \bf{0.434 (0.10)} & 7.948 (5.29)\\
 &  & 40\% & - & \bf{3.742 (0.45)} & 25.168 (3.96) & 6.938 (6.43) & 26.469 (5.00)\\
 &  & 50\% & - & \bf{5.957 (0.63)} & 36.216 (3.05) & 23.472 (7.22) & 36.284 (2.77)\\[8pt]
& MAR & 10\% & - & 1.948 (0.17) & 2.483 (0.24) & 0.197 (0.03) & \bf{0.168 (0.04)} \\
&  & 20\% & - & 2.181 (0.14) & 6.130 (1.68) & 0.246 (0.04) & \bf{0.185 (0.03)} \\
&  & 30\% & - & 2.657 (0.29) & 11.834 (1.28) & \bf{0.340 (0.10)} & 6.495 (5.06) \\[8pt]
& MNAR1 & 10\% & - & 26.022 (0.44) & 26.514 (0.53) & \bf{3.261 (0.14)} & 4.963 (1.05) \\
&  & 20\% & - & 33.406 (0.50) & 35.853 (1.29) & \bf{2.853 (0.19)} & 6.562 (8.24) \\
&  & 30\% & - & 26.842 (0.72) & 39.057 (2.24) & \bf{2.095 (0.31)} & 40.053 (2.89) \\[8pt]
& MNAR2 & 10\% & - & 32.759 (0.66) & 33.161 (0.79) & \bf{4.880 (0.18)} & 16.871 (2.00) \\
&  & 20\% & - & 104.249 (1.67) & 109.296 (3.24) & \bf{97.496 (2.95)} & 109.614 (2.98) \\
\bottomrule
\end{tabular}}
\end{adjustbox}
\end{table*}

\begin{table*}[htbp]
\centering
\caption{\normalfont CER (standard deviation in brackets) of different methods}
\label{rkpod_table_syn_exp_cer}
\begin{adjustbox}{center}
\resizebox{0.85\columnwidth}{!}{
\begin{tabular}{@{}lcccccccc@{}}
\toprule
&\makecell{Missing\\mechanism}  &  \makecell{Missing\\proportion}  &  Mice & $k$-POD & \makecell{Reg. $k$-POD\\(group lasso)} & \makecell{Reg. $k$-POD\\($l_0$)} \\
\midrule
$p=10$ &MCAR & 10\% &  0.136 (0.01)  & 0.148 (0.02) & 0.123 (0.01)  & \bf{0.123 (0.01)} \\
 &  & 20\% &  0.224 (0.02)  & 0.236 (0.02) & 0.193 (0.01)  & \bf{0.186 (0.01)} \\
  &  & 30\%  & 0.281 (0.01)  & 0.302 (0.01) & 0.250 (0.01)  & \bf{0.241 (0.01)} \\
 &  & 40\% & 0.310 (0.01)  & 0.337 (0.01) & 0.290 (0.01)  & \bf{0.285 (0.01)} \\
 &  & 50\% & 0.334 (0.00)  & 0.349 (0.01) & \bf{0.315 (0.01)}  & 0.345 (0.01) \\[8pt]
& MAR & 10\% & 0.097 (0.01) & 0.122 (0.01) & 0.093 (0.01) & \bf{0.090 (0.00)} \\
 &  & 20\%  & 0.139 (0.01) & 0.166 (0.01) & 0.125 (0.00) & \bf{0.124 (0.00)} \\
 &   & 30\%  & 0.176 (0.01) & 0.199 (0.01) & 0.162 (0.01) & \bf{0.161 (0.00)} \\[8pt]
& MNAR1 & 10\%  & 0.178 (0.01)  & 0.176 (0.02) & 0.151 (0.01)  & \bf{0.149 (0.01)} \\
 &  & 20\% & 0.228 (0.00)  & 0.271 (0.02) & 0.212 (0.01)  & \bf{0.202 (0.01)} \\
 &   & 30\% & 0.300 (0.00)  & 0.312 (0.01) & 0.255 (0.01)  & \bf{0.251 (0.01)} \\[8pt]
& MNAR2 & 10\% & 0.145 (0.00)  & 0.148 (0.01) & 0.130 (0.00)  & \bf{0.130 (0.00)} \\
 &  & 20\%  & 0.257 (0.02)  & 0.236 (0.02) & 0.242 (0.01)  & \bf{0.210 (0.01)} \\
 &   & 30\%  & 0.330 (0.00)  & 0.323 (0.01) & 0.426 (0.03)  & \bf{0.292 (0.03)} \\
 \\
$p=100$ & MCAR & 10\%  & 0.109 (0.01) & 0.118 (0.01) & 0.094 (0.00) & \bf{0.089 (0.01)} \\
 &  & 20\%  & 0.135 (0.01) & 0.175 (0.02) & \bf{0.109 (0.01)} & 0.113 (0.00) \\
 &  & 30\%  & 0.165 (0.01) & 0.288 (0.02) & \bf{0.138 (0.00)} & 0.245 (0.04) \\
 &  & 40\% & \bf{0.203 (0.01)} & 0.357 (0.01) & 0.248 (0.05) & 0.375 (0.03) \\
 & & 50\% & \bf{0.249 (0.01)} & 0.376 (0.01) & 0.359 (0.02) & 0.376 (0.01) \\[8pt]
& MAR & 10\%  & 0.109 (0.01) & 0.118 (0.01) & \bf{0.089 (0.01)} & 0.092 (0.01) \\
&  & 20\%  & 0.131 (0.01) & 0.192 (0.02) & \bf{0.116 (0.00)} & 0.122 (0.01) \\
&  & 30\%  & 0.161 (0.01) & 0.257 (0.01) & \bf{0.145 (0.01)} & 0.229 (0.04) \\[8pt]
& MNAR1 & 10\%  & 0.129 (0.01) & 0.132 (0.01) & \bf{0.098 (0.00)} & 0.104 (0.01) \\
&  & 20\% & 0.150 (0.01) & 0.190 (0.02) & \bf{0.118 (0.01)} & 0.176 (0.07) \\
&  & 30\%  & 0.175 (0.01) & 0.300 (0.02) & \bf{0.145 (0.01)} & 0.311 (0.02) \\[8pt]
& MNAR2 & 10\%  & 0.149 (0.01) & 0.158 (0.01) & \bf{0.110 (0.01)} & 0.136 (0.01) \\
&  & 20\%  & \bf{0.238 (0.01)} & 0.294 (0.01) & 0.304 (0.02) & 0.313 (0.02) \\
 \bottomrule
\end{tabular}}
\end{adjustbox}
\end{table*}

\begin{table*}[htbp]
\centering
\caption{\normalfont Predictive CER (standard deviations in brackets) of different methods}
\label{rkpod_table_syn_exp_predictcer}
\begin{adjustbox}{center}
\resizebox{\columnwidth}{!}{
\begin{tabular}{@{}lccccccccc@{}}
\toprule
&\makecell{Missing\\mechanism}  &  \makecell{Missing\\proportion}  &  \makecell{Complete-case\\analysis} &  Mice & $k$-POD & \makecell{Reg. $k$-POD\\(group lasso)} & \makecell{Reg. $k$-POD\\($l_0$)} \\
\midrule
$p=10$ &MCAR & 10\% & 0.075 (0.02) & 0.065 (0.02)  & 0.081 (0.02) & 0.042 (0.01)  & \bf{0.042 (0.01)} \\
 &  & 20\% & 0.201 (0.04) & 0.131 (0.03)  & 0.142 (0.03) & 0.060 (0.02)  & \bf{0.046 (0.01)} \\
  &  & 30\% & 0.278 (0.03) & 0.183 (0.02)  & 0.218 (0.03) & 0.075 (0.02)  & \bf{0.043 (0.01)} \\
 &  & 40\% & 0.335 (0.03) & 0.210 (0.02)  & 0.267 (0.02) & 0.102 (0.02)  & \bf{0.062 (0.04)} \\
 &  & 50\% & - & 0.240 (0.02)  & 0.284 (0.02) & \bf{0.081 (0.03)}  & 0.249 (0.04) \\[8pt]
& MAR & 10\% & 0.074 (0.02) & 0.076 (0.02) & 0.094 (0.02) & 0.048 (0.01) & \bf{0.047 (0.01)} \\
&  & 20\% & 0.220 (0.03) & 0.220 (0.01) & 0.197 (0.04) & 0.046 (0.01) & \bf{0.045 (0.01)} \\
&   & 30\% & 0.274 (0.03) & 0.228 (0.01) & 0.240 (0.02) & 0.050 (0.01) & \bf{0.045 (0.01)} \\[8pt]
& MNAR1 & 10\% & 0.074 (0.02) & 0.076 (0.02)  & 0.094 (0.02) & 0.048 (0.01)  & \bf{0.047 (0.01)} \\
 &  & 20\% & 0.220 (0.03) & 0.220 (0.01)  & 0.197 (0.04) & 0.046 (0.01)  & \bf{0.045 (0.01)} \\
 &   & 30\% & 0.274 (0.03) & 0.228 (0.01)  & 0.240 (0.02) & 0.050 (0.01)  & \bf{0.045 (0.01)} \\[8pt]
& MNAR2 & 10\% & 0.065 (0.01) & 0.060 (0.01)  & 0.073 (0.01) & 0.048 (0.01)  & \bf{0.048 (0.01)} \\
 &  & 20\% & 0.175 (0.03) & 0.136 (0.05)  & 0.128 (0.02) & \bf{0.053 (0.01)}  & 0.060 (0.01) \\
 &   & 30\% & 0.323 (0.06) & 0.203 (0.01)  & 0.247 (0.02) & 0.248 (0.02)  & \bf{0.124 (0.05)} \\
\\
$p=100$ & MCAR & 10\% & - & 0.087 (0.01) & 0.100 (0.01) & 0.071 (0.01) & \bf{0.071 (0.01)} \\
 &  & 20\% & - & 0.091 (0.01) & 0.127 (0.02) & \bf{0.072 (0.01)} & 0.074 (0.01) \\
 &  & 30\% & - & 0.097 (0.01) & 0.228 (0.02) & \bf{0.066 (0.01)} & 0.162 (0.06) \\
 &  & 40\% & - & \bf{0.112 (0.02)} & 0.313 (0.02) & 0.154 (0.08) & 0.328 (0.04) \\
 & & 50\% & - & \bf{0.142 (0.02)} & 0.353 (0.01) & 0.314 (0.04) & 0.356 (0.01) \\[8pt]
& MAR & 10\% & - & 0.091 (0.01) & 0.094 (0.01) & \bf{0.068 (0.01)} & 0.068 (0.01) \\
&  & 20\% & - & 0.090 (0.01) & 0.141 (0.02) & \bf{0.069 (0.01)} & 0.069 (0.02) \\
&  & 30\% & - & 0.097 (0.01) & 0.202 (0.02) & \bf{0.070 (0.01)} & 0.154 (0.06) \\[8pt]
& MNAR1 & 10\% & - & 0.108 (0.01) & 0.118 (0.01) & \bf{0.080 (0.01)} & 0.089 (0.01) \\
&  & 20\% & - & 0.110 (0.01) & 0.146 (0.03) & \bf{0.079 (0.01)} & 0.135 (0.09) \\
&  & 30\% & - & 0.110 (0.01) & 0.250 (0.02) & \bf{0.083 (0.01)} & 0.262 (0.03) \\[8pt]
& MNAR2 & 10\% & - & 0.124 (0.01) & 0.143 (0.01) & \bf{0.089 (0.01)} & 0.116 (0.02) \\
&  & 20\% & - & \bf{0.231 (0.03)} & 0.317 (0.03) & 0.316 (0.04) & 0.323 (0.03) \\
 \bottomrule
\end{tabular}}
\end{adjustbox}
\end{table*}

\newpage
\subsection{Simulations on a real-world dataset with artificial missingness}
\label{rkpod_sec_realdata_complete}

At the end of this section, we evaluate the performance of the proposed method via simulations on a real-world dataset with artificial missingness. 
%Since we focus on the performance of estimating the cluster centers, the ground truth can be obtained only when the dataset includes no missingness. Then, we construct incomplete datasets in the same way as numerical experiments, that is, we artificially set missingness on original complete datasets. 
We consider a microarray genomics dataset \textit{Lymphoma}, which can be downloaded from \url{https://www.stat.cmu.edu/~jiashun/Research/software/GenomicsData/Lymphoma/}. It consists of 4026 gene expressions ($p=4026$), collected over 62 samples ($n=62$). Out of the 62 samples, 42 are Diffuse Large B-Cell Lymphoma (DLBCL), 9 are Follicular Lymphoma (FL), and 11 are Chronic Lymphocytic Leukemia (CLL) cell samples ($k=3$). 
The original dataset is complete and includes no missingness. We consider the MCAR mechanism with missing proportion from 10\% to 50\%, the MAR mechanism with missing proportion from 10\% to 30\%, and the MNAR1 and MNAR2 mechanisms with missing proportion from 10\% to 20\%. The generation of missingness for MCAR and MNAR mechanisms is the same as introduced in Section~\ref{rkpod_sec_setup}. For the MAR mechanism, we fix the 40th feature to be complete, which is one of the most \textit{influential} features according to analysis of existing literature, and the missingness of other features depends on the values of the 40th feature. 

In this case, since $p$ is much larger than $n$, there is no complete data point when artificial missingness is added, and the Complete-case analysis method is no longer applicable. Moreover, we cannot use the multiple imputation method, such as mice, because the computation time would be extremely long and not acceptable in practice. Therefore, we only compare the proposed method to the $k$-POD clustering.
To calculate the MSE for evaluation, the ground truth of cluster centers $\bm{M}^{\ast}$ is needed. 
According to existing literature \citep{sun2012regularized,jin2016influential}, for the \textit{Lymphoma} dataset there exists a small subset of influential features, with which the $k$-means clustering can give a better clustering result. For example, the CER of classical $k$-means with all features is about 0.3, while that with 44 influential features is 0.05, which means that $\bm{M}^{\ast}$ is more likely to be sparse. 
Therefore, we use the result of \cite{jin2016influential} as an approximation of $\bm{M}^{\ast}$. 

\begin{figure}[htbp]
\captionsetup[subfigure]{justification=centering}
    \centering
    \begin{subfigure}{0.24\textwidth}
        \includegraphics[width=\textwidth]{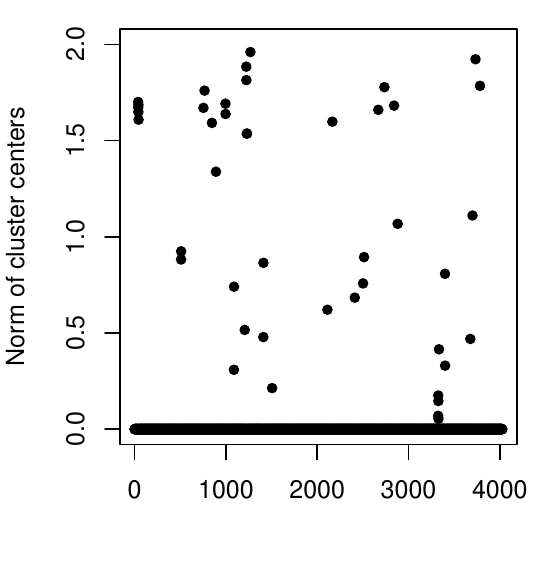}
        \caption*{Ground truth\\ \ }
    \end{subfigure}
    \begin{subfigure}{0.24\textwidth}
        \includegraphics[width=\textwidth]{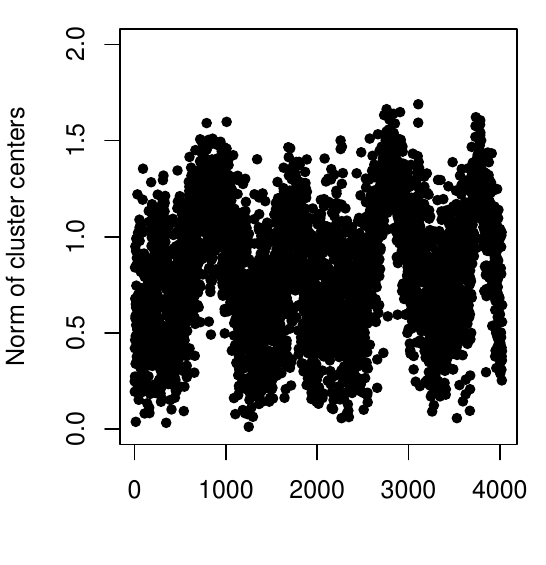}
        \caption*{$k$-POD\\ \ }
    \end{subfigure}
    \begin{subfigure}{0.24\textwidth}
        \includegraphics[width=\textwidth]{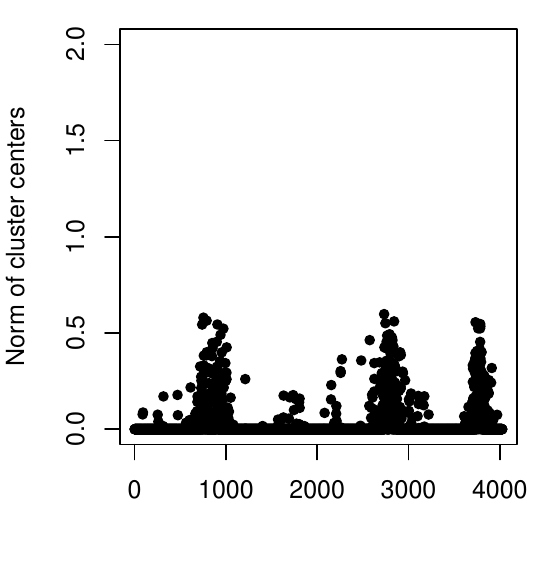}
        \caption*{Reg. $k$-POD\\(group lasso)}
    \end{subfigure}
    \begin{subfigure}{0.24\textwidth}
        \includegraphics[width=\textwidth]{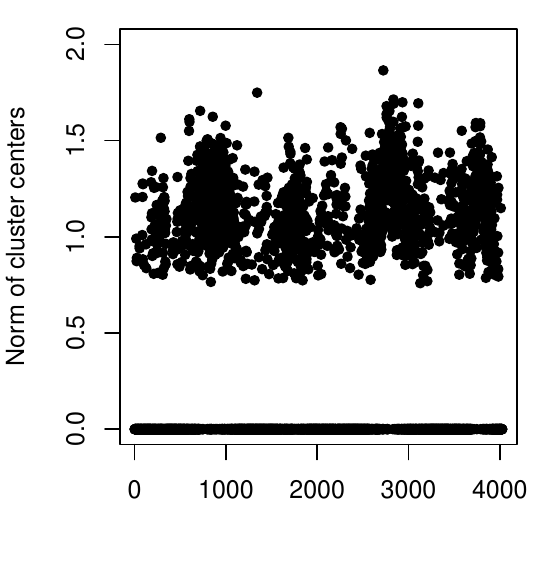}
        \caption*{Reg. $k$-POD\\($l_0$)}
    \end{subfigure}
    \caption{\normalfont \doublespacing The estimated cluster centers of different methods for \textit{Lymphoma} dataset under MCAR mechanism with 30\% missing proportion. The x-axis is the feature index. The y-axis is the $l_2$ norm of cluster centers in each feature. }
    \label{rkpod_fig_lymphoma_CenterNorm}
\end{figure}

Table~\ref{rkpod_table_realdata_mse_cer_lymphoma} illustrates the results of MSE for estimated cluster centers and CER for estimated membership of different methods. The reported values are the average of ten repetitions with standard deviations in brackets. 
It can be seen that the proposed method, especially the group lasso type generally outperforms the $k$-POD clustering on both MSE and CER in various settings. 
Figure~\ref{rkpod_fig_lymphoma_CenterNorm} illustrates the norm of estimated cluster centers in each feature for \textit{Lymphoma} dataset under MCAR mechanism with 30\% missing proportion. It can be seen that the results of the proposed method are more sparse than that of $k$-POD clustering. 
Moreover, since the $l_0$ type of proposed method is based on the hard threshold, there remain a lot of features, which leads to similar clustering performance to $k$-POD clustering. The group lasso type does not only select relevant features but also shrinks them, which leads to better performance. 

\begin{table*}[htbp]
\centering
\caption{\normalfont MSE and CER (standard deviations in brackets) of different methods for \textit{Lymphoma} datasets}
\label{rkpod_table_realdata_mse_cer_lymphoma}
\begin{adjustbox}{center}
\resizebox{\columnwidth}{!}{
\begin{tabular}{@{}cccccccc@{}}
\toprule
 \multirow{3}{*}{\makecell{Missing\\mechanism}} &  \multirow{3}{*}{\makecell{Missing\\proportion}}  & \multicolumn{3}{c}{MSE} & \multicolumn{3}{c}{CER} \\
 \cmidrule(r){3-5}\cmidrule(r){6-8}
 & & $k$-POD & \makecell{Reg. $k$-POD\\(group lasso)} & \makecell{Reg. $k$-POD \\($l_0$)} & $k$-POD & \makecell{Reg. $k$-POD\\(group lasso)} & \makecell{Reg. $k$-POD \\($l_0$)} \\
\midrule
 MCAR & 10\%   & 2077.987 (88.49) & \bf{73.249 (0.54)}  & 1565.534 (139.37) & 0.290 (0.01) & \bf{0.135 (0.01)}  & 0.284 (0.01) \\
  & 20\%  & 2193.763 (189.35) & \bf{72.843 (0.34)}  & 1176.306 (246.66) & 0.293 (0.01) & \bf{0.130 (0.01)}  & 0.274 (0.06) \\
  & 30\%  & 2254.447 (154.07) & \bf{72.533 (0.06)}  & 1196.070 (175.43) & 0.290 (0.01) & \bf{0.123 (0.01)}  & 0.276 (0.08) \\
  & 40\%   & 2299.094 (154.84) & \bf{73.615 (0.49)}  & 1042.052 (193.14) & 0.281 (0.02) & \bf{0.145 (0.01)}  & 0.220 (0.12) \\
 & 50\%  & 2448.054 (216.05) & \bf{72.856 (0.27)}  & 828.744 (121.03) & 0.308 (0.02) & \bf{0.131 (0.01)}  & 0.180 (0.11) \\[8pt]
 MAR & 10\% & 2092.131 (8.36) & \bf{73.674 (0.31)}  &  1625.762 (165.52) & 0.278 (0.00) & \bf{0.156 (0.01)} & 0.296 (0.01) \\
 & 20\% & 2182.586 (90.95) & \bf{73.252 (0.25)} & 1197.600 (254.64) & 0.287 (0.01) & \bf{0.157 (0.02)} & 0.298 (0.01) \\
 & 30\% & 2284.245 (153.49) & \bf{72.774 (0.13)} & 1563.565 (281.55) & 0.309 (0.02) & \bf{0.177 (0.03)} & 0.312 (0.02) \\[8pt]
MNAR1 & 10\%  & 1677.725 (7.37) & \bf{75.813 (0.22)}  & 976.441 (108.01) & 0.285 (0.00) & \bf{0.163 (0.00)}  & 0.258 (0.08)\\
  & 20\%  & 2209.57 (48.64) & \bf{73.421 (0.59)}  & 1220.936 (157.73) & 0.284 (0.01) & \bf{0.136 (0.02)}  & 0.261 (0.08)\\[8pt]
 MNAR2 & 10\%  & 1837.678 (0.00) & \bf{73.456 (0.00)}  & 1050.854 (120.44) & 0.300 (0.00) & \bf{0.156 (0.00)}  & 0.274 (0.08)\\
 \bottomrule
\end{tabular}}
\end{adjustbox}
\end{table*}

\newpage
\section{Applications}
\label{rkpod_sec_application}

In this section, we evaluate the performance of the proposed method on real-world incomplete datasets. Since the ground truth of cluster centers of the complete dataset is unknown and cannot be approximated as in Section~\ref{rkpod_sec_experiments}, we mainly concern with the practical effects of clustering. 
We consider two single-cell RNA sequence datasets: 
\begin{itemize}
    \item \textit{Usoskin} dataset contains 622 neuronal cells ($n=622$) that are divided into four sensory subtypes ($k=4$): peptidergic nociceptors (PEP), non-peptidergic nociceptors (NP), neurofilament containing (NF) and tyrosine hydroxylase containing (TH). We here use a subset of this dataset and corresponding labels provided by \cite{usoskin2015unbiased}, which consists of 452 genes ($p=452$). The total missing proportion is about 73\%. 
    \item \textit{Treutlein} dataset contains 265 cells ($n=265$) that are in different transcriptional states during direct reprogramming process from mouse embryonic fibroblasts to induced neuronal cells. We here use a subset of this dataset and corresponding assignment of states provided by \cite{treutlein2016dissecting}, which consists of 396 genes ($p=396$) and 7 types of states ($k=7$), roughly including the initial state (MEF), induced state, intermediate states, early and terminal neuron states, as well as those cells that fail to reprogram. The total missing proportion is about 44\%. 
\end{itemize}
For both datasets, since $p$ and the missing proportion are large, there is no complete data point left and thus the complete-case analysis method is no longer applicable. Moreover, the multiple imputation method takes extremely long time. 
Nevertheless, some imputation methods specially for scRNA-seq data have been developed and widely used, such as \textit{scImpute} \citep{li2018accurate}, we will apply the \textit{zero imputation} and \textit{scImpute} methods to original data and then perform $k$-means clustering on imputed data in this section, where the R package {\fontfamily{qcr}\selectfont scImpute} will be used. Also, we will consider the $k$-POD clustering as a peer method for comparison. 

Table~\ref{rkpod_table_realdata_cer_incomplete} summarizes the averaged CER of 30 repetitions of different methods with standard deviation in brackets, and shows that the group lasso type of proposed method has the lowest CER and outperforms other methods on both datasets. 
This coincides with the results of numerical experiments, where the group lasso type of proposed method shows more stable and better performance in more complicated cases (large $p$ and complicated missingness mechanism with a large proportion of missingness), because of the adjustment on both noise and relevant features. 

\begin{figure}[htbp]
\captionsetup[subfigure]{justification=centering}
    \centering
    \begin{subfigure}{0.22\textwidth}
        \includegraphics[width=\textwidth]{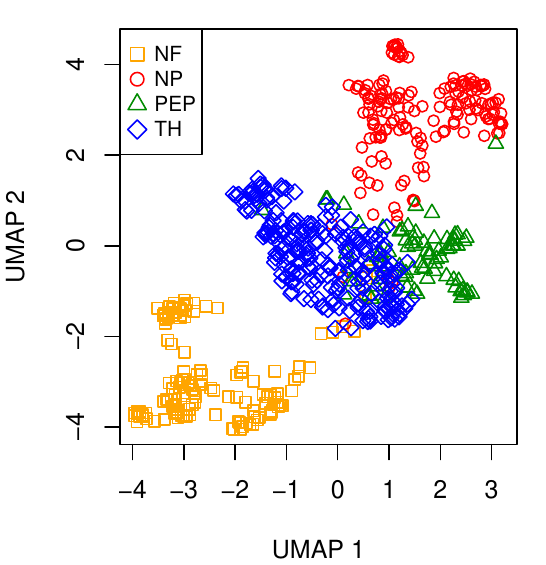}
        \centering{\caption*{Ground truth\\ \ }}
    \end{subfigure}\begin{subfigure}{0.22\textwidth}
        \includegraphics[width=\textwidth]{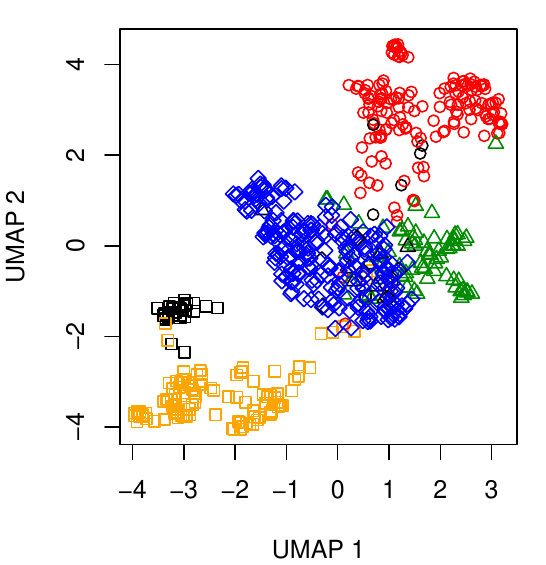}
        \caption*{Reg. $k$-POD\\(group lasso)}
    \end{subfigure}
    \begin{subfigure}{0.22\textwidth}
        \includegraphics[width=\textwidth]{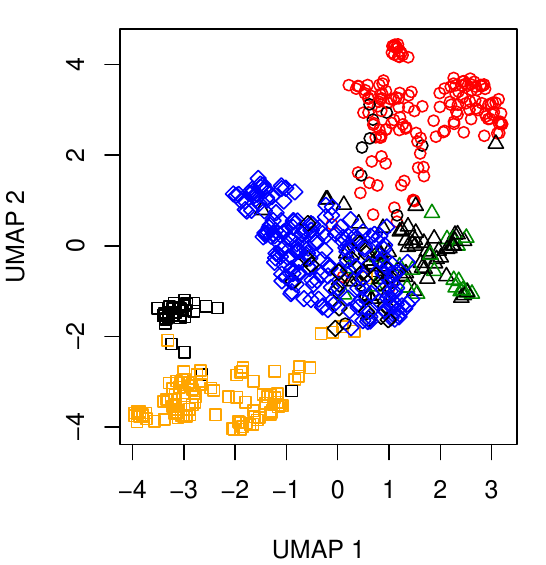}
        \caption*{Reg. $k$-POD\\($l_0$)}
    \end{subfigure}\par
    \begin{subfigure}{0.22\textwidth}
        \includegraphics[width=\textwidth]{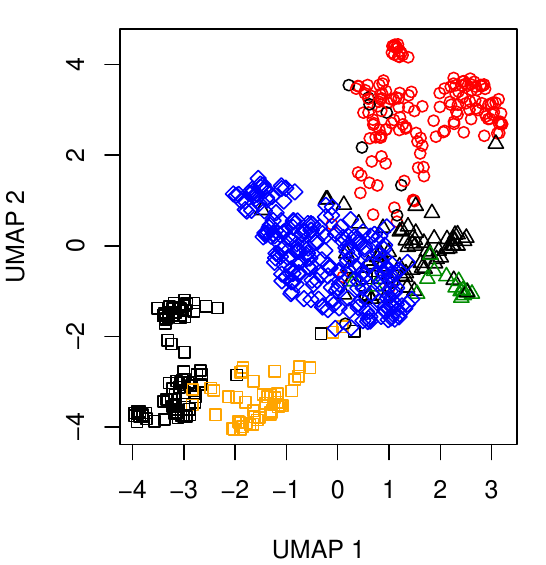}
        \centering{\caption*{Zero imputation}}
    \end{subfigure}
    \begin{subfigure}{0.22\textwidth}
        \includegraphics[width=\textwidth]{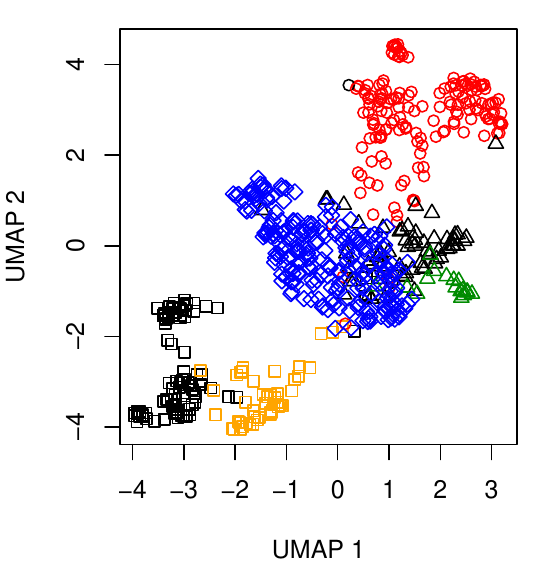}
        \centering{\caption*{scImpute}}
    \end{subfigure}
    \begin{subfigure}{0.22\textwidth}
        \includegraphics[width=\textwidth]{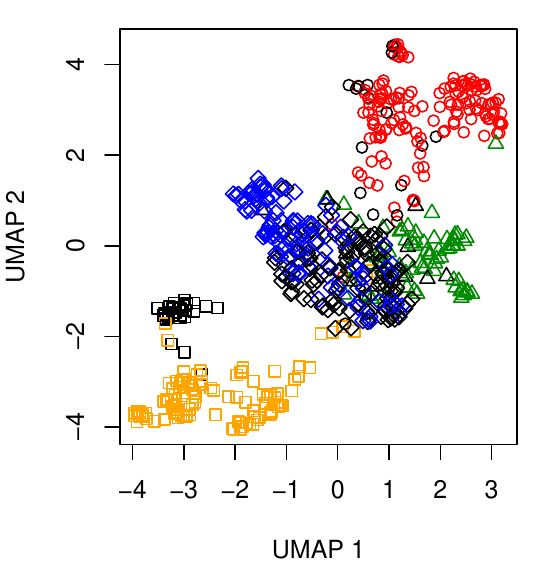}
        \caption*{$k$-POD}
    \end{subfigure}
    \caption{\normalfont The visualization of clustering results using UMAP for cells in \textit{Usoskin} dataset. The shape of points is the true label. The four colors represent the correctly estimated labels, while black means mis-clustered. }
    \label{rkpod_fig_Usoskin}
\end{figure}

For \textit{Usoskin} dataset, we provide the visualization of clustering results in Figure~\ref{rkpod_fig_Usoskin} by using UMAP \citep{becht2019dimensionality}, where the shape of points represents the ground truth label and the color (red, blue, green, orange) represents the estimated label given by different methods, while the black means mis-clustered cells. 
It shows that the group lasso type of proposed method gives a relatively more separated partition for all 622 cells, whose clustering result is closest to the ground truth of the cluster structure among these cells, expect for incorrectly grouping several NF cells to be PEP (around 50 mis-clustered cells). However, peer methods have over 150 cells mis-clustered, where imputation methods almost confuse PEP and NF cells, and the $k$-POD method fails to discriminate TH cells and PEP cells. 

\begin{figure}[htbp]
\captionsetup[subfigure]{justification=centering}
    \centering
    \begin{subfigure}{0.22\textwidth}
        \includegraphics[width=\textwidth]{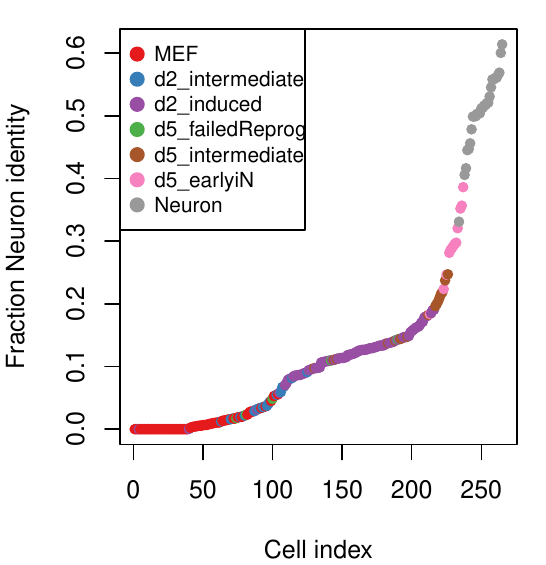}
        \centering{\caption*{Ground truth\\ \ }}
    \end{subfigure}\begin{subfigure}{0.22\textwidth}
        \includegraphics[width=\textwidth]{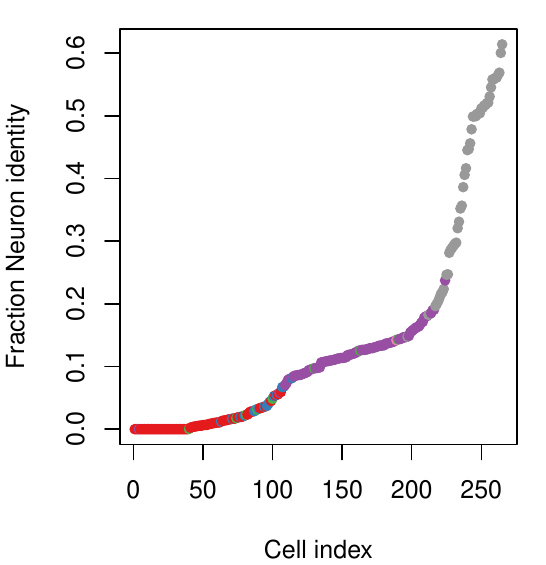}
        \caption*{Reg. $k$-POD\\(group lasso)}
    \end{subfigure}
    \begin{subfigure}{0.22\textwidth}
        \includegraphics[width=\textwidth]{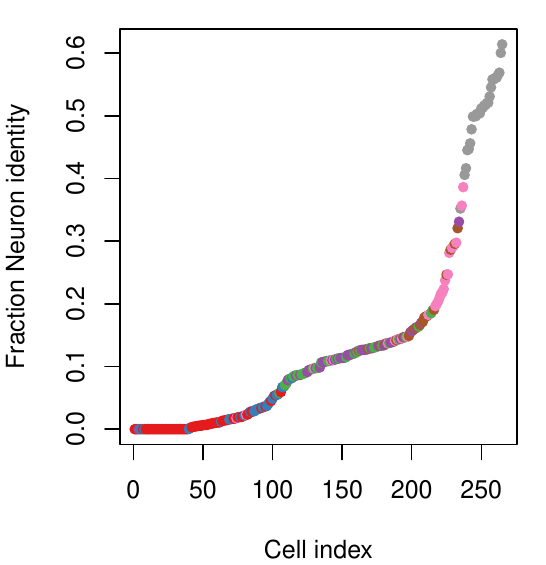}
        \caption*{Reg. $k$-POD\\($l_0$)}
    \end{subfigure}\par
    \begin{subfigure}{0.22\textwidth}
        \includegraphics[width=\textwidth]{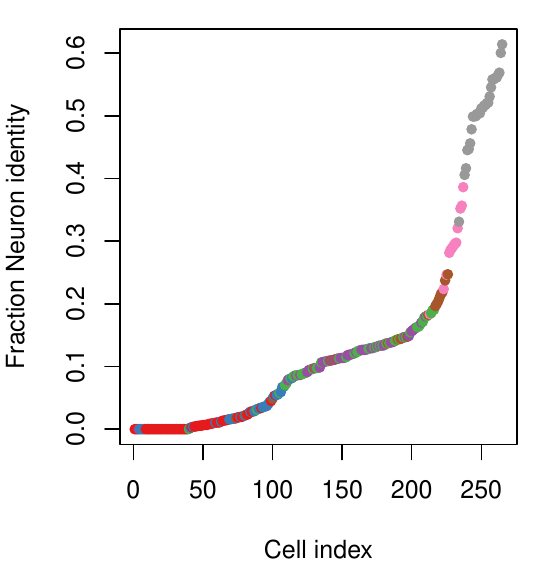}
        \centering{\caption*{Zero imputation}}
    \end{subfigure}
    \begin{subfigure}{0.22\textwidth}
        \includegraphics[width=\textwidth]{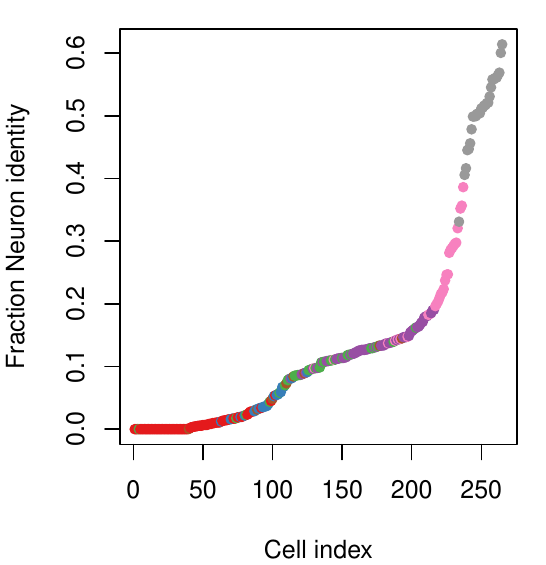}
        \centering{\caption*{scImpute}}
    \end{subfigure}
    \begin{subfigure}{0.22\textwidth}
        \includegraphics[width=\textwidth]{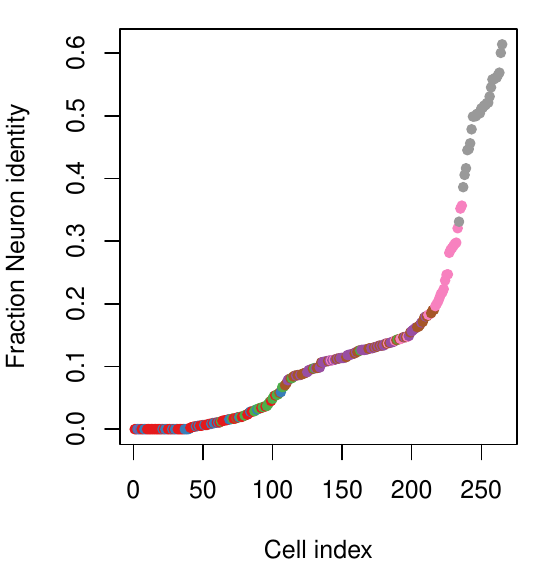}
        \caption*{$k$-POD}
    \end{subfigure}
    \caption{\normalfont The conversion of estimated states for cells in \textit{Treulein} datasets. The y-axis represents the identity of a cell to the terminal neuron state, and the x-axis represents the cell index ordered by the identity. The color of points represents the estimated state.  }
    \label{rkpod_fig_treulein}
\end{figure}

For \textit{Treutlein} dataset, we provide the estimated states from initial MEF to terminal neuron in Figure~\ref{rkpod_fig_treulein}, where the y-axis represents the degree of identity of a cell to the terminal neuron state, and the x-axis represents the cell index ordered by the identity (the identity degree of each cell is provided by \cite{treutlein2016dissecting}). The color of each point represents the estimated state by different methods. 
It shows that the proposed method (group lasso type) partitions states of the conversion most clearly, which correctly distinguishes initial MEF state, induced state and terminal neuron state, except for mis-clustering the early neuron state. Even though peer methods can separate early and terminal neuron states, they almost mix up the induced state and other intermediate states. Whereas, distinguishing the induced state from other states is a key step in determining the expression threshold of target genes required to productively initiate the reprogramming process, which shows the practical utility of the proposed method in real-world applications.

\begin{table*}[htbp]
\centering
\caption{\normalfont CER (standard deviations in brackets) of different methods for real-world incomplete datasets}
\label{rkpod_table_realdata_cer_incomplete}
\begin{adjustbox}{center}
\resizebox{\columnwidth}{!}{
\begin{tabular}{@{}lcccccc@{}}
\toprule
 Dataset &  \makecell{Zero imputation} & \makecell{scImpute} & $k$-POD & \makecell{Reg. $k$-POD (group lasso)} & \makecell{Reg. $k$-POD ($l_0$)} \\
\midrule
Usoskin & 0.138 (0.00) & 0.118 (0.00) & 0.198 (0.05) & \bf{0.064 (0.01)}  & 0.167 (0.03) \\
Treutlein & 0.110 (0.00) & 0.091 (0.00) & 0.126 (0.02) & \bf{0.084 (0.01)} & 0.136 (0.02)\\
 \bottomrule
\end{tabular}}
\end{adjustbox}
\end{table*}

\clearpage
\section{Discussions}
\label{rkpod_sec_discussion}

In this paper, we proposed a regularized $k$-POD clustering method for high-dimensional missing data. The proposed loss function consists of the existing $k$-POD loss and a regularization penalty term on cluster centers. A simple and fast majorization-minimization algorithm is employed for optimization, which includes an imputation step based on cluster means and a clustering step based on imputed data. 
The proposed method gives a feature-sparse estimator for cluster centers that can be less biased than the existing $k$-POD method in high-dimensional cases, showing its capability to mitigate bias in $k$-means-type clustering for high-dimensional missing data, and meanwhile, to maintain the computational efficiency and flexibility. 
Moreover, comparisons with existing methods based on simulations verified the effectiveness of our method in reducing bias and improving clustering performance. Applications to real-world scRNA-seq data demonstrated its practical utility. 

There are still some limitations in this study. 
First, the effectiveness of bias reduction relies on recognizing true noise features correctly. The effect of feature selection of the proposed method is only verified through simulations and lacks theoretical guarantees, which is an important direction for our future work. Moreover, in practice, it is hard to appropriately define true noise features, making the identification rather challenging in real-world applications. 
Second, when the missing proportion is large or the missing mechanism is complex, as illustrated in our simulations, the proposed method is less effective. A possible explanation is that our method essentially relies on \textit{mean imputation} (within each cluster separately), which, even if the cluster assignment is correct, is unbiased only when missingness is completely at random. To handle more complex cases, the data-generating process and missingness mechanisms need to be considered, which requires making reasonable assumptions based on domain-specific knowledge. 
Third, since the proposed method is a $k$-means-type clustering method, it still retains limitations common to $k$-means, such as for data with overlapping or imbalanced clusters.

\section*{SUPPLEMENTARY INFORMATION}
Supplementary material is available online, organized as follows: (a) Details of Algorithm~\ref{alg_rkm} are summarized in Section~A. (b) Section~B is the derivation of BIC used in the main paper. (c) All proofs are in Section~C. (d) More details and results of numerical experiments are given in Section~D. The R code for implementation is available at \url{https://github.com/GXguanxin/rkpod}. 

\section*{FUNDING}
This research was supported by China Scholarship Council (NO.202108050077 to XG) and JSPS, Japan KAKENHI Grant (JP20K19756, JP20H00601, and JP24K14855 to YT). 

\section*{CONFLICT OF INTEREST STATEMENT}
None declared.

%%%%%%%%%%%%%%%%%%%%%%%%%%%%%%%%%%%%%%%%%%%%%%%%%%%%%%%%%%%%%%%%%%%%%%%%%%%%%%%%%%%%%%%%%%%%%%%%%%%%%%%%%%%%%%%%%%%%%%%%%%%%%%%%%%%%%%%%%%%%%%%%%%%%%%%%%%%%%%%%%%%%%%%%%%%%%%%%%%%%%%%%%%%%%%%%%%%%%%%%%%%%%%%%%%%%%%%%%%%%%%%%%%%%%%%%%%%%%%%%%%%%%%%%%%%%%%%%%%%%%%%%%%%%%%%%%%%%%%%%%%%%%%%%%%%%%%%%%%%%%%%%%%%%%%%%%%%%%%%%%%%%%%%%%%%%%%%%%%%%%%

%USE THE BELOW OPTIONS IN CASE YOU NEED AUTHOR YEAR FORMAT.
\bibliographystyle{abbrvnat}
\bibliography{reference}

\end{document}

\title[]{Supplementary materials for ``Regularized $k$-POD: Sparse $k$-means clustering for high-dimensional missing data"}

\author[]{Xin Guan}
\author[]{Yoshikazu Terada}

\abstract{}

\maketitle

\doublespacing

\vspace{-30pt}
\section{Details of Algorithm~2}

In this section, we provide technical details of Algorithm~2 in the main paper. 

\subsection{Derivation of updating cluster centers for $J=J_0$}

For $J=J_0$, given $\bm{U}^{(r+1)}$, the update $\bm{M}^{(r+1)}$ is given by the solution of
\begin{align*}
    \min_{\bm{M}} \| \widehat{\bm{X}} - \bm{U}^{(r+1)}\bm{M} \|_F^2 + \lambda \sum_{j=1}^{p} \mathds{1}(\|\bm{\mu}_{(j)}\|>0).
\end{align*}
Because $ \| \widehat{\bm{X}} - \bm{U}^{(r+1)}\bm{M} \|_F^2=\sum_{j=1}^{p} \|\hat{\bm{x}}_{(j)} - \bm{U}^{(r+1)}\bm{\mu}_{(j)}\|^2 $, we can separately solve the minimization problem in each feature, that is, for each $j=1,\dots,p$, 
    \begin{align*}
        \min_{\bm{\mu}_{(j)}} \|\hat{\bm{x}}_{(j)} - \bm{U}^{(r+1)}\bm{\mu}_{(j)}\|^2 + \lambda \mathds{1}(\|\bm{\mu}_{(j)}\|>0).
    \end{align*}
If the solution $\hat{\bm{\mu}}_{(j)}\neq \bm{0}_k$, then $\mathds{1}(\|\hat{\bm{\mu}}_{(j)}\|>0)=1$ and the KKT condition implies that
    \begin{align*}
        \hat{\bm{\mu}}_{(j)}=(\bm{U}^{(r+1),T} \bm{U}^{(r+1)})^{-1}\bm{U}^{(r+1),T}\hat{\bm{x}}_{(j)}. 
    \end{align*}
If the solution $\hat{\bm{\mu}}_{(j)}= \bm{0}_k$, then the corresponding value of objective function is $\|\hat{\bm{x}}_{(j)}\|^2$, which should be smaller than the objective function at any non-zero point. Therefore, there must be
    \begin{align*}
        \|\hat{\bm{x}}_{(j)} - \bm{U}^{(r+1)}\bm{v}_{(j)}\|^2 + \lambda \geq \|\hat{\bm{x}}_{(j)}\|^2,
    \end{align*}
where $\bm{v}_{(j)}=(\bm{U}^{(r+1),T} \bm{U}^{(r+1)})^{-1}\bm{U}^{(r+1),T}\hat{\bm{x}}_{(j)}$. We obtained Eq.~(5) of Algorithm 2.

\subsection{Derivation of updating cluster centers for $J=J_1$}

For $J=J_1$, given $\bm{U}^{(r+1)}$, the update $\bm{\mu}^{(r+1)}$ is given by the solution of 
\begin{align*}
    \min_{\bm{M}}\|\widehat{\bm{X}} - \bm{U}^{(r+1)}\bm{M}\|_F^2 + \lambda \sum_{j=1}^{p} w_j \|\bm{\mu}_{(j)}\|.
\end{align*}
We denote the above objective function by $f(\bm{M})$ as that in the main paper. 
Since it is not easy to derive an explicit solution, we instead apply the MM algorithm again to obtain $\bm{M}^{(r+1)}$. As introduced in the main paper, at any point $\bm{M}^{(r_s)}$ ($s\in\mathbb{N}$) we consider the following function 
\begin{align*}
    h(\bm{M}\mid \bm{M}^{(r_s)}) = \|\widehat{\bm{X}} - \bm{U}^{(r+1)}\bm{M}\|_F^2 + \lambda \sum_{j=1}^{p} w_j \left(  \frac{\|\bm{\mu}_{(j)}\|^2}{2\|\bm{\mu}_{(j)}^{(r_s)}\|} + \frac{1}{2}\|\bm{\mu}_{(j)}^{(r_s)}\|\right). 
\end{align*}
Based on the basic equality, we have for each $j=1,\dots,p$, 
\begin{align*}
      \frac{\|\bm{\mu}_{(j)}\|^2}{2\|\bm{\mu}_{(j)}^{(r_s)}\|} + \frac{1}{2}\|\bm{\mu}_{(j)}^{(r_s)}\| \geq \|\bm{\mu}_{(j)}\|,
\end{align*}
where the equality holds if and only if $\bm{\mu}_{(j)}^{(r_s)}=\bm{\mu}_{(j)}$. 
It follows that 
\begin{align*}
    h(\bm{M}\mid \bm{M}^{(r_s)})\geq f(\bm{M}) \quad\text{   and   }\quad  h(\bm{M}^{(r_s)}\mid \bm{M}^{(r_s)}) = f(\bm{M}^{(r_s)}),
\end{align*}
which means that the domination condition and tangency condition are satisfied and $h(\bm{M}\mid \bm{M}^{(r_s)})$ majorizes $f(\bm{M})$ at any $\bm{M}^{(r_s)}$. Now we can apply the MM algorithm in the following way. Starting from $\bm{M}^{(r_0)}$, the $(s+1)$-th iteration includes: (i) construct the majorization function $h(\bm{M}\mid \bm{M}^{(r_s)})$ with current $\bm{M}^{(r_s)}$; (ii)~update $\bm{M}^{(r_{s+1})}$ by minimizing $h(\bm{M}\mid \bm{M}^{(r_s)})$, the solution of which can be easily derived by KKT condition in each feature, that is, for any $j=1,\dots,p$,
\begin{align*}
    \bm{\mu}_{(j)}^{(r_{s+1})}=\left( \bm{U}^{(r+1),T}\bm{U}^{(r+1)} + \frac{\lambda w_j}{2 \|\bm{\mu}_{(j)}^{(r_s)} \|}\cdot \bm{I}_k \right)^{-1}  \bm{U}^{(r+1),T}  \hat{\bm{x}}_{(j)},
\end{align*}
where $\bm{I}_k$ is the identity matrix with the size of $k\times k$. 
This procedure ensures that $f(\bm{M}^{(r_{s+1})})\leq f(\bm{M}^{(r_{s})})$ for any $s\in\mathbb{N}$. 
As stated in the main paper, there is no need to exactly minimize $f(\bm{M})$. Instead, reducing $f(\bm{M})$ is enough. Therefore, to simplify the computation, we only conduct once iteration about $s$, that is, we start from $\bm{M}^{(r_0)}=\bm{M}^{(r)}$ and update the 
$j$-th column of $\bm{M}^{(r+1)}$ by 
\begin{align*}
    \bm{\mu}_{(j)}^{(r+1)}=\left( \bm{U}^{(r+1),T}\bm{U}^{(r+1)} + \frac{\lambda w_j}{2 \|\bm{\mu}_{(j)}^{(r)} \|}\cdot \bm{I}_k \right)^{-1}  \bm{U}^{(r+1),T}  \hat{\bm{x}}_{(j)},
\end{align*}
which is Eq.~(6) of Algorithm~2. 

\subsection{Discussion on Algorithm~2 for $J=J_1$}

As explained in the main paper, updating $\bm{M}^{(r+1)}$ for $J=J_1$ is equivalent to the group lasso regression. Specifically, minimizing $f(\bm{M})$ is equivalent to minimizing $f_j(\bm{\mu}_{(j)})$ for each $j=1,\dots,p$, where 
\begin{align*}
    f_j(\bm{\mu}_{(j)})=\| \hat{\bm{x}}_{(j)} - \bm{U}^{(r+1)}\bm{\mu}_{(j)} \|^2+\lambda w_j \|\bm{\mu}_{(j)}\|. 
\end{align*}
It can be viewed as a regression model of response $\hat{\bm{x}}_{(j)}$ on design matrix $\bm{U}^{(r+1)}$ with a group lasso penalty $\|\bm{\mu}_{(j)}\|$, where the number of groups is one. For simplification of notations, we write $\bm{y}$ for $\hat{\bm{x}}_{(j)}$, write $\bm{U}$ for $\bm{U}^{(r+1)}$ and write $\bm{\beta}$ for $\bm{\mu}_{(j)}$. 

Following the method of \cite{yang2015fast}, we can construct a majorization function $\tilde{h}_j(\bm{\beta} \mid \bm{\beta}^{(s)})$ for $f_j(\bm{\beta})$ at any point $\bm{\beta}^{(s)}$ via quadratic approximation of the first term of $f_j(\bm{\beta})$. Let $l(\bm{\beta})=\| \bm{y}- \bm{U}\bm{\beta} \|^2$. Because
\begin{align*}
    l(\bm{\beta}) \leq  l(\bm{\beta}^{(s)}) + (\bm{\beta}-\bm{\beta}^{(s)})^T\nabla l(\bm{\beta}^{(s)}) + \frac{1}{2}(\bm{\beta}-\bm{\beta}^{(s)})^T \bm{H} (\bm{\beta}-\bm{\beta}^{(s)}), 
\end{align*}
where $\nabla l(\bm{\beta}^{(s)})=-2\bm{U}^{T}(\bm{y} - \bm{U}\bm{\beta}^{(s)})$ and $\bm{H}=2\bm{U}^{T}\bm{U}$, 
we can define 
\begin{align*}
    \tilde{h}_j(\bm{\beta} \mid \bm{\beta}^{(s)})=l(\bm{\beta}^{(s)}) + ( \bm{\beta}-\bm{\beta}^{(s)}  )^T\cdot \left(-2\bm{U}^{T}\right) \cdot (\bm{y} - \bm{U}\bm{\beta}^{(s)}) + \frac{\gamma}{2} \| \bm{\beta}-\bm{\beta}^{(s)} \|^2 + \lambda w_j \|\bm{\beta}\|,
\end{align*}
where $\gamma = 2\max_l \sum_{i=1}^{n}u_{il}$ is the largest size of clusters associated with $\bm{U}$. 
Then we have $\tilde{h}_j(\bm{\beta} \mid \bm{\beta}^{(s)})\geq f_j(\bm{\beta})$ for any $\bm{\beta} \in \mathbb{R}^k$ and $\tilde{h}_j(\bm{\beta}^{(s)} \mid \bm{\beta}^{(s)}) = f_j(\bm{\beta}^{(s)})$, which means that $\tilde{h}_j(\bm{\beta} \mid \bm{\beta}^{(s)})$ is a majorization function of $f_j(\bm{\beta})$ at the point $\bm{\beta}^{(s)}$. 
Moreover, the minimizer of $\tilde{h}_j(\bm{\beta} \mid \bm{\beta}^{(s)})$ is give by 
\begin{align*}
    \bm{\beta}^{(s+1)}
    = \tilde{\bm{\beta}}  \cdot \left( 1 - \frac{\lambda w_j/\gamma}{ \| \tilde{\bm{\beta}}  \|  } \right)_+ ,
\end{align*}
where $\tilde{\bm{\beta}}=\bm{\beta}^{(s)}+(2/\gamma)\cdot\bm{U}^T (\bm{y}-\bm{U}\bm{\beta}^{(s)})$ is the gradient descent update of $l(\bm{\beta})$ and $(\cdot)_+=\max(\cdot,0)$. 
Therefore, we propose Algorithm~\ref{alg_rkm_quadratic} for $J=J_1$ based on the quadratic approximation. 

\begin{algorithm}
\doublespacing
\caption{\ Regularized $k$-means clustering using quadratic approximation}
\textbf{Input}: complete data matrix $\widehat{\bm{X}}$, number of clusters $k$. \\
\textbf{Parameters}: regularized parameter $\lambda$, weights $\{w_j\}$
\begin{algorithmic}
 \State Initialize $\bm{M}^{(0)}$ 
 \While{Loss function Eq.~(4) does not converge}
  \State a: Given $\bm{M}^{(r)}$, update $\bm{U}^{(r+1)}$ by: for any $i=1,\dots,n$
  \begin{align*}
      u_{il^{\ast}}^{(r+1)}=\left\{
      \begin{array}{ll}
          1 & \text{  if  }  l^{\ast}=\arg\min_{1\leq l \leq k} \| \hat{\bm{x}}_{i} - \bm{\mu}_l^{(r)} \|^2 \\
          0 & \text{  else}
      \end{array}
      \right. 
  \end{align*}
  \State b: Given $\bm{U}^{(r+1)}$, update $\bm{M}^{(r+1)}$ by: for any $j=1,\dots,p$ 
  \begin{align*}
        &\bm{\mu}_{(j)}^{(r+1)}= \tilde{\bm{v}}_{(j)} \cdot \left( 1- \frac{\lambda w_j/\gamma}{ \| \tilde{\bm{v}}_{(j)} \|  }\right)_+, \\
        \text{where    }  &\tilde{\bm{v}}_{(j)}= \bm{\mu}_{(j)}^{(r)}+(2/\gamma)\cdot \bm{U}^{(r+1),T}\cdot \left(\hat{\bm{x}}_{(j)}-\bm{U}^{(r+1)}\bm{\mu}_{(j)}^{(r)}\right)\\
        &\gamma=2\cdot \max \left\{ \text{diag}\left(\bm{U}^{(r+1),T}\bm{U}^{(r+1)}\right) \right\}
  \end{align*}
 \EndWhile
\end{algorithmic}
\textbf{Output}: $\bm{U}^{(r+1)}$ and $\bm{M}^{(r+1)}$
\label{alg_rkm_quadratic}
\end{algorithm}

\begin{figure}[!h]
    \setlength{\tempwidth}{.22\linewidth}
\settoheight{\tempheight}{\includegraphics[width=\tempwidth]{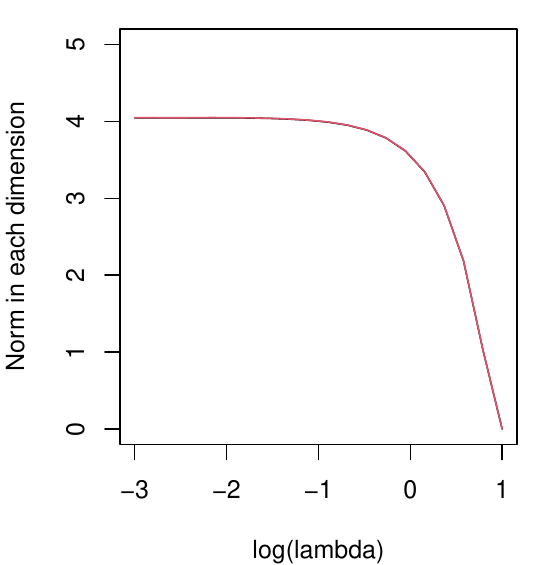}}%
\centering
\hspace{\baselineskip}
\columnname{$p=10$, relevant}\hfil
\columnname{$p=10$, noise}\hfil
\columnname{$p=100$, relevant}\hfil
\columnname{$p=100$, noise}\\
\rowname{Alg. 2}
    \includegraphics[width=0.23\textwidth]{fig_rkm_g_ridgeVSquadratic_p10_relevant_ridge.pdf}
    \includegraphics[width=0.23\textwidth]{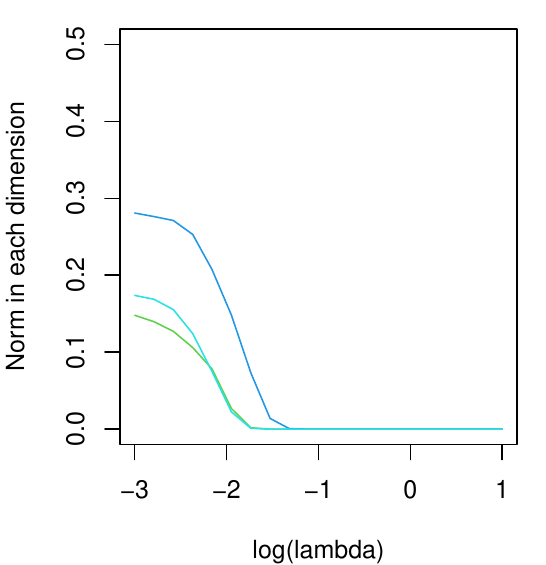}
    \includegraphics[width=0.23\textwidth]{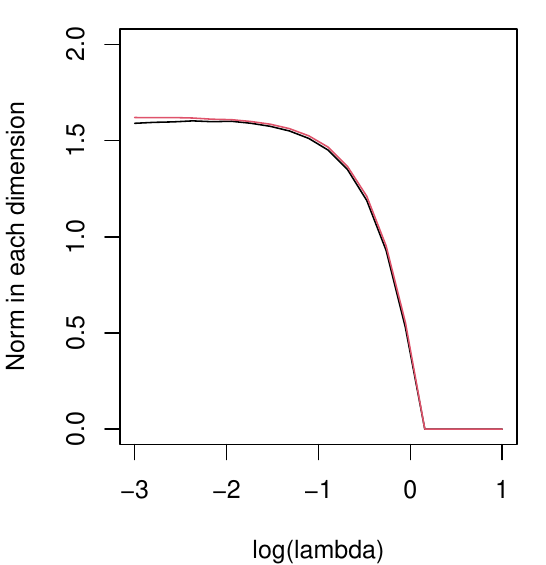}
    \includegraphics[width=0.23\textwidth]{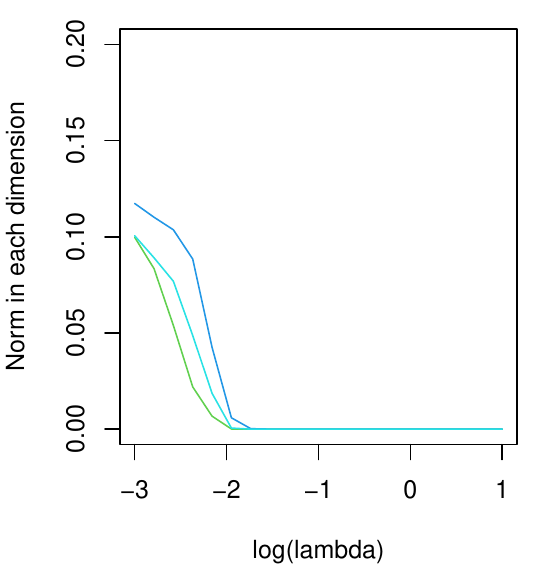}\\
\rowname{Alg. 3}
    \includegraphics[width=0.23\textwidth]{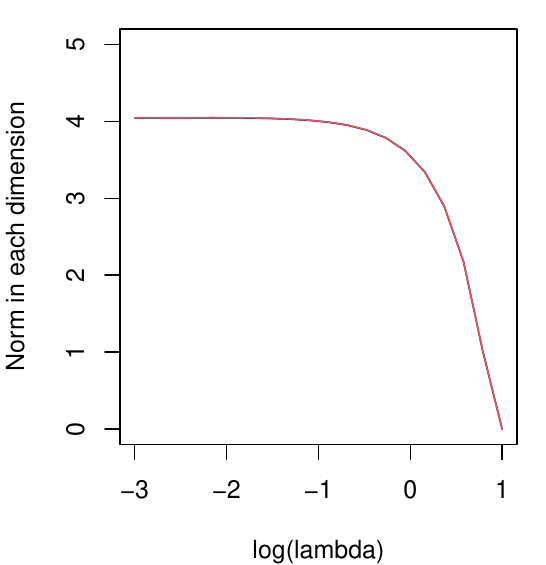}
    \includegraphics[width=0.23\textwidth]{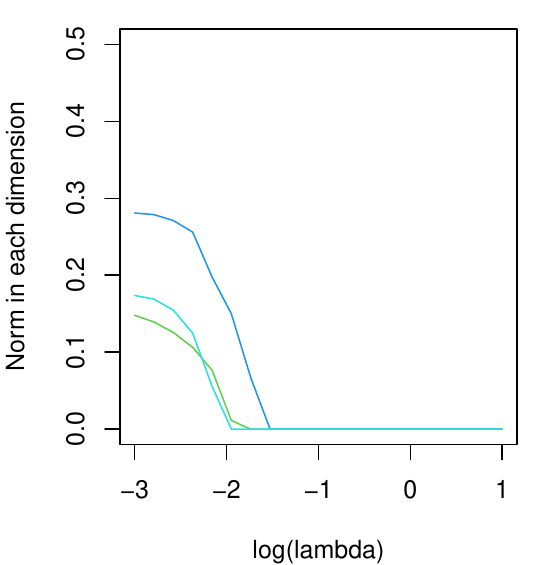}
    \includegraphics[width=0.23\textwidth]{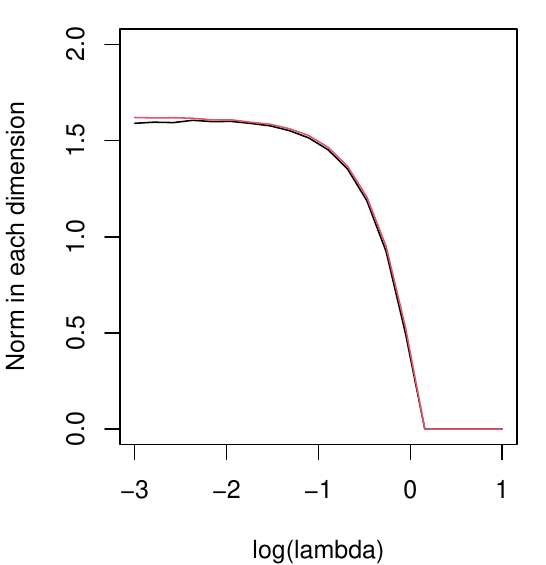}
    \includegraphics[width=0.23\textwidth]{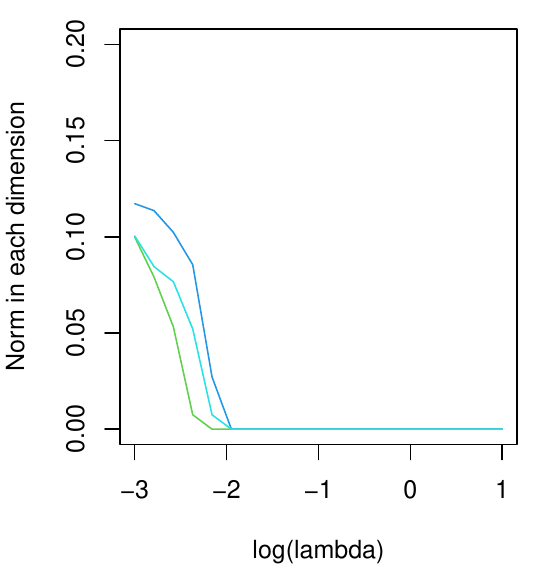}
    \caption{\doublespacing\normalfont Regularization paths of Algorithm~2 (top) and Algorithm~\ref{alg_rkm_quadratic} (bottom). The x-axis is the $\log(\lambda)$ and the y-axis is $\|\bm{\mu}_{(j)}\|$. The four columns are for two relevant features in case of $p=10$, three noise features in case of $p=10$, two relevant features in case of $p=100$ and three noise features in case of $p=100$. }
    \label{fig_rkpod_compare_Quadratic}
\end{figure}

\begin{figure}[!h]
\captionsetup[subfigure]{justification=centering}
    \centering
    \begin{subfigure}{0.28\textwidth}
        \includegraphics[width=\textwidth]{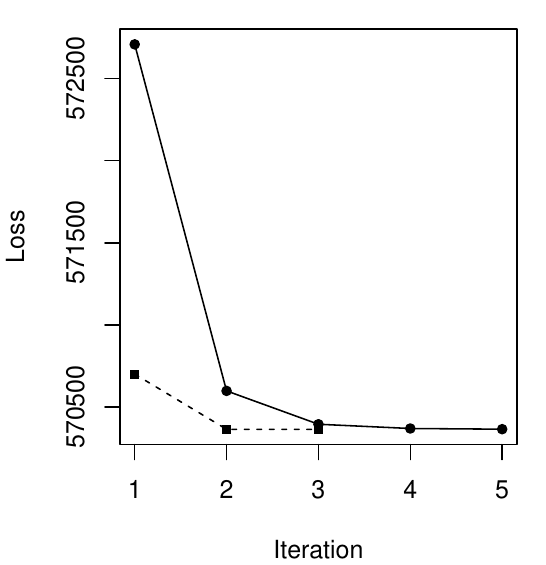}
        \centering{\caption*{(a)}}
    \end{subfigure}
    \begin{subfigure}{0.28\textwidth}
        \includegraphics[width=\textwidth]{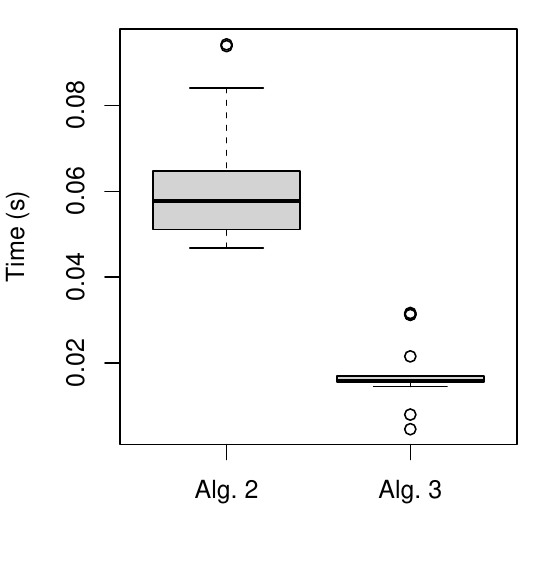}
        \centering{\caption*{(b)}}
    \end{subfigure}
    \caption{\doublespacing\normalfont (a) Convergence of Algorithm~2 (solid) and Algorithm~\ref{alg_rkm_quadratic} (dotted) in the case of $p=100$. \\ (b) Comparison of computational time. }
    \label{fig_rkpod_quad_losstime}
\end{figure}

Next, we compare Algorithm~2 and Algorithm~\ref{alg_rkm_quadratic} via numerical experiments on synthetic complete datasets. 
Figure~\ref{fig_rkpod_compare_Quadratic} illustrates regularization paths of these two algorithms on datasets with $p=10$ and $p=100$, and Figure~\ref{fig_rkpod_quad_losstime} shows the convergence and computational time in the case of $p=100$. 
It can be seen that the paths of two algorithms are almost the same, while Algorithm~\ref{alg_rkm_quadratic} needs fewer iterations and thus less computational time.

\clearpage
\section{Derivation of BIC}
In this section, we provide technical details of deriving the expression of BIC given in Section~2.2.3 of the main paper. 
We first consider the classification likelihood \citep{fraley2002model} to formulate the $k$-means likelihood. Let $\{\bm{x}_i\}_{i=1}^n$ be data points independently draw from the same distribution, $\bm{U}=(u_{il})_{n\times k}\in \{0,1\}^{n\times k}$ be the indicators of membership of $\{\bm{x}_i\}_{i=1}^{n}$ and $\bm{U}\bm{1}_k=\bm{1}_n$. Denote by $\phi_p(\cdot\mid \bm{\mu}_l)$ the density function of Gaussian distribution in $\mathbb{R}^p$ with mean vector $\bm{\mu}_l=(\mu_{l1},\dots,\mu_{lp})$ and covariance matrix $\sigma^2 \bm{I}_p$, where $\sigma^2$ is fixed. Write $\bm{M}=(\mu_{lj})_{k\times p}$. The classification likelihood of $\bm{x}_i$, $i=1,\dots,n$ is given by 
\begin{align*}
    \ell (\bm{x}_i\mid \bm{U},\bm{M} )&=\prod_{l=1}^{k}\left\{\phi_p(\bm{x}_i\mid \bm{\mu}_l)\right\}^{u_{il}} \\
    &=\prod_{l=1}^{k}\left\{ (2\pi\sigma^2)^{-\frac{p}{2}} \exp\left(-\frac{\sum_{j=1}^{p}(x_{ij}-\mu_{lj})^2  }{2\sigma^2}\right)\right\}^{u_{il}}
\end{align*}
Now we consider the missing data. 
Assume that for $\bm{x}_i=(x_{i1},\dots,x_{ip})$, any element $x_{ij}$ is missing completely at random (MCAR) and $\bm{x}_i$ would be partially observed. 
As introduced in the main paper, we use a binary random variable $r_{ij}$ to indicate whether $x_{ij}$ is observed. That is, $r_{ij}=1$ if $x_{ij}$ is observed, 0 otherwise. Write $\bm{r}_i=(r_{i1},\dots,r_{ip}) \in \{0,1\}^{p}$. The MCAR mechanism means that $\bm{r}_i$ is independent with $\bm{x}_i$.  
Because the covariance matrix is $\sigma^2\bm{I}_p$, we have $\phi_p(\bm{x}_i\mid \bm{\mu}_l)=\prod_{j=1}^{p} \phi(x_{ij}\mid \mu_{lj})$, where $\phi(\cdot\mid \mu_{lj})$ is the density function of Gaussian distribution in $\mathbb{R}$ with mean $\mu_{lj}$ and variance $\sigma^2$. 
Then the likelihood of $\bm{x}_i$ can be written as 
\begin{align*}
    \ell(\bm{x}_i \mid \bm{U},\bm{M})&=\prod_{l=1}^{k}\left\{\prod_{j=1}^{p}\phi(x_{ij}\mid \mu_{lj})\right\}^{u_{il}}\\
    &=\prod_{l=1}^{k}\left\{\prod_{j:r_{ij}=1}\phi(x_{ij}\mid \mu_{lj}) \cdot \prod_{j:r_{ij}=0}\phi(x_{ij}\mid \mu_{lj})  \right\}^{u_{il}}\\
    &=\prod_{l=1}^{k}\left\{\prod_{j:r_{ij}=1}\phi(x_{ij}\mid \mu_{lj}) \right\}^{u_{il}} \cdot \prod_{l=1}^{k}\left\{\prod_{j:r_{ij}=0}\phi(x_{ij}\mid \mu_{lj}) \right\}^{u_{il}}. 
\end{align*}
The likelihood of partially observed part, denoted by $\bm{x}_i^{obs}$, is thus equivalent to the density of marginal distribution of $\{x_{ij}\mid  r_{ij}=1,\; \forall j=1,\dots,p\}$, which is given by 
\begin{align*}
    \ell(\bm{x}_i^{obs} \mid \bm{U},\bm{M},\bm{r}_i) = \prod_{l=1}^{k}\left\{\prod_{j:r_{ij}=1}\phi(x_{ij}\mid \mu_{lj}) \right\}^{u_{il}}. 
\end{align*}
Therefore, the likelihood of partially observed sample $\{\bm{x}_1^{obs},\dots, \bm{x}_n^{obs}\}$ is given by 
\begin{align*}
    \ell_n (\bm{x}_1^{obs},\dots,\bm{x}_n^{obs} \mid \bm{R}, \bm{U},\bm{M} )=\prod_{i=1}^{n}\prod_{l=1}^{k}\left\{(2\pi\sigma^2)^{-\frac{\|\bm{r}_i\|}{2}} \exp \left(-\frac{\sum_{j=1}^{p}r_{ij}(x_{ij}-\mu_{lj})^2 }{2\sigma^2}\right)  \right\}^{u_{il}}.
\end{align*}
Then we have 
\begin{align*}
    \text{BIC}=\sum_{i=1}^{n}\|\bm{r}_i\| \log (2\pi\sigma^2) + \frac{1}{\sigma^2}\sum_{i=1}^{n}\sum_{j=1}^{p}\sum_{l=1}^{k} r_{ij}u_{il}(x_{ij}-\mu_{lj})^2  + \log(n)\cdot df. 
\end{align*}
The first term is a fixed constant, and when $\sigma^2=1$, the second term is equivalent to $\|\mathcal{P}_{\Omega}(\bm{X}-\bm{UM})\|_F^2$, then we can write BIC to be 
\begin{align*}
    \text{BIC}=\|\mathcal{P}_{\Omega}(\bm{X}-\bm{UM})\|_F^2+ \log (n) \cdot df,
\end{align*}
where $df$ is the number of independent parameters, which is simply $kd$ with $d=\sum_{j=1}^{p}\mathds{1}(\|\bm{\mu}_{(j)}\|>0)$. Note that $df$ can be further approximated by using the effective degree of freedom as discussed in \cite{hofmeyr2020degrees}.

\section{Proof of Proposition 2.1} 

 (a) 
    For $J(\cdot)=J_0(\cdot)$, estimating $\widehat{\bm{M}}$ is equivalent to solving 
    \begin{align*}
        \min_{\bm{\mu}_{(j)}}\quad \sum_{i=1}^{n} r_{ij} (x_{ij} - \bm{u}_i \bm{\mu}_{(j)})^2 + \lambda \mathds{1} (\|\bm{\mu}_{(j)}\|>0)
    \end{align*}
    for each $j=1,\dots,p$, where $\bm{U}$ is associated with $\bm{M}$ and $\bm{u}_i$ is the $i$-th row of $\bm{U}$.  

    If the minimizer $\hat{\bm{\mu}}_{(j)}\neq (0,0,\dots,0)^T$, then $\mathds{1} (\|\hat{\bm{\mu}}_{(j)}\|>0)=1$ and the optimality according to KKT condition implies that 
    \begin{align*}
        \bm{0}_k=-2\sum_{i=1}^{n} r_{ij} \hat{\bm{u}}_i^T (x_{ij} -  \hat{\bm{u}}_i \hat{\bm{\mu}}_{(j)}). 
    \end{align*}
    It follows that for all $l=1,\dots,k$, 
    \begin{align*}
        \hat{\mu}_{lj}=\frac{\sum_{i=1}^{n} \hat{u}_{il}r_{ij}x_{ij}  }{ \sum_{i=1}^{n}\hat{u}_{il}r_{ij} }. 
    \end{align*}
    According to the definition of $\bar{\mu}_{lj}$, we have $\hat{\mu}_{lj}=\bar{\mu}_{lj}$. 

    If $\hat{\bm{\mu}}_{(j)}= (0,0,\dots,0)^T$, then $\mathds{1} (\|\hat{\bm{\mu}}_{(j)}\|>0)=0$ and the optimality according to KKT condition implies that for any $\bm{V}\in \mathbb{R}^{k\times p}$, its $j$-th column $\bm{v}_{(j)}\in \mathbb{R}^{k}$ satisfies
    \begin{align*}
        \sum_{i=1}^{n}r_{ij}(x_{ij}-\bm{u}_i^{(v)} \bm{v}_{(j)})^2 + \lambda \geq \sum_{i=1}^{n}r_{ij}(x_{ij}-\hat{\bm{u}}_i \hat{\bm{\mu}}_{(j)})^2, 
    \end{align*}
    where $\bm{u}_i^{(v)}$ is the $i$-th row of $\bm{U}^{(v)}$, and $\bm{U}^{(v)}$ is the membership matrix associated with $\bm{V}$. 
    Because 
    \begin{align*}
        \sum_{i=1}^{n}r_{ij}(x_{ij}-\bm{u}_i^{(v)} \bm{v}_{(j)})^2 + \lambda \leq \sum_{i=1}^{n}r_{ij}(x_{ij}-\hat{\bm{u}}_i \bm{v}_{(j)})^2 + \lambda
    \end{align*}
    and 
    \begin{align*}
        \sum_{i=1}^{n}r_{ij}(x_{ij}-\hat{\bm{u}}_i \hat{\bm{\mu}}_{(j)})^2 = \sum_{i=1}^{n}r_{ij}(x_{ij}-0)^2 = \sum_{i=1}^{n}r_{ij} x_{ij}^2,
    \end{align*}
    by taking $\bm{v}_{(j)}=\bar{\bm{\mu}}_{(j)}=(\bar{\mu}_{1j},\dots,\bar{\mu}_{kj})^T$, we have 
    \begin{align*}
        \sum_{i=1}^{n}r_{ij}(x_{ij}-\hat{\bm{u}}_i \bar{\bm{\mu}}_{(j)})^2 + \lambda \geq \sum_{i=1}^{n}r_{ij} x_{ij}^2. 
    \end{align*}
    According to the definitions of $\text{WCSS}_j(\widehat{\mathcal{C}})$, $\hat{q}_j$ and $\bar{\sigma}_j^2$ in Section~2.3 of the main paper, we obtain
    \begin{align*}
        \lambda \geq n\cdot \hat{q}_j \bar{\sigma}_j^2 - n\cdot \text{WCSS}_j(\widehat{\mathcal{C}}), 
    \end{align*}
    which completes the proof of (a). 

    (b) 
    For $J(\cdot)=J_1(\cdot)$ with weights $\{w_j\}_{j=1}^{p}$, estimating $\widehat{\bm{M}}$ is equivalent to solving 
    \begin{align*}
        \min_{\bm{\mu}_{(j)}}\quad \sum_{i=1}^{n} r_{ij} (x_{ij} - \bm{u}_i \bm{\mu}_{(j)})^2 + \lambda w_j \|\bm{\mu}_{(j)}\|  
    \end{align*}
    for each $j=1,\dots,p$, where $\bm{U}$ is associated with $\bm{M}$ and $\bm{u}_i$ is the $i$-th row of $\bm{U}$.  

    If the minimizer $\hat{\bm{\mu}}_{(j)}\neq (0,0,\dots,0)^T$, then the optimality according to KKT condition implies that 
    \begin{align}
    \begin{aligned}
        \bm{0}_k&=-2\sum_{i=1}^{n} r_{ij} \hat{\bm{u}}_i^T (x_{ij} -  \hat{\bm{u}}_i \hat{\bm{\mu}}_{(j)}) + \lambda w_j \frac{\hat{\bm{\mu}}_{(j)}}{\|\hat{\bm{\mu}}_{(j)}\|}. 
    \end{aligned}
    \label{eq_proof_proposition_1}
    \end{align}
    That is, for each $l=1,\dots,k$, we have
    \begin{align*}
        \sum_{i=1}^{n}r_{ij}\hat{u}_{il} (x_{ij}-\hat{\mu}_{lj}) = \frac{\lambda w_j}{2\|\hat{\bm{\mu}}_{(j)}\|}\cdot \hat{\mu}_{lj}. 
    \end{align*}
    Recall that 
    \begin{align*}
        \bar{\mu}_{lj}=\frac{\sum_{i=1}^{n} \hat{u}_{il}r_{ij}x_{ij} }{ \sum_{i=1}^{n}\hat{u}_{il}r_{ij} },
    \end{align*}
    then we obtain
    \begin{align*}
        \hat{\mu}_{lj}=\left(\frac{\lambda w_j}{ 2\|\hat{\bm{\mu}}_{(j)}\|\cdot \sum_{i=1}^{n}\hat{u}_{il}r_{ij} }+1\right)^{-1}\cdot \bar{\mu}_{lj}. 
    \end{align*}
    
    Moreover, since Eq.~(\ref{eq_proof_proposition_1}) is equivalent to 
    \begin{align*}
        \bm{0}_k=-2 \widehat{\bm{U}}^T \left\{\bm{x}_{(j)} \circ \bm{r}_{(j)} - \big(\widehat{\bm{U}}\hat{\bm{\mu}}_{(j)}\big)\circ \bm{r}_{(j)} \right\} + \lambda w_j \frac{\hat{\bm{\mu}}_{(j)}}{\|\hat{\bm{\mu}}_{(j)}\|}.
    \end{align*}
    It follows that 
    \begin{align}
        \left\| \widehat{\bm{U}}^T \left\{\bm{x}_{(j)} \circ \bm{r}_{(j)} - \big(\widehat{\bm{U}}\hat{\bm{\mu}}_{(j)}\big)\circ \bm{r}_{(j)} \right\}  \right\| = \frac{\lambda w_j}{2}. 
    \label{eq_proof_proposition_2}
    \end{align}
    Because the term within $\|\cdot\|$ is a vector in $\mathbb{R}^k$, the $l$-th component of which is 
    \begin{align*}
        \sum_{i=1}^{n} \hat{u}_{il} \left\{ x_{ij}r_{ij} - (\hat{\bm{u}}_i \hat{\bm{\mu}}_{(j)})\cdot r_{ij} \right\} &= \sum_{\bm{x}_i \in \widehat{C}_l} \left\{ x_{ij}r_{ij} - (\hat{\bm{u}}_i \hat{\bm{\mu}}_{(j)})\cdot r_{ij} \right\} \\
        &= \sum_{\bm{x}_i \in \widehat{C}_l} \left( x_{ij}r_{ij} - \hat{\mu}_{lj}r_{ij} \right),
    \end{align*}
    then we have 
    \begin{align*}
        \left\{ \sum_{\bm{x}_i \in \widehat{C}_l} \left( x_{ij}r_{ij} - \hat{\mu}_{lj}r_{ij} \right) \right\}^2 &= \left\{ \sum_{\bm{x}_i \in \widehat{C}_l} x_{ij}r_{ij}  - \left( \sum_{\bm{x}_i \in \widehat{C}_l} r_{ij}\right)\cdot \hat{\mu}_{lj} \right\}^2 \\
        &= \left( \sum_{\bm{x}_i \in \widehat{C}_l} r_{ij}\right)^2 \cdot \left\{ \frac{\sum_{\bm{x}_i \in \widehat{C}_l} x_{ij}r_{ij}}{\sum_{\bm{x}_i \in \widehat{C}_l} r_{ij}} - \hat{\mu}_{lj}\right\}^2.
    \end{align*}
    Thereby, we obtain
    \begin{align*}
        \frac{1}{n^2}\left\| \widehat{\bm{U}}^T \left\{\bm{x}_{(j)} \circ \bm{r}_{(j)} - \big(\widehat{\bm{U}}\hat{\bm{\mu}}_{(j)}\big)\circ \bm{r}_{(j)} \right\} \right\|^2 &= \sum_{l=1}^{k} \left( \frac{1}{n}\sum_{\bm{x}_i \in \widehat{C}_l} r_{ij}\right)^2 \cdot \left\{ \frac{\sum_{\bm{x}_i \in \widehat{C}_l} x_{ij}r_{ij}}{\sum_{\bm{x}_i \in \widehat{C}_l} r_{ij}} - \hat{\mu}_{lj}\right\}^2.
    \end{align*}
    Furthermore, since $r_{ij}\in \{0,1\}$ and $\widehat{C}_l\subset \{\bm{x}_i\}_{i=1}^{n}$, then $\sum_{\bm{x}_i \in \widehat{C}_l} r_{ij} \leq n$, which follows that $\left( n^{-1}\sum_{\bm{x}_i \in \widehat{C}_l} r_{ij}\right)^2 \leq  n^{-1}\sum_{\bm{x}_i \in \widehat{C}_l} r_{ij}$. Then, using the definition of $\bar{\mu}_{lj}$ leads to 
    \begin{align*}
        &\sum_{l=1}^{k} \left( \frac{1}{n}\sum_{\bm{x}_i \in \widehat{C}_l} r_{ij}\right) \cdot \left( \bar{\mu}_{lj} - \hat{\mu}_{lj}\right)^2 \\ 
        &=\frac{1}{n} \sum_{l=1}^{k} \left( \sum_{\bm{x}_i  \in \widehat{C}_l} r_{ij}\right)\cdot \left( \bar{\mu}_{lj}^2 + \hat{\mu}_{lj}^2 -2\bar{\mu}_{lj}\hat{\mu}_{lj} \right) \\
        &=\frac{1}{n} \sum_{l=1}^{k} \left\{ \sum_{i=1}^{n} \mathds{1}(\bm{x}_i \in \widehat{C}_l, r_{ij}=1)\right\}\cdot \left( \hat{\mu}_{lj}^2 - \bar{\mu}_{lj}^2 - 2\bar{\mu}_{lj}\hat{\mu}_{lj} + 2\bar{\mu}_{lj}^2 \right) \\
        &= \frac{1}{n} \sum_{i=1}^{n} \sum_{l=1}^{k} \mathds{1}(\bm{x}_i \in \widehat{C}_l, r_{ij}=1) \hat{\mu}_{lj}^2 - \frac{1}{n}\sum_{i=1}^{n} \sum_{l=1}^{k} \mathds{1}(\bm{x}_i \in \widehat{C}_l, r_{ij}=1) \bar{\mu}_{lj}^2 \\
        &\quad -\frac{2}{n} \sum_{i=1}^{n} \sum_{l=1}^{k} \mathds{1}(\bm{x}_i  \in \widehat{C}_l, r_{ij}=1) \hat{\mu}_{lj} x_{ij} + \frac{2}{n} \sum_{i=1}^{n} \sum_{l=1}^{k} \mathds{1}(\bm{x}_i \in \widehat{C}_l, r_{ij}=1) \bar{\mu}_{lj} x_{ij}\\
        &= \frac{1}{n} \sum_{i=1}^{n} \sum_{l=1}^{k} \mathds{1}(\bm{x}_i \in \widehat{C}_l, r_{ij}=1)\cdot \left\{ \left(x_{ij}-\hat{\mu}_{lj}\right)^2 - \left(x_{ij}-\bar{\mu}_{lj}\right)^2\right\}\\
        &= \underbrace{\frac{1}{n} \sum_{i=1}^{n} \sum_{l=1}^{k} \mathds{1}(\bm{x}_i \in \widehat{C}_l) r_{ij}\left(x_{ij}-\hat{\mu}_{lj}\right)^2}_{\text{(I)}} - \underbrace{\frac{1}{n} \sum_{i=1}^{n} \sum_{l=1}^{k} \mathds{1}(\bm{x}_i \in \widehat{C}_l) r_{ij}\left(x_{ij}-\bar{\mu}_{lj}\right)^2}_{\text{(II)}}. 
    \end{align*}
    It means that 
    \begin{align*}
        \frac{1}{n^2}\left\| \widehat{\bm{U}}^T \left\{\bm{x}_{(j)} \circ \bm{r}_{(j)} - \big(\widehat{\bm{U}}\hat{\bm{\mu}}_{(j)}\big)\circ \bm{r}_{(j)} \right\} \right\|^2  
        \leq \text{(I)} - \text{(II)},
    \end{align*}
    and thus, it suffices to bound the two parts. 
    
    For $\text{(I)}$, for this fixed $j$, let $\widehat{\bm{V}}\in \mathbb{R}^{k\times p}$ be the sparse modification of $\widehat{\bm{M}}$ with its $j$-th column being zero, that is, $\hat{\bm{v}}_{(j)}=\bm{0}_k$ and $\hat{\bm{v}}_{(j')}=\hat{\bm{\mu}}_{(j')}$ for any $j'\neq j$. 
    Because $\widehat{\bm{M}}$ minimizes $\widehat{L}_n(\bm{M})$ and the partition $\widehat{\mathcal{C}}=\{\widehat{C}_1,\dots,\widehat{C}_k\}$ is determined by $\widehat{\bm{M}}$, then we have
    \begin{align*}
        \widehat{L}_n(\widehat{\bm{M}}) &=\sum_{i=1}^{n} \min_{l=1,\dots,k}\| \bm{x}_i\circ \bm{r}_i - \hat{\bm{\mu}}_l \circ \bm{r}_i \|^2 + \lambda \cdot J_1(\widehat{\bm{M}})\\
        &= \sum_{i=1}^{n} \sum_{l=1}^{k} \mathds{1}(\bm{x}_i\in \widehat{C}_l) \| \bm{x}_i\circ \bm{r}_i - \hat{\bm{\mu}}_l \circ \bm{r}_i \|^2 + \lambda \cdot J_1(\widehat{\bm{M}})\\
        &\leq \sum_{i=1}^{n} \sum_{l=1}^{k} \mathds{1}(\bm{x}_i\in \widehat{C}_l) \| \bm{x}_i\circ \bm{r}_i - \hat{\bm{v}}_l \circ \bm{r}_i \|^2 + \lambda \cdot J_1(\widehat{\bm{V}}). 
    \end{align*}
    Moreover, according to the definition of group lasso penalty $J_1(\cdot)$, we have $J_1(\widehat{\bm{V}})\leq J_1(\widehat{\bm{M}})$, as $\widehat{\bm{V}}$ equals to $\widehat{\bm{M}}$ except for the $j$-th column. 
    Thereby, we obtain
    \begin{align*}
        \sum_{i=1}^{n} \sum_{l=1}^{k} \mathds{1}(\bm{x}_i\in \widehat{C}_l) \| \bm{x}_i\circ \bm{r}_i - \hat{\bm{\mu}}_l \circ \bm{r}_i \|^2 
        \leq 
        \sum_{i=1}^{n} \sum_{l=1}^{k} \mathds{1}(\bm{x}_i\in \widehat{C}_l) \| \bm{x}_i\circ \bm{r}_i - \hat{\bm{v}}_l \circ \bm{r}_i \|^2. 
    \end{align*}
    Further denote the $(l,j)$-th entry of 
    $\widehat{\bm{V}}$ by $\hat{v}_{lj}$, then we have 
    \begin{align*}
        \sum_{i=1}^{n} \sum_{l=1}^{k} \mathds{1}(\bm{x}_i\in \widehat{C}_l) (x_{ij}r_{ij} - \hat{\mu}_{lj}r_{ij})^2 
        &\leq \sum_{i=1}^{n} \sum_{l=1}^{k} \mathds{1}(\bm{x}_i\in \widehat{C}_l) (x_{ij}r_{ij} - \hat{v}_{lj}r_{ij})^2 \\
        &= \sum_{i=1}^{n} \sum_{l=1}^{k} \mathds{1}(\bm{x}_i\in \widehat{C}_l) (x_{ij}r_{ij} - 0)^2\\
        &=\sum_{i=1}^{n}r_{ij}x_{ij}^2 =n\hat{q}_j\bar{\sigma}_j^2. 
    \end{align*}
    This implies
    \begin{align}
        \text{(I)}\leq \hat{q}_j\bar{\sigma}_j^2. 
    \label{eq_proof_proposition_3}
    \end{align}
    For $\text{(II)}$, we have 
    \begin{align*}
        \text{(II)}=\frac{1}{n}\sum_{i=1}^{n} \sum_{l=1}^{k} \mathds{1}(\bm{x}_i \in \widehat{C}_l) r_{ij}\left(x_{ij}-\bar{\mu}_{lj}\right)^2 
        \geq \frac{1}{n}\sum_{i=1}^{n}\min_{l=1,\dots,k} r_{ij}(x_{ij}-\bar{\mu}_{lj})^2
    \end{align*} 
    The right hand is actually $Q_j(\bar{\bm{\mu}}_{(j)})$ defined in Section~2.3 of the main paper, which must be no less than than the minima of function $Q_j$, that is $\widehat{Q}_j$. This implies
    \begin{align}
        \text{(II)} \geq \widehat{Q}_j.  
    \label{eq_proof_proposition_4}
    \end{align}
    Combining Eq.~(\ref{eq_proof_proposition_3}) and Eq.~(\ref{eq_proof_proposition_4}), we obtain 
    \begin{align*}
       \frac{1}{n^2}\left\| \widehat{\bm{U}}^T \left\{\bm{x}_{(j)} \circ \bm{r}_{(j)} - \big(\widehat{\bm{U}}\hat{\bm{\mu}}_{(j)}\big)\circ \bm{r}_{(j)} \right\} \right\|^2   &\leq  \hat{q}_j \bar{\sigma}_j^2 - \widehat{Q}_j.
    \end{align*}
    By using Eq.~(\ref{eq_proof_proposition_2}), we have for this given $j$, it must hold that $(\lambda w_j)/(2n)\leq \sqrt{\hat{q}_j \bar{\sigma}_j^2 - \widehat{Q}_j}$. 
    Consequently, if a feature $j=1,\dots,p$ satisfies 
    \begin{align*}
        \frac{\lambda w_j}{2n} > \sqrt{\hat{q}_j \bar{\sigma}_j^2 - \widehat{Q}_j},
    \end{align*}
    then there must be $\hat{\bm{\mu}}_{(j)}=(0,0,\dots,0)^T$, which completes the proof of (b).

\section{More details and results of simulations}
In this section, we provide more details and results of simulations in Section~3 of the main paper. 

\subsection{Supplementary for Section~3.1}

In Section~3 of the main paper, we consider four types of procedures for generating missingness. For MAR and MNAR1 mechanisms, different parameters used to meet the total proportion of missingness are summarized in Table~\ref{rkpod_table_generate_missing_para}. 

\begin{table*}[htbp]
\centering
\caption{\normalfont \centering Different parameters to meet total missing proportion}
\label{rkpod_table_generate_missing_para}
\begin{adjustbox}{center}
\resizebox{0.5\columnwidth}{!}{
\begin{tabular}{@{}lccccc@{}}
\toprule
 \multirow{2}{*}{Dataset} & \multirow{2}{*}{\makecell{Missing\\Proportion}} &  \multicolumn{2}{c}{MAR} & \multicolumn{2}{c}{MNAR1} \\
 \cmidrule(r){3-4} \cmidrule(r){5-6}
 & & $\psi_1$ & $\psi_2$  & $\phi_1$ & $\phi_2$ \\
\midrule
$p=10$ & 10\% & 1.80 & 3.0 & 1.5 & 3.0 \\
 & 20\% & 0.55 & 3.0 & 0.6 & 3.0 \\
 & 30\% & 0.25 & 3.0 & 0.3 & 3.0 \\[8pt]
\makecell{$p=100$, $a=0.8$} & 10\% & 2.0 & 2.0 & 2.5 & 2.0 \\
 & 20\% & 0.8 & 2.0 & 0.9 & 2.0 \\
 & 30\% & 0.4 & 2.0 & 0.45 & 2.0 \\[8pt]
\makecell{$p=100$, $a=1$} & 10\% & 2.5 & 2.0 & 2.5 & 2.0 \\
 & 20\% & 0.9 & 2.0 & 0.9 & 2.0 \\
 & 30\% & 0.45 & 2.0 & 0.45 & 2.0 \\
 \bottomrule
\end{tabular}}
\end{adjustbox}
\end{table*}

\subsection{Supplementary for Section~3.2}

We compare the random initialization with non-random initialization for $l_0$ type of proposed method. 
Specifically, we consider the \textit{sparse initialization}, which is also used in \cite{raymaekers2022regularized}. First, based on the estimated cluster centers of $k$-POD clustering, we rank all $p$ features in a decreasing order by the $l_2$ norms of $k$-POD estimator in each feature. Then by retaining only the leading 1\%, 2\%, 5\%, 10\%, 15\%, 20\%, 30\%, 40\%, 50\%, 100\% features, we can get 10 sparse versions of $k$-POD estimator. These 10 sparse estimators would serve as 10 initialization points for the proposed method. 
For the random initialization, we use 100 initialization points. 

Table~\ref{rkpod_table_syn_exp_sparse_ini} illustrates the comparison results between random initialization and sparse initialization. 
For the random initialization, since we consider two strategies, only the best results are reported. For the sparse initialization, when $p=10$, we use the sequence $\{ 10\%, 20\%, \dots, 100\% \}$ to generate the 10 sparse initialization points. Moreover, for the dataset with $p=100$, the setting of $d=10$ and $a=0.8$ is used. 

It can be seen that the sparse initialization generally provides comparable results, especially in the case of $p=100$, while it only uses 10 initial points and needs less computational time. 
%when $p=100$, the random initialization and sparse initialization are generally comparable, while the random initialization outperforms when $p=10$. Moreover, when the missing proportion is not large, the random initialization performs better/comparable, while under a large proportion of missingness, the sparse initialization relatively improves the performance. 
Therefore, the sparse initialization can be used as a faster substitute for random initialization when the number of features is large, as it requires fewer initialization points. 
%In addition, as noted in the main paper, despite of the improvement, the $l_0$ type of proposed method still performs worse than group lasso type in general. 

\begin{table*}[htbp]
\centering
\caption{\normalfont \centering Comparison of random initialization and sparse initialization for $l_0$ type of proposed method}
\label{rkpod_table_syn_exp_sparse_ini}
\begin{adjustbox}{center}
\resizebox{0.9\columnwidth}{!}{
\begin{tabular}{@{}lcccccc@{}}
\toprule
\multirow{2}{*}{Dataset} & \multirow{2}{*}{\makecell{Missing\\mechanism}} &  \multirow{2}{*}{\makecell{Missing\\proportion}}  &  \multicolumn{2}{c}{MSE} & \multicolumn{2}{c}{CER} \\
\cmidrule(r){4-5}\cmidrule(r){6-7} 
& & &  random & sparse & random & sparse  \\
\midrule
$p=10$ & MCAR & 10\% & \bf{0.025 (0.01)} & 0.110 (0.03) & \bf{0.123 (0.01)} & 0.124 (0.01) \\
 &  & 20\% & \bf{0.079 (0.03)} & 0.296 (0.07) & \bf{0.186 (0.01)} & 0.190 (0.00)  \\
 &   & 30\% & \bf{0.097 (0.00)} & 0.557 (0.10) & \bf{0.241 (0.01)} & 0.242 (0.01)  \\
 &  & 40\% & \bf{1.139 (2.46)} & 2.406 (5.38) & \bf{0.285 (0.01)} & 0.297 (0.03) \\
 &  & 50\% & 22.601 (6.93) & \bf{4.466 (7.09)} & \bf{0.345 (0.01)} & 0.353 (0.04) \\[8pt]
$p=100$ & MCAR & 10\% & 0.134 (0.02) & \bf{0.131 (0.02)} & 0.089 (0.01) & \bf{0.086 (0.00)}\\
 &  & 20\% & 0.153 (0.03) & \bf{0.149 (0.03)} & 0.113 (0.00) & \bf{0.108 (0.01)}\\
 &  & 30\% & 7.948 (5.29) & \bf{2.285 (3.65)} & 0.245 (0.04) & \bf{0.177 (0.04)}\\
 &  & 40\% & 26.469 (5.00) & \bf{18.428 (8.16)} & 0.375 (0.03) & \bf{0.303 (0.05)}\\
 &  & 50\% & 36.284 (2.77) & \bf{26.843 (4.60)} & 0.376 (0.01) & \bf{0.329 (0.01)}\\[8pt]
Usoskin & MNAR & 73\% & - & - & 0.167 (0.03) & \bf{0.133 (0.04)} \\
\bottomrule
\end{tabular}}
\end{adjustbox}
\end{table*}

\subsection{Supplementary for Section~3.3}

\begin{figure}[htpb]
    \setlength{\tempwidth}{.22\linewidth}
\settoheight{\tempheight}{\includegraphics[width=\tempwidth]{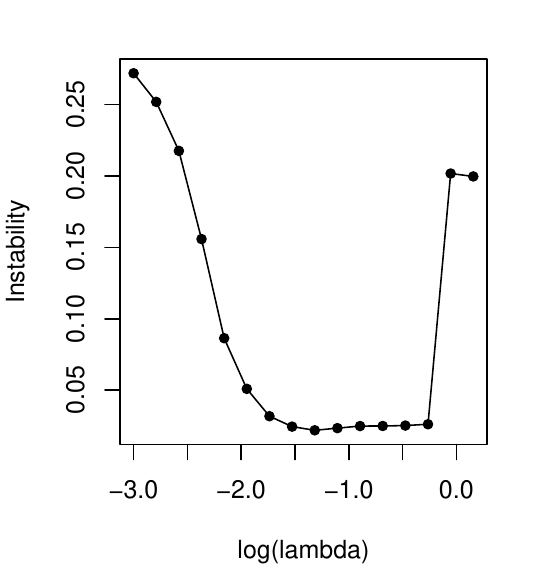}}%
\centering
\rowname{group lasso}
    \includegraphics[width=0.23\textwidth]{fig_tuning_g_p100mcar_instability.pdf}
    \includegraphics[width=0.23\textwidth]{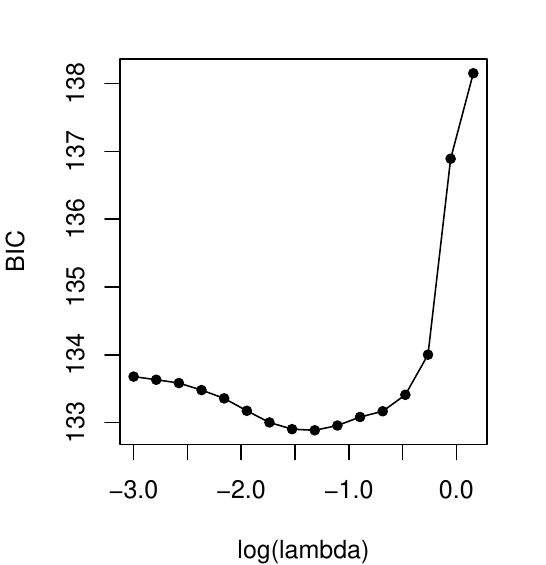}
    \includegraphics[width=0.23\textwidth]{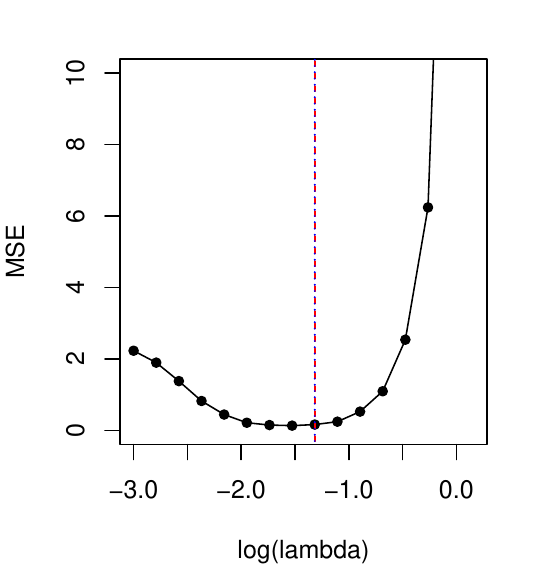}
    \includegraphics[width=0.23\textwidth]{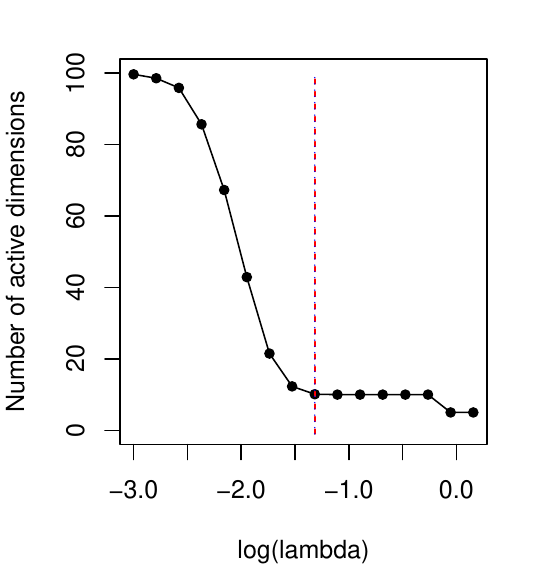}\\
\rowname{$l_0$}
    \includegraphics[width=0.23\textwidth]{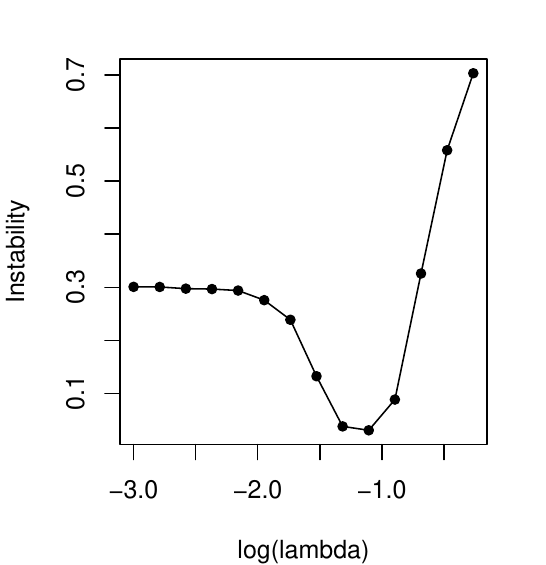}
    \includegraphics[width=0.23\textwidth]{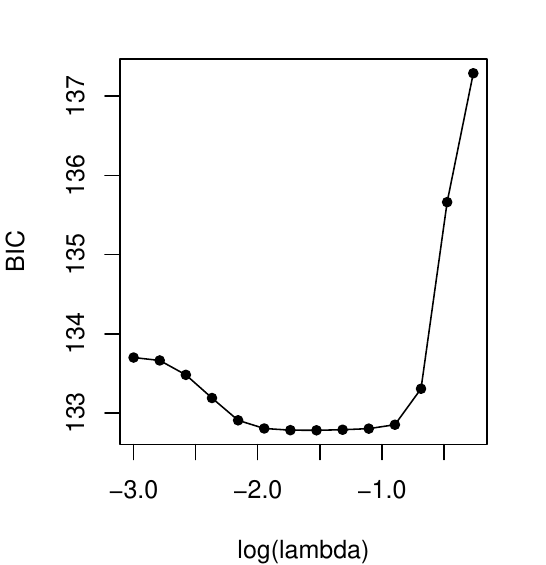}
    \includegraphics[width=0.23\textwidth]{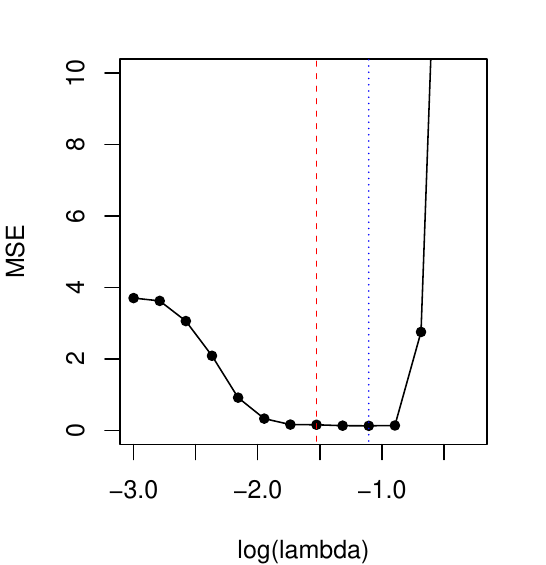}
    \includegraphics[width=0.23\textwidth]{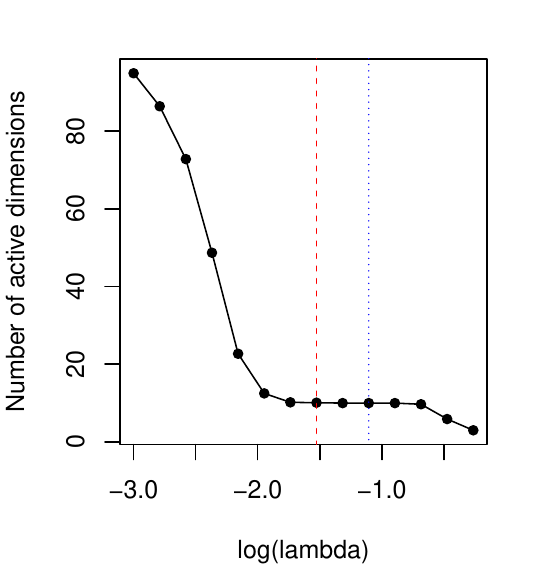}
    \caption{\normalfont\doublespacing Comparison of instability and BIC criteria for selecting $\lambda$. The top and bottom rows are for group lasso and $l_0$ types of proposed method, respectively. The red dashed lines denote the choice of BIC, while the blue dotted lines denote instability.}
    \label{fig_tuning}
\end{figure}

We first analyze the sensitivity of the regularization parameter based on the case of $p=100$. Figure~\ref{fig_tuning} illustrates the results of instability and BIC under MCAR mechanism with missing proportion 30\%, where the reported values are the average of 10 repetitions. It can be seen that a suitable $\lambda$ can reduce the value of MSE and provide a reasonable set of features that contribute to clustering. Moreover, the instability is more sensitive to $\lambda$ than BIC. 

\begin{table*}[!h]
\centering
\caption{\normalfont MSE (number of active features in brackets) of proposed method using different criteria for selecting $\lambda$ ($p=10$)}
\label{table_tuning_p10}
\begin{adjustbox}{center}
\resizebox{0.65\columnwidth}{!}{
\begin{tabular}{@{}cccccc@{}}
\toprule
 \multirow{2}{*}{\makecell{Missing\\mechanism}} &  \multirow{2}{*}{\makecell{Missing\\proportion}}  &  \multicolumn{2}{c}{\makecell{group lasso}} & \multicolumn{2}{c}{\makecell{$l_0$}} \\
\cmidrule(r){3-4}\cmidrule(r){5-6} 
 & &  Instability & BIC & Instability & BIC  \\
\midrule
  MCAR & 10\%  & \textbf{0.118 (3)} & 1.508 (10) & \textbf{0.038 (2)}  & 1.324 (7) \\
 & 20\% & \textbf{0.872 (6)} & 2.767 (10) & \textbf{0.079 (2)} & 4.677 (9) \\
 & 30\% & \textbf{1.853 (7)} & 8.467 (10) & \textbf{0.097 (2)} & 16.547 (9) \\
 & 40\% & \textbf{3.160 (7)} & 26.199 (10) & \textbf{1.139 (2)} & 24.100 (8) \\
 & 50\% & \textbf{4.732 (3)} & 30.416 (10) & \textbf{22.601 (4)} & 31.611 (9) \\[8pt]
 MAR & 10\% & \textbf{0.364 (3)} & 1.764 (10) & \textbf{0.203 (2)} & 1.335 (5) \\
 & 20\% & \textbf{0.298 (2)} & 5.501 (10) & \textbf{0.117 (2)} & 5.022 (8) \\
 & 30\% & \textbf{0.484 (2)} & 2.861 (8) & \textbf{0.115 (2)} & 4.487 (5) \\[8pt]
 MNAR1 & 10\% & \textbf{1.151 (5)} & 5.100 (10) & \textbf{0.462 (2)} & 5.576 (10) \\
 & 20\% & \textbf{3.932 (2)} & 12.476 (10) & \textbf{0.283 (2)} & 15.486 (10) \\
 & 30\% & \textbf{2.301 (4)} & 21.715 (10) & \textbf{0.210 (2)} & 21.032 (10) \\[8pt]
 MNAR2 & 10\% & \textbf{2.006 (3)} & 6.322 (10) & \textbf{0.691 (2)} & 6.384 (10) \\
 & 20\% & \textbf{4.901 (10)} & 21.431 (10) & \textbf{2.346 (2)} & 21.598 (10) \\
 & 30\% & \textbf{24.829 (3)} & 45.131 (10) & \textbf{9.733 (2)} & 47.213 (10)\\
\bottomrule
\end{tabular}}
\end{adjustbox}
\end{table*}

We then report the comparison of the instability and BIC criteria for selecting $\lambda$ in the case of $p=10$ in Table~\ref{table_tuning_p10}. 
It can be seen that the instability is stable in various settings, which is similar to the case of $p=100$, while BIC almost fails. The main reason is that there are only two relevant features in this case, the decrease of active features has more influence on increasing the loss than decreasing the degree of freedom. 
%Since the number of true informative features in this case is $d=2$, , which makes the BIC criterion almost fail in this case. However, the instability criterion follows the spirit of cross-validation and is based on the clustering alignment, which shows more stable performance in various settings. 

\subsection{Supplementary for Section~3.4}

In this section, we compare the performance of the proposed method with other methods in the case $p=100$ and $a=1$, which is an easier clustering task with more separable cluster centers. 
Table~\ref{rkpod_table_syn_exp_mse_p100a1}, Table~\ref{rkpod_table_syn_exp_cer_p100a1} and Table~\ref{rkpod_table_syn_exp_predictcer_p100a1} illustrate the comparison of MSE, CER and predictive CER, respectively. 

\begin{table*}[htbp]
\centering
\caption{\normalfont MSE (standard deviations in brackets) of different methods ($p=100$ and $a=1$)}
\label{rkpod_table_syn_exp_mse_p100a1}
\begin{adjustbox}{center}
\resizebox{0.8\columnwidth}{!}{
\begin{tabular}{@{}cccccccc@{}}
\toprule
\makecell{Missing\\mechanism}  &  \makecell{Missing\\proportion}   &  Mice & $k$-POD & \makecell{Reg. $k$-POD\\(group lasso)} & \makecell{Reg. $k$-POD\\($l_0$)} \\
\midrule
 MCAR & 10\%  & 1.286 (0.09)  & 1.430 (0.09) & 0.126 (0.02)  & \bf{0.109 (0.02)} \\
  & 20\%  & 1.462 (0.10)  & 1.870 (0.14) & 0.206 (0.04)  & \bf{0.156 (0.03)} \\
  & 30\%  & 1.788 (0.11)  & 3.063 (0.49) & 0.407 (0.10)  & \bf{0.280 (0.08)} \\
  & 40\%  & 2.272 (0.14)  & 19.121 (2.43) & \bf{1.918 (0.30)}  & 2.675 (1.60) \\
 & 50\%  & \bf{3.267 (0.23)}  & 36.512 (3.54) & 5.546 (2.91)  & 25.073 (4.03) \\[8pt]
 MAR & 10\%  & 1.338 (0.13) & 1.516 (0.14) & 0.150 (0.04) & \bf{0.131 (0.03)} \\
  & 20\%  & 1.517 (0.11) & 1.842 (0.16) & 0.140 (0.03) & \bf{0.126 (0.02)} \\
  & 30\%  & 1.771 (0.14) & 3.117 (0.73) & 0.204 (0.05) & \bf{0.164 (0.03)} \\[8pt]
 MNAR1 & 10\%  & 25.983 (0.58)  & 26.039 (0.52) & 3.073 (0.16)  & \bf{1.873 (0.13)} \\
  & 20\%  &  32.579 (0.70)  & 33.187 (0.73) & 3.109 (0.17)  & \bf{1.738 (0.33)} \\
  & 30\%  &  25.673 (0.56)  & 27.698 (0.83) & 2.139 (0.20)  & \bf{1.324 (0.37)} \\[8pt]
 MNAR2 & 10\%  & 31.768 (0.62)  & 31.161 (0.61) & 4.696 (0.18)  & \bf{2.693 (0.22)} \\
  & 20\%  & 101.579 (0.97) & 99.327 (1.280) & \bf{40.286 (0.04)}  & 99.507 (1.31) \\
\bottomrule
\end{tabular}}
\end{adjustbox}
\end{table*}

\begin{table*}[htbp]
\centering
\caption{\normalfont CER (standard deviations in brackets) of different methods ($p=100$ and $a=1$)}
\label{rkpod_table_syn_exp_cer_p100a1}
\begin{adjustbox}{center}
\resizebox{0.8\columnwidth}{!}{
\begin{tabular}{@{}cccccccc@{}}
\toprule
\makecell{Missing\\mechanism}  &  \makecell{Missing\\proportion}  &   Mice & $k$-POD & \makecell{Reg. $k$-POD\\(group lasso)} & \makecell{Reg. $k$-POD\\($l_0$)} \\
\midrule
 MCAR & 10\%  & \bf{0.044 (0.00)}  & 0.046 (0.00) & 0.050 (0.00)  & 0.049 (0.00) \\
  & 20\%  & \bf{0.064 (0.00)}  & 0.072 (0.00) & 0.091 (0.00)  & 0.092 (0.00) \\
  & 30\%  & \bf{0.092 (0.00)}  & 0.124 (0.02) & 0.147 (0.00)  & 0.147 (0.00) \\
  & 40\% & \bf{0.126 (0.00)}  & 0.287 (0.02) & 0.186 (0.00)  & 0.236 (0.02) \\
 & 50\% & \bf{0.170 (0.01)}  & 0.364 (0.01) & 0.259 (0.01)  & 0.356 (0.02) \\[8pt]
 MAR & 10\%  & 0.052 (0.00) & 0.056 (0.01) & \bf{0.047 (0.00)} & 0.051 (0.00) \\
  & 20\%  & 0.063 (0.00) & 0.074 (0.01) & \bf{0.063 (0.00)} & 0.064 (0.01) \\
  & 30\%  & 0.086 (0.01) & 0.127 (0.02) & 0.088 (0.00) & \bf{0.086 (0.01)} \\[8pt]
 MNAR1 & 10\%  & 0.063 (0.00)  & 0.058 (0.00) & \bf{0.053 (0.00)}  & 0.056 (0.00) \\
  & 20\%  & 0.079 (0.00)  & 0.082 (0.01) & \bf{0.079 (0.01)}  & 0.091 (0.01) \\
  & 30\%  & \bf{0.102 (0.00)}  & 0.150 (0.02) & 0.139 (0.01)  & 0.152 (0.01) \\[8pt]
 MNAR1 & 10\%  & 0.064 (0.00)  & 0.056 (0.00) & \bf{0.051 (0.01)}  & 0.056 (0.00) \\
  & 20\% & 0.124 (0.00) & \bf{0.117 (0.01)}  & 0.746 (0.00) & 0.149 (0.01) \\
\bottomrule
\end{tabular}}
\end{adjustbox}
\end{table*}

\begin{table*}[htbp]
\centering
\caption{\normalfont Predictive CER (standard deviations in brackets) of different methods ($p=100$ and $a=1$)}
\label{rkpod_table_syn_exp_predictcer_p100a1}
\begin{adjustbox}{center}
\resizebox{0.8\columnwidth}{!}{
\begin{tabular}{@{}cccccccc@{}}
\toprule
\makecell{Missing\\mechanism}  &  \makecell{Missing\\proportion}  &  Mice & $k$-POD & \makecell{Reg. $k$-POD\\(group lasso)} & \makecell{Reg. $k$-POD\\($l_0$)} \\
\midrule
 MCAR & 10\%  & 0.030 (0.01)  & 0.033 (0.01) & \bf{0.027 (0.01)}  & 0.028 (0.01) \\
  & 20\%  & 0.030 (0.01)  & 0.035 (0.01) & \bf{0.026 (0.01)}  & 0.027 (0.01) \\
  & 30\%  & 0.032 (0.01)  & 0.043 (0.01) & 0.029 (0.01)  & \bf{0.028 (0.01)} \\
  & 40\%  & 0.032 (0.01)  & 0.190 (0.02) & \bf{0.027 (0.01)}  & 0.043 (0.02) \\
 & 50\%  & \bf{0.036 (0.01)}  & 0.280 (0.02) & 0.043 (0.03)  & 0.234 (0.04) \\[8pt]
 MAR & 10\%  & 0.030 (0.01) & 0.031 (0.01) & 0.024 (0.01) & \bf{0.023 (0.01)} \\
  & 20\%  & 0.028 (0.01) & 0.036 (0.01) & 0.028 (0.01) & \bf{0.028 (0.01)} \\
  & 30\%  & 0.032 (0.01) & 0.042 (0.01) & 0.025 (0.01) & \bf{0.024 (0.01)} \\[8pt]
 MNAR1 & 10\%  & 0.039 (0.01)  & 0.042 (0.01) & \bf{0.028 (0.01)}  & 0.029 (0.01) \\
  & 20\%  & 0.040 (0.01)  & 0.045 (0.01) & \bf{0.034 (0.01)}  & 0.035 (0.01) \\
  & 30\%  & 0.036 (0.01)  & 0.052 (0.01) & \bf{0.034 (0.01)}  & 0.036 (0.01) \\[8pt]
 MNAR2 & 10\%  & 0.046 (0.01)  & 0.046 (0.01) & \bf{0.033 (0.01)}  & 0.037 (0.01) \\
  & 20\%  & \bf{0.091 (0.01)} & 0.096 (0.02)  & 0.303 (0.02) & 0.099 (0.02) \\
\bottomrule
\end{tabular}}
\end{adjustbox}
\end{table*}

%%%%%%%%%%%%%%%%%%%%%%%%%%%%%%%%%%%%%%%%%%%%%%%%%%%%%%%%%%%%%%%%%%%%%%%%%%%%%%%%%%%%%%%%%%%%%%%%%%%%%%%%%%%%%%%%%%%%%%%%%%%%%%%%%%%%%%%%%%%%%%%%%%%%%%%%%%%%%%%%%%%%%%%%%%%%%%%%%%%%%%%%%%%%%%%%%%%%%%%%%%%%%%%%%%%%%%%%%%%%%%%%%%%%%%%%%%%%%%%%%%%%%%%%%%%%%%%%%%%%%%%%%%%%%%%%%%%%%%%%%%%%%%%%%%%%%%%%%%%%%%%%%%%%%%%%%%%%%%%%%%%%%%%%%%%%%%%%%%%%%%

%USE THE BELOW OPTIONS IN CASE YOU NEED AUTHOR YEAR FORMAT.
\bibliographystyle{abbrvnat}
\bibliography{reference}